\documentclass[12pt, draftclsnofoot, onecolumn]{IEEEtran}
\usepackage{epsfig,graphicx,subfigure,psfrag,amsmath,cases}
\usepackage{latexsym,amsmath,amssymb,epsfig,subfigure,algorithm,mathtools}
\usepackage{algorithmic}
\usepackage{color}
\usepackage{url}
\usepackage{scrtime}
\usepackage{cite}
\usepackage{epstopdf} 
\usepackage{subfigure}
\usepackage{bbding}
\usepackage{multicol}
\usepackage{makecell}

\author{{Zhiqiang Wei,~\IEEEmembership{Member,~IEEE,} Dongfang Xu,~\IEEEmembership{Member,~IEEE,} Shuangyang Li,~\IEEEmembership{Member,~IEEE,} Shenghui Song,~\IEEEmembership{Senior Member,~IEEE,} Derrick Wing Kwan Ng,~\IEEEmembership{Fellow,~IEEE}, and Giuseppe Caire,~\IEEEmembership{Fellow,~IEEE}}\\
\thanks{Z. Wei is with the School of Mathematics and Statistics, Xi'an Jiaotong University, Xi'an 710049, China (e-mail: zhiqiang.wei@xjtu.edu.cn);}
\thanks{D. Xu and S. Song are with the Division of Integrative Systems and Design, The Hong Kong University of Science and Technology, Clear Water Bay, 999077, Hong Kong, China (e-mail: eedxu@ust.hk, eeshsong@ust.hk);}
\thanks{S. Li and G. Caire are with the Faculty of Electrical Engineering and Computer Science,  Technische Universit\"{a}t Berlin, Berlin 10587, Germany, (e-mail: shuangyang.li@tu-berlin.de, caire@tu-berlin.de);}
\thanks{D. W. K.~Ng is with the School of Electrical Engineering and Telecommunications, University of New South Wales, Sydney, NSW
	2052, Australia (e-mail: w.k.ng@unsw.edu.au).}}

\title{Resource Allocation Design for Next-Generation Multiple Access: A Tutorial Overview}

\newtheorem{T-Prob}{Transformed Problem}

\DeclareMathOperator{\Tr}{Tr}

\DeclareMathOperator{\maxo}{maximize}
\DeclareMathOperator{\mino}{minimize}
\DeclareMathOperator{\diag}{\mathrm{diag}}

\newcommand{\abs}[1]{\lvert#1\rvert}

\allowdisplaybreaks

\begin{document}
\maketitle
\begin{abstract}
Multiple access is the cornerstone technology for each generation of wireless cellular networks, which fundamentally determines the method of radio resource sharing and significantly influences both the system performance and transceiver complexity. 
Meanwhile, resource allocation design plays a crucial role in multiple access, as it can manage both encompassing radio resources and interference, and it is critical for providing high-speed and reliable communication services to multiple users.
Given that resource allocation design is intrinsically scenario-specific and the optimization tools for resource allocation design are typically varied, in this paper, we present a comprehensive tutorial overview for junior researchers in this field, aiming to offer a foundational guide for resource allocation design in the context of next-generation multiple access (NGMA).
Our discussion spans a broad range of fundamental topics: from typical system models, through intriguing problem formulation in resource allocation design, to the exploration of various potential optimization solution methodologies. 
 Initially, we identify three types of channels in future wireless cellular networks over which NGMA will be implemented, namely: natural channels, reconfigurable channels, and functional channels.
Natural channels are traditional uplink and downlink communication channels; reconfigurable channels are defined as channels that can be proactively reshaped via emerging platforms or techniques, such as intelligent reflecting surface (IRS), unmanned aerial vehicle (UAV), and movable/fluid antenna (M/FA); and functional channels support not only communication but also other functionalities simultaneously, with typical examples including integrated sensing and communication (ISAC) and joint computing and communication (JCAC) channels.
Then, we introduce NGMA models applicable to these three types of channels that cover most of the practical communication scenarios of future wireless communications.
Subsequently, we articulate the key optimization technical challenges inherent in the resource allocation design for NGMA, categorizing them into rate-oriented, power-oriented, and reliability-oriented resource allocation designs.
The corresponding optimization approaches for solving the formulated resource allocation design problems are then presented.
Finally, simulation results are presented and discussed to elucidate the practical implications and insights derived from resource allocation designs in NGMA.
\end{abstract}


\section{Introduction}
The sixth-generation (6G) wireless networks are envisioned to provide massive, immersive, and reliable communications, which are expected to revolutionize our daily lives.
To this end, the IMT-2030 framework has highlighted three groundbreaking usage scenarios: ubiquitous connectivity, integrated sensing and communication, and integrated artificial intelligence (AI) and communication \cite{IMT-2030}. These scenarios lay the groundwork for a sophisticated and intelligent digital world, serving as a pivotal data foundation.
Moreover, these innovative services open doors to novel business models and applications, simultaneously introducing unprecedented challenges in the evolution of 6G wireless communication technologies. 
 In particular, massive communication is expected to accommodate an extraordinary connectivity density ranging from  $10^6$ to $10^7$ $\mbox{devices/km}^2$ \cite{IMT-2030}, encompassing a variety of ubiquitous wideband and Internet-of-Things (IoT) devices across a broad spectrum of coverage and mobility.
Beyond these capabilities, immersive communication demands stringent requirements on peak and experienced data rates, latency, and connectivity capacity, which are crucial for supporting a wide variety of applications in the entertainment, education, and manufacturing sectors.

Radio access technology plays a crucial role in sharing limited radio resources among multiple users in a controllable manner to achieve satisfactory quality-of-service (QoS).
This is vital for providing high-speed and reliable services to massive wideband users and IoT devices.
Multiple access technology is the most fundamental aspect of the physical layer and defines each generation of wireless cellular networks.
Dated back to the early 80s, in the first generation (1G) cellular networks, frequency-division multiple access (FDMA) was adopted which assigns different frequency bands to various users, separated by guard bands to mitigate any potential inter-user interference (IUI).
 In the 1990s, the second-generation (2G) wireless networks adopted time-division multiple access (TDMA), allocating the system bandwidth based on a time-sharing scheme.
Besides, code-division multiple access (CDMA)\cite{Verdu1999,YangCDMA}, utilized in both 2G and the third-generation (3G), enabled users to transmit data across the entire bandwidth simultaneously by exploiting the unique properties of code sequences to distinguish among users.
As time evolves, in the fourth-generation (4G) and the fifth-generation (5G), orthogonal frequency division multiplexing (OFDM) waveform \cite{Wong1999,Strohmer2003} has been predominantly adopted owing to its higher spectral efficiency than FDMA and the low implementation complexity of frequency domain equalization (FDE).
Also, orthogonal frequency-division multiple access (OFDMA) \cite{Kwan_AF_2010,DerrickEEOFDMA,DerrickEESWIPT,DerrickLimitedBackhaul,ZhiqiangOFDMA} further enhances this by allowing different users to occupy different subcarriers, thus providing high flexibility in resource allocation (RA) design. 
Furthermore, by equipping multiple antennas at transceivers, spatial division multiple access (SDMA) \cite{yang2017noma,Ngo2011} has been recognized as an efficient strategy for supporting multi-user communication, which has gained increased popularity with the advancement of multiple-input multiple-output (MIMO) technology in recent years. 

From an information-theoretical perspective, the number of orthogonal system resources can be interpreted as the system's degrees-of-freedom (DoF)\cite{Tse2005},  enabling the transmission of multiple data streams across different DoFs without mutual interference.
Traditional multiple access protocols, whether in time, frequency, spatial, or code domains, are categorized as orthogonal multiple access (OMA) schemes.
In particular, these schemes are designed to allocate distinct system DoFs to different users or data streams exclusively, primarily to minimize interference, thereby facilitating interference-free transmission and simplifying channel equalization at the receiver's end.
However,  despite its advantages, OMA demands a certain level of system signaling overhead to maintain this orthogonality, such as time synchronization overhead for TDMA and the required accurate channel state information (CSI) at the transmitter side for SDMA.
Moreover, OMA struggles to support massive communications and ubiquitous connectivity due to the limited system DoFs.
In contrast, NOMA serves as a more versatile generalization of OMA, with the former introducing a manageable level of IUI by permitting multiple users to share the same system DoF \cite{Ding2015b,LiuSWIPT,Lei2016NOMA,9679390}. Specifically, NOMA leverages advanced multi-user detection (MUD) techniques to distinguish users at the receiver\cite{MUDCDMA,Hanzo2003},  representing a more flexible approach to resource allocation than OMA, which strictly avoids DoF sharing among users\cite{PerGainWei}.
Through DoF sharing, NOMA can accommodate more users than its orthogonal counterpart with limited system resources, i.e., time-frequency resource block, spatial DoF, or radio frequency (RF) chains \cite{XuMassiveMIMONOMA,DingSignalAlignment,wei2018multibeam,WeiBeamWidthControl}.
Moreover, from the optimization point of view, by relaxing the orthogonality constraint of OMA, NOMA enables a more flexible management of radio resources and offers an efficient manner to improve spectral efficiency via resource allocation design\cite{Wei2017}.
Typical NOMA schemes include power-domain and code-domain NOMA, which characterize the approach for DoF sharing. 
Specifically, power domain NOMA exploits the discrepancy in users' transmit/receive power levels for multi-user multiplexing, combined with superposition coding at the transmitter and successive interference cancellation (SIC) at the receivers to harness IUI\cite{PerGainWei}.
In contrast, code-domain NOMA shares the system DoF among multiple users via non-orthogonal coding/spreading, facilitating user separation at the receiver.
This includes low-density spreading (LDS)\cite{Razavi2012}, sparse code multiple access (SCMA) \cite{WeijieSCMA,WeijieSCMAII}, and pattern division multiple access (PDMA)\cite{ShanzhiPDMA}, etc. 
On the other hand, rate splitting multiple access (RSMA) \cite{BrunoJSAC,YijieRSMA}, a generalization of power/code domain NOMA, partially decodes interference and partially treats the remainder as noise, based on rate splitting principles.
Despite the gradual standardization of NOMA and RSMA in 5G, their related research has attracted significant attention and efforts in academia\cite{Access2015}.
Indeed, the notion of NOMA and RSMA might serve as a foundation for addressing the unique challenges of 6G, including massive communication and ubiquitous connectivity.

As we navigate the evolving service demands of 6G and the technological advancements in multiple access from 1G through 5G, a critical question emerges:  ``what will the NGMA look like \cite{YuanweiNGMA}?"
One potential new feature of NGMA could be innovative multiple access schemes that can enhance the connectivity density, spectral efficiency, and/or energy efficiency.
In particular, NGMA may incorporate, solely or in combination, existing OMA, NOMA, and RSMA to address the enormous challenges and to meet the heterogeneous service demands of next-generation wireless networks. 
Secondly, NGMA is anticipated to exploit cutting-edge infrastructures and enabling technologies, such as intelligent reflecting surface (IRS) \cite{ZhiqiangOFDMA,XianghaoJSAC,9183907}, UAV-assisted communication \cite{8663615,8644086}, and movable/fluid antenna (M/FA) systems\cite{LipengMovableAntenna,10318134}.
These technologies could offer additional design DoF via adaptive channel reconfiguration and thus improve system performance.
Thirdly, NGMA aims not only to satisfy the stringent communication requirements but also to realize versatile functionalities, including sensing \cite{Fan2020}, computing \cite{Yuyi2017}, and learning \cite{YandongCST}.
In light of these considerations, this paper introduces a comprehensive framework and resource allocation strategy for NGMA, addressing its application in natural, reconfigurable, and functional channels, which cover a broad spectrum of practical scenarios.

In practice, to unleash the potential of effective multiple access schemes, meticulous resource allocation design is of utmost importance.
It can make the best use of limited communication resources based on the information available at the resource allocator, such as the CSI, QoS, power budget, and the number of antennas/subcarriers, to improve overall system performance.
The resource allocation designs for OMA, NOMA, and RSMA schemes have been extensively studied\cite{Kwan_AF_2010,DerrickEEOFDMA,DerrickEESWIPT,DerrickLimitedBackhaul,ZhiqiangOFDMA,XuMassiveMIMONOMA,DingSignalAlignment,wei2018multibeam,WeiBeamWidthControl,Shuangyang_FTN_NOMA}.
While it is not our goal to exhaustively list all the existing resource allocation designs for OMA, NOMA, and RSMA detailed in the literature, our focus is to present a holistic overview of the predominant models and approaches in resource allocation design for NGMA, addressing its application in natural channels, reconfigurable channels, and functional channels.
In natural channels, the primary resource allocation for NGMA lies in the optimal deployment of limited resources towards achieving specific objectives, such as maximizing the sum-rate, minimizing transmit power, or enhancing communication reliability \cite{WeiProceeding}.
In the context of reconfigurable channels, resource allocation design explores the best methods of channel reconfiguration in conjunction with the usage of system resources for NGMA.
This often necessitates a joint design, such as joint active and passive beamforming design \cite{QingqingIRS} in IRS-assisted systems and joint trajectory and resource allocation design\cite{YuanxinUAVNOMA} in UAV-enabled systems.
Besides, next-generation wireless networks are anticipated to support additional capabilities of sensing, computing, and/or learning alongside communication.
This anticipation has spurred the development of integrated sensing and communication (ISAC) systems \cite{Fan2020} \cite{ZhiqiangISAC}, integrated computing and communication systems\cite{Yuyi2017}, and integrated learning and communication systems\cite{YandongCST}.
The key challenges in resource allocation for NGMA, particularly within these functional channels, revolve around efficiently managing user access while delicately balancing the trade-offs between communication and these auxiliary functionalities.

\begin{figure}[t] 
\centering
\includegraphics[width=6in]{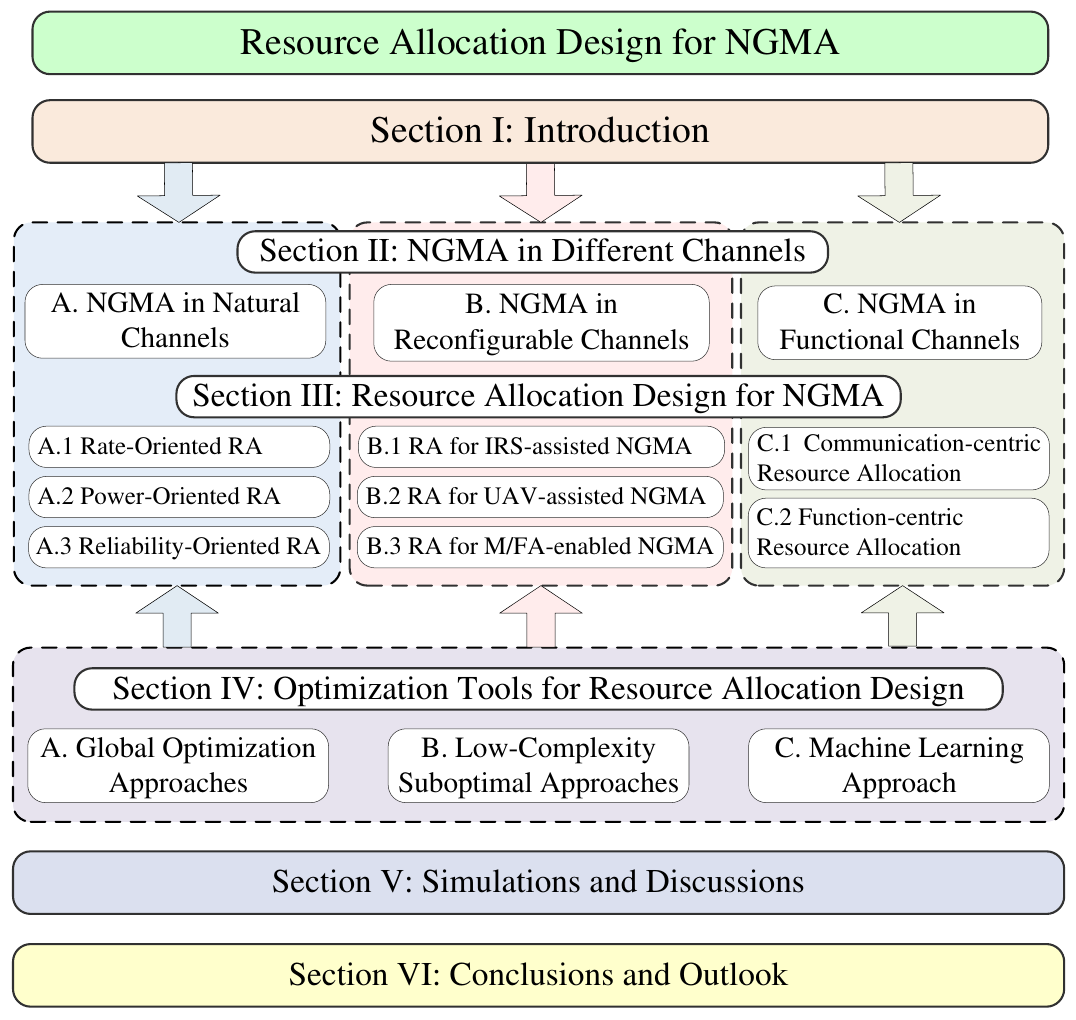}
\caption{ Conceptual framework and sectional outline of the paper.}
\label{Fig.Flowchart}
\end{figure}

{We note that implementing NGMA in uplinks and downlinks is crucial from both practical and research perspectives.
Firstly, multiple access is essential for both uplinks and downlinks as modern communication systems usually operate bi-directionally.
Secondly, NGMA and its resource allocation design in uplinks and downlinks present distinctive challenges. 
In uplink NGMA, each user terminal has limited radio resources and the resource allocation is inherently distributed, while downlink NGMA enables centralized resource allocation with a total resource constraint at the base station.
Thirdly, NGMA should be designed jointly for both uplinks and downlinks with a proper resource allocation strategy for duplexing such that the interference between uplinks and downlinks is minimized. 
For instance, uplink and downlink transmissions can be operated in different frequency bands or different time slots to avoid cross-link interference, corresponding to frequency division duplexing (FDD) and time division duplexing (TDD), respectively.}

The main objective of this tutorial paper is to provide a framework for NGMA and its resource allocation design.
Section I introduces the unified framework for NGMA in different channels and provides some exemplary system models.
Section II focuses on the foundational problem formulations of resource allocation design for NGMA in different types of channels. 
Section III introduces various advanced optimization tools for resource allocation design, highlighting their tradeoffs between computational complexity, convergence, and optimality.
Some simulations and discussions are provided in Section V.
Section VI concludes this work and provides an outlook on resource allocation design for NGMA. An illustrative diagram, depicted in Fig. \ref{Fig.Flowchart}, guides the reader through the logical progression and interconnectedness of the topics discussed within this paper.

\begin{table*}[t]
\centering
\caption{List of key acronyms} 
\begin{tabular}{l|c||l|c}
\hline
{Common term} & {Acronym} &  {Common term} & {Acronym}  
\\
\hline
Sixth-generation & 6G &  Artificial intelligence & AI  
\\ 
Internet-of-Things & IoT & Next-generation multiple access & NGMA  
\\ 
Orthogonal multiple access & OMA  & Non-orthogonal multiple access & NOMA  
\\ 
Frequency-division multiple access & FDMA & Inter-user interference & IUI
\\ 
Time-division multiple access & TDMA & Code-division multiple access & CDMA
\\ 
Rate splitting multiple access & RSMA & Delay-division multiple access & DDMA
\\ 
Orthogonal frequency division multiplexing & OFDM & Orthogonal frequency-division multiple access & OFDMA
\\ 
Spatial division multiple access & SDMA & Multiple-input multiple-output & MIMO
\\ 
Degrees-of-freedom & DoF & Quality-of-service & QoS
\\ 
Single-input single-output & SISO & Channel state information & CSI
\\ 
Unmanned aerial vehicle & UAV & Intelligent reflecting surface & IRS
\\ 
Integrated sensing and communication & ISAC & Joint computing and communication & JCAC
\\ 
Movable/fluid antenna & M/FA & Low-density spreading & LDS
\\ 
Uniform planar array & UPA & Uniform linear array & ULA
\\ 
Line-of-sight & LoS & Additive white Gaussian noise & AWGN
\\ 
Non-convex objective function & NOF & Variable coupling & VC
\\ 
Fractional constraint & FC & Binary constraint & BC
\\ 
Semi-infinite constraint & SMIC & Unit-modulus constraint & UMC
\\ 
Equality constraint & EC & Branch-and-bound & BnB 
\\ 
Successive convex approximation & SCA & Semidefinite relaxation & SDR
\\ 
Block coordinate descent & BCD & Monotonic optimization & MO
\\ \hline
\end{tabular}
\label{major_acronyms}
\end{table*}

The following notations are adopted in this paper. Boldface capital and lowercase letters are reserved for matrices and vectors, respectively. $\mathbb{C}^{M\times N}$ denotes the set of all $M\times N$ matrices with complex entries; $\mathbb{R}^{M\times N}$ denotes the set of all $M\times N$ matrices with real entries; $\mathbb{B}^{M\times N}$ denotes the set of all $M\times N$ matrices with binary entries; ${\left( \cdot \right)^{\mathrm{T}}}$ and ${\left( \cdot \right)^{\mathrm{H}}}$ denote the transpose and the Hermitian transpose of a vector or a matrix, respectively;
$\abs{\cdot}$ denotes the absolute value of a complex scalar or the cardinality of a set, $\left\|\cdot\right\|$ denotes the Euclidean norm of a vector, and $\left\|\cdot\right\|_{\rm F}$ denotes the Frobenius norm of a matrix.
$\Tr\left( \cdot \right)$ and $\det\left( \cdot \right)$ denote the trace and determinant of a matrix; $\mathrm{Rank}\left( \cdot \right)$ denotes the rank of a matrix; $\diag\{\mathbf{x}\}$ denotes a diagonal matrix whose main diagonal elements are given by its input vector $\mathbf{x}$; $\mathbf{I}_{M}$ denotes the $M \times M$ identity matrix; $\mathbf{0}_{M}$ denotes the $M \times 1$ zero vector with all zero entries;
$\left\{\mathbf{X}\right\}_{ij}$ returns the entry in the $i$-th row and $j$-th column of matrix $\mathbf{X}$; $\left\{\mathbf{x}\right\}_{i}$ returns the $i$-th entry of vector $\mathbf{x}$.
$\otimes$ and $\frac{\partial f(x)}{\partial x}$ denote the Kronecker product and the differential operator, respectively.
The real-valued Gaussian distribution with mean $\boldsymbol{\mu}$ and covariance matrix $\boldsymbol{\Sigma}$ is denoted by ${\cal N}(\boldsymbol{\mu},\boldsymbol{\Sigma})$, and the circularly symmetric complex Gaussian distribution with mean $\boldsymbol{\mu}$ and covariance matrix $\boldsymbol{\Sigma}$ is denoted by ${\cal CN}(\boldsymbol{\mu},\boldsymbol{\Sigma})$.
The key acronyms and notations adopted in this paper are summarized in Table \ref{major_acronyms} and Table \ref{major_Notation}.


\begin{table*}[t]
\centering
\caption{List of key notations} 
\begin{tabular}{l|c||l|c}
\hline
{Physical meaning} & Notation & {Physical meaning} & Notation  
\\ 
\hline
{Number of data streams} & {$D$} &  {Number of users} & {$K$}  
\\ 
{Number of resource elements at the transmitter} & {$N$} &  {Number of resource elements at the receiver} & {$M$}  
\\ 
{Channel matrix} & {$\mathbf{H}$} &  {Data stream} & {$\mathbf{s}$}  
\\
{Received signal} & {$\mathbf{y}$} &  {Noise} & {$\mathbf{z}$}  
\\ 
{Transmit power of user $k$} & {$p_k$} &  {User scheduling matrix} & {$\mathbf{\Pi}$}   
\\ 
Achievable rate of user $k$ & {$R_k$}  & Channel reconfiguration variable & {$\mathbf{\Phi}$} 
\\ 
3D Cartesian coordinates of the UAV & {$\mathbf{r}_0$} & 3D Cartesian coordinates of user $k$ & {$\mathbf{r}_k$}
\\  
Vertical angle-of-departure of user $k$ & {$\theta_k$} & Horizontal angle-of-departure of user $k$ & {$\varphi_k$}
\\ 
SIC decoding order of users $k$ and $r$ & {$\alpha_{k,r}$} & Received SINR of user $k$ & {$\Gamma_k$}
\\ 
Channel matrix between the BS and the IRS & {$\mathbf{F}$} & Phase shift matrix of the IRS $k$ & {$\mathbf{\Psi}$}
\\ 
Radar transmitted signal & {${\bf X}_0$} & Task symbols concerning the computation & {${\bf d}$}
\\ \hline
\end{tabular}
\label{major_Notation}
\end{table*}

\section{Next-Generation Multiple Access in Different Channels}
In this section, we introduce a unified framework accompanied by various representative system models for NGMA across natural, reconfigurable, and functional channels.
This comprehensive approach encompasses a broad spectrum of NGMA scenarios anticipated in future wireless networks.
By presenting these models, we aim to provide a holistic overview that encapsulates the extensive variety of NGMA categories, ensuring a thorough understanding of the challenges and prospects awaiting in the advancement of wireless network technologies.

\subsection{NGMA in Natural Channels}
In this subsection, we first introduce the unified framework for NGMA in natural channels, which subsumes OMA, NOMA, and RSMA as special cases.
For each multiple access scheme, the corresponding transceiver architectures, the available system resources, and the resource allocation design are discussed.
To further illustrate the applicability of the framework, we present four exemplary NGMA systems, including OFDMA, NOMA, DDMA, and RSMA.

\subsubsection{A Unified Framework for NGMA in Natural Channels}

The fundamental input-output relationship of NGMA in natural channels is given by
\begin{equation}\label{Eqn:NaturalChannelModel}
	\mathbf{y} = \mathbf{Hg}(\mathbf{s}) + \mathbf{z},
\end{equation}
where $\mathbf{g}(\cdot)$ is a mapping function that will be discussed later, and $\mathbf{s}\in \mathbb{C}^{D \times 1}$ contains a total of $D$ data streams of all $K$ users such that $\mathbf{s} \sim \mathcal{CN}(\mathbf{0},\mathbf{I}_D)$, as widely adopted in the literature\cite{Wei2017}.
The subset $\mathbf{s}_{\Omega_k} \in \mathbb{C}^{D_k \times 1}$ of $\bf s$ contains $D_k = \left|\Omega_k\right|$ data streams of user $k$, $\forall k \in \{1,\ldots,K\}$, where $\sum_{k=1}^{K} D_k = D$ and $\cup_{k = 1}^K {\Omega_k} = \left\{1,\ldots, D\right\}$.
%
%
The channel matrix $\mathbf{H} \in \mathbb{C}^{M \times N}$ models the propagation between transmitter and receiver, where $N$ and $M$ are determined by the number of available system resource elements at the transmitter and receiver, respectively, such as the number of antennas, time slots, or subcarriers, etc.
In practice, the channel matrix $\mathbf{H}$ needs to be defined in a proper domain such that it does not violate the intrinsic physics of wave propagation.
For instance, if the channel matrix is defined in the spatial domain, we usually consider a narrowband assumption for the system \cite{YijieRSMA}, such that each entry of $\mathbf{H}$ can characterize the flat fading coefficient between each pair of transceiver antennas. 
Vectors $\mathbf{y}\in \mathbb{C}^{M \times 1}$ and $\mathbf{z}\in \mathbb{C}^{M \times 1}$ denote the received signal and noise at the receiver side, respectively, where the additive white Gaussian noise (AWGN) follows $\mathbf{z} \sim \mathcal{CN}(\mathbf{0}, \sigma^2\mathbf{I}_M)$ with the noise power of $\sigma^2$.
If the channel matrix is defined in the frequency domain, a multicarrier communication system is usually considered with $M=N$ denoting the number of subcarriers, where $\mathbf{H}$ denotes the frequency domain effective channel between transceiver\cite{Al-Imari2011LDSOFDM}.
For concise notation in this paper, the notations $\mathbf{y}$, $\mathbf{H}$, $\mathbf{z}$, and $\mathbf{s}$ will be consistently employed to denote entities with analogous meanings, albeit with varying specifications and dimensions across different scenarios.

\begin{table*}[t]
	\scriptsize
	\vspace{-5mm}
	\caption{The domain of channel matrix $\mathbf{H}$, modulation waveform, and corresponding multiple access schemes} \label{HDomain_NGMA}
	\vspace{-5mm}
	\begin{center}
		\begin{tabular}{ c | c | c }
			\hline			
			Domain of channel matrix $\mathbf{H}$   & Modulation waveform & Typical multiple access schemes  \\ \hline
		    Frequency domain &  FDM or OFDM & FDMA, OFDMA, NOMA \\
                Time domain &  TDM & TDMA \\
                Spatial domain & Any narrowband  modulation waveform   & SDMA, RSMA \\
                Delay-Doppler domain & OTFS & DDMA \\
                \hline
		\end{tabular}
	\end{center}
	\vspace{-5mm}
\end{table*}

On the other hand, the channel matrix $\mathbf{H}$ can be defined in different domains depending on the adopted modulation waveform and on which domain NGMA is performed, as shown in Table \ref{HDomain_NGMA}.
Note that the channel matrix $\mathbf{H}$ can be defined in multiple domains, such as the frequency-spatial domain for MIMO-OFDM systems.
In such a case, NGMA can be performed in the frequency domain, spatial domain, or cross the frequency-spatial domain, depending on the structure of $\mathbf{H}$.
Despite the domain of $\mathbf{H}$, its rank $\mathrm{Rank}\left(\mathbf{H}\right)$ determines the maximum available DoFs of the considered communication systems, which is the theoretical maximum number of users that can be accommodated in an interference-free manner.
Moreover, the eigenvalue distribution of $\mathbf{H}$ determines the channel disparity in a certain domain, facilitating the exploration of multi-user diversity for efficient resource allocation design.
When $D \le \mathrm{Rank}\left(\mathbf{H}\right)$, the system is known as underloaded \cite{DerrickEEOFDMA}, where the system resources are sufficient to accommodate all the data streams of all the $K$ users. 
In this case, OMA exhibits a handy approach to serve all the $K$ users, since each data stream can be allocated in at least one available system DoF and enjoys an interference-free transmission. 
In contrast, when $D > \mathrm{Rank}\left(\mathbf{H}\right)$, the system is overloaded \cite{PerGainWei} and the users access the channel in a non-orthogonal manner, i.e., NOMA or RSMA.
In such a case, advanced multi-user interference mitigation techniques need to be adopted to improve system performance, such as SIC\cite{Wei2017} and message passing-based multiuser detection \cite{BayestehSCMA}.

In \eqref{Eqn:NaturalChannelModel}, the vector function $\mathbf{g}(\cdot): \mathbb{C}^{D \times 1} \xrightarrow{} \mathbb{C}^{N \times 1}$
maps the information symbols $\mathbf{s}$ of all the $K$ users to the $N$ system resources at the transmitter side.
In particular, this mapping function captures a specific approach of multiple access as well as resource allocation among multiple users, which plays a key role in determining the system performance.
In other words, the joint consideration of multiple access and resource allocation is to design the mapping function, $\mathbf{g}(\cdot)$, for a given channel matrix $\mathbf{H}$, such that the transmission and the channel properties match with each other so as to achieve certain purposes, e.g., interference management, rate maximization, and power minimization.
In practice, advanced channel estimation technologies are generally required to acquire the channel matrix $\mathbf{H}$ \cite{BigueshMMSE,zhao2017multiuser,Sun2019}.
In certain scenarios, such as slow-fading channels, accurate CSI can be acquired at the transmitter with a reasonable amount of system overhead.
In such cases, the estimated accurate CSI would be treated as perfect CSI to simplify the design of multiple access and resource allocation.
In contrast, for the cases in which acquiring accurate CSI is not possible, only imperfect CSI is available.
As such, different robust resource allocation design approaches should be performed such that the QoS of all users can be guaranteed to a certain extent, e.g.,  \cite{WeiProceeding}.

For a given $\mathbf{H}$, OMA schedules to serve one user in each DoF, while NOMA schedules to serve more than one user in each DoF.
Also, hybrid NOMA strategically combines the approaches by serving a portion of the users through OMA while allocating the remaining users via NOMA \cite{ZengHybridNOMA}.
In practice, the formulation of resource allocation design problems can be tailored to specific system design goals, depending on the chosen multiple access scheme.
Specifically, OMA imposes strict constraints to ensure resource orthogonality at the expense of underutilization of limited system resources. 
In contrast, NOMA is more flexible in terms of resource allocation design and can degenerate to  OMA configuration if the power or rate allocated to certain NOMA users is set to zero. Nonetheless, the receiver in an OMA system benefits from a reduced complexity compared to that of a NOMA system\cite{PerGainWei}.
Meanwhile, hybrid NOMA synergizes the advantages of both NOMA and OMA, enabling IUI-free transmission for certain users while simultaneously supporting a larger number of users through non-orthogonal multiplexing.
In contrast, RSMA multiplexes common and private data streams via precoding in the spatial domain based on a proper rate splitting strategy at the transmitter side\cite{BrunoJSAC}. 
At the RSMA receiver, the common data stream is decoded by treating the interference as noise, and the private data streams are decoded successively based on the SIC framework.
This structure offers high flexibility in exploiting the spatial DoF and resource allocation, which bridges spatial-division multiple access (SDMA) and NOMA in an efficient manner\cite{BrunoJSAC}.

We note that the input-output relationship in \eqref{Eqn:NaturalChannelModel} can be applied to both uplink and downlink transmission.
In downlink communication systems, the BS has the full information of $\mathbf{s}$ and $\mathbf{H}$ and thus the mapping function $\mathbf{g}(\cdot)$ can be designed in a centralized manner at the BS.
As in uplink systems, the mapping function $\mathbf{g}(\cdot)$ needs to be designed in a distributed manner, since each user only has its own specific CSI (either perfect or imperfect CSI) and information symbols.

    \begin{figure}[t] 
    \centering
    \includegraphics[width=5.4in]{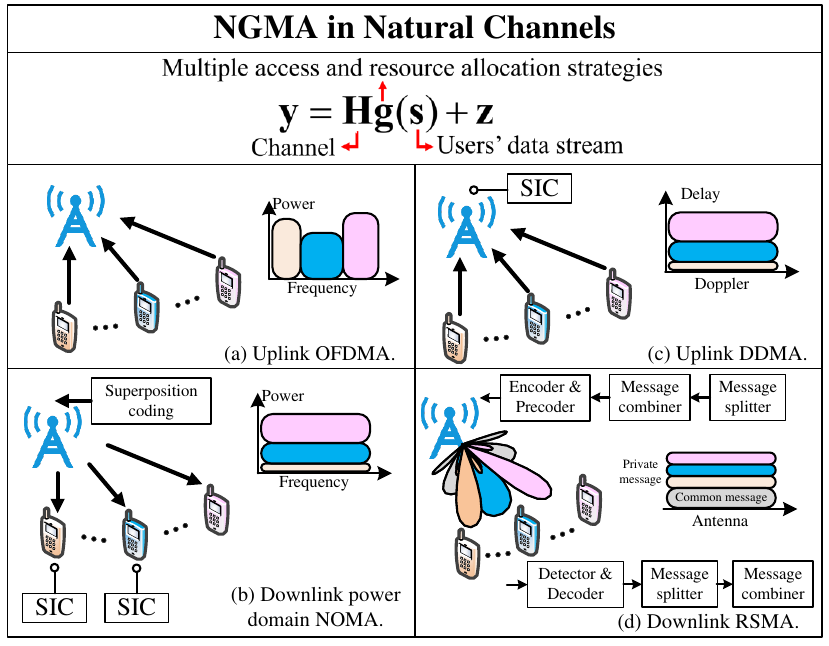}
    \caption{NGMA in natural channels.}
    \label{Fig.NGMA_Natural}
    \end{figure}
    
\subsubsection{Uplink OFDMA}
Consider an uplink system, where a single-antenna BS serves $K$ single-antenna users.
All the transceivers are equipped with an OFDM modulator or demodulator \cite{DerrickEEOFDMA} and all the users are accessing the channel concurrently based on OFDMA, {as shown in Fig. \ref{Fig.NGMA_Natural} (a)}.
In this case, $M$ in \eqref{Eqn:NaturalChannelModel} is the total number of subcarriers adopted in OFDM modulation.
Between user $k$ and the BS, there would be a frequency domain channel matrix $\mathbf{H}_k \in \mathbb{C}^{M\times M}$.
Thus, we have $N=MK$ and the channel matrix in \eqref{Eqn:NaturalChannelModel} can be rewritten as
$\mathbf{H} = \left[\mathbf{H}_1,\ldots,\mathbf{H}_K\right]$.
The data streams $\mathbf{s}$ and the received signals $\mathbf{y}$ are also defined in the frequency domain.
In this case, the total number of data streams $D \le M$ as the system DoF is limited by the number of subcarriers adopted at the BS.

For the considered uplink OFDMA system, the input-output relationship in \eqref{Eqn:NaturalChannelModel} can be specified as 
\begin{equation}\label{Eqn:NaturalChannelOFDMA}
	\mathbf{y} = \sum_{k=1}^{K} \mathbf{H}_k \mathbf{\Pi}_k \mathbf{\Lambda}_k \mathbf{s}_{\Omega_k} + \mathbf{z}, 
\end{equation}
where $\mathbf{\Lambda}_k = \diag\{\sqrt{p_{k1}},\ldots,\sqrt{p_{kD_k}}\}$ and $p_{kd_k} \ge 0$ is the power allocation for the $d_k$-th data stream of user $k$.
Note that the total transmit power of user $k$ should be limited by its power budget $P_{k,\mathrm{max}}$, i.e.,  $\sum_{d_k=1}^{D_k} p_{kd_k} \le P_{k,\mathrm{max}}$.
The binary matrix $\mathbf{\Pi}_k \in \mathbb{B}^{M \times D_k}$ is the user scheduling matrix.
$\{\mathbf{\Pi}_k\}_{md_k} = 1$ denotes that the $d_k$-th data stream of user $k$ is scheduled to the $m$-th subcarrier, otherwise $\{\mathbf{\Pi}_k\}_{md_k} = 0$.
Without loss of generality, we assume that each data stream is allocated to one subcarrier, i.e., $\sum_{m = 1}^{M}\{\mathbf{\Pi}_k\}_{md_k} = 1$, $\forall d_k$, and thus the total number of subcarriers allocated to user $k$ is $\sum_{d_k = 1}^{D_k}\sum_{m = 1}^{M}\{\mathbf{\Pi}_k\}_{md_k} = D_k$.
For OFDMA, we need to limit $\sum_{k=1}^{K}\sum_{d_k=1}^{D_k}\{\mathbf{\Pi}_k\}_{md_k} \le 1$ to guarantee that at most one data stream of a user can be scheduled to each subcarrier.

In practice, when a static multipath propagation environment and a proper cyclic prefix length are considered, the frequency domain channel of each user $\mathbf{H}_k$ is a diagonal matrix, which lays the foundation for interference-free parallel data stream transmission and facilitates low-complexity single-tap FDE at the BS.
In contrast, in high-mobility channels when the Doppler is not well compensated, the frequency domain channel $\mathbf{H}_k$ is generally not diagonal, leading to severe inter-carrier interference (ICI) in OFDM and IUI in OFDMA systems\cite{Zhiqiang_magzine}.
In this case, multiple access schemes designed in the delay-Doppler domain might be more beneficial \cite{Chong2022achievable}, as will be detailed in the later sections.
For now, we restrict ourselves to OFDM transmission in a static environment, i.e., ICI-free, in which case OFDMA is appealing for multi-user communication.
In this case, the received signal in subcarrier $m$, $\forall m \in \{1,\ldots,M\}$, at the BS is given by 
\begin{equation}\label{Eqn:NaturalChannelSCOFDMA}
	\{\mathbf{y}\}_m = \sum_{k=1}^{K} \sum_{d_k = 1}^{D_k}\{\mathbf{\Pi}_k\}_{md_k}\{\mathbf{H}_k\}_{mm} \sqrt{p_{kd_k}} \{\mathbf{s}_{\Omega_k}\}_{d_k} + \{\mathbf{z}\}_m, \forall m.
\end{equation}
Due to $\sum_{k=1}^{K}\sum_{d_k=1}^{D_k}\{\mathbf{\Pi}_k\}_{md_k} \le 1$, the multiple data streams of each user can be easily retrieved at the BS based on a simple single-tap FDE, and thus the achievable rate of user $k$ can be formulated as:
\begin{equation}\label{Eqn:NaturalChannelOFDMARate}
	{R}_k = \sum_{d_k = 1}^{D_k}\sum_{m = 1}^{M}\{\mathbf{\Pi}_k\}_{md_k}\log_2\left(1+\frac{\left|\{\mathbf{H}_k\}_{mm}\right|^2 p_{kd_k}}{\sigma^2}\right), \forall k = 1,\ldots, K.
\end{equation}
The resource allocation design challenge is to schedule the data streams of all users and allocate power to each data stream based on the available CSI, for achieving the design goal while taking into the system resource constraints.
Note that the user scheduling matrices $\mathbf{\Pi}_k$, $\forall k$, need to be designed at the BS in a centralized manner\cite{DerrickEEOFDMA}, while the power allocation $\mathbf{\Lambda}_k$ for user $k$ can be designed distributively at each user, based on its individual CSI.

\subsubsection{Downlink Power Domain NOMA}
Consider a downlink system where a single-antenna BS serves $K$ single-antenna users, {as shown in Fig. \ref{Fig.NGMA_Natural} (b)}.
Both the BS and users still adopt OFDM modulation and demodulation by dividing the system bandwidth into $M$ subcarriers.
In this downlink system, the channel matrix in \eqref{Eqn:NaturalChannelModel} is constituted by $\mathbf{H} = \left[\mathbf{H}_1,\ldots,\mathbf{H}_K\right]^{\mathrm{T}}\in \mathbb{C}^{MK \times M}$ and the received signal in \eqref{Eqn:NaturalChannelModel} is given by $\mathbf{y} = \left[\mathbf{y}^{\rm T}_1,\ldots,\mathbf{y}^{\rm T}_K\right]^{\mathrm{T}}\in \mathbb{C}^{MK \times 1}$, i.e., $N=MK$, where $\mathbf{H}_k \in \mathbb{C}^{M \times M}$ and $\mathbf{y}_k \in \mathbb{C}^{M \times 1}$ denote the frequency domain channel matrix between the BS and user $k$ and the received frequency domain signal at user $k$, respectively.
In contrast to OFDMA, we consider an overloaded scenario and assume that the BS can schedule only one data stream to transmit for each user, i.e., $K = D>M$ and $D_k=1$, $\forall k$.
Thus, power domain NOMA can be adopted to accommodate all the $K$ users to form multi-carrier NOMA\cite{Wei2017}.
For power-domain NOMA, superimposition coding is adopted at the transmitter \cite{Wei2017,PerGainWei} to multiplex more than one user on each subcarrier while SIC decoding is employed at the user side to mitigate IUI.
In this case, the input-output relationship in \eqref{Eqn:NaturalChannelModel} can be specified as 
\begin{equation}\label{Eqn:NaturalChannelNOMA}
	\mathbf{y}_k = \mathbf{H}_k\mathbf{\Pi}\mathbf{\Lambda}\mathbf{s} + \mathbf{z}_k, \forall k,
\end{equation}
where $\mathbf{s}\in \mathbb{C}^{K \times 1}$ collects the information symbols for all the $K$ users, $\mathbf{z}_k \sim \mathcal{CN}(\mathbf{0}, \sigma^2_k\mathbf{I}_M)$ denotes the noise at user $k$ with noise power of $\sigma^2_k$, and $\mathbf{\Pi} \in \mathbb{B}^{M \times K}$ denotes the user scheduling matrix at the BS.
The diagonal matrix $\mathbf{\Lambda}$ is defined by $\mathbf{\Lambda} = \diag\{\sqrt{p_1},\ldots,\sqrt{p_K}\} \in \mathbb{R}^{K \times K}$, where $p_{k} \ge 0$ is the power allocation for user $k$ at the BS.
$\{\mathbf{\Pi}\}_{mk} = 1$ denotes that user $k$ is scheduled to subcarrier $m$, otherwise $\{\mathbf{\Pi}\}_{mk} = 0$.
More explicitly, the received signal in subcarrier $m$ at user $k$ is given by
\begin{equation}\label{Eqn:NaturalChannelSCNOMA}
	\{\mathbf{y}_k\}_m = \{\mathbf{H}_k\}_{mm}\left(\sum_{k=1}^{K}\{\mathbf{\Pi}\}_{mk} \sqrt{p_k}\{\mathbf{s}\}_k\right) + \{\mathbf{z}_k\}_m.
\end{equation}

For power domain NOMA, since more than one user can be scheduled on each subcarrier, i.e., $\sum_{k=1}^{K}\{\mathbf{\Pi}\}_{mk} \ge 1$, achieving IUI-free transmission is not a practical expectation without exploiting extra DoF from other domains. 
Instead, the power domain NOMA relies on power allocation and SIC decoding by exploiting the channel discrepancies such that the inherent interference can be properly harnessed.
According to the principle of power domain NOMA, on subcarrier $m$, the power allocation and the SIC decoding order should be determined by the channel gain order\cite{Wei2017}.
In particular, denote the user set scheduled on subcarrier $m$ as $\mathcal{K}_m = \{k|\{\mathbf{\Pi}\}_{mk} = 1,\forall k\}$ and without loss of generality their channel gains are sorted $\{\mathbf{H}_k\}_{mm} \ge \{\mathbf{H}_{k'}\}_{mm}$, $\forall k < k' \in \mathcal{K}_m$.
Accordingly, if both users $k$ and $k'$ are scheduled on subcarrier $m$, a higher power is allocated to the user with a lower channel gain, i.e., ${p}_k \le p_{k'}$. 
Also, the user with a higher channel gain always first decodes the information of the user with a lower channel gain before attempting to decode its own information symbol \cite{Tse2005}.
The achievable rate for user $k$ is given by
\begin{equation}\label{Eqn:NaturalChannelSCNOMARate}
	R_k = \sum_{m = 1}^{M}\{\mathbf{\Pi}\}_{mk}\log_2\left(1+\frac{\{\mathbf{H}_k\}_{mm}p_k}{\sum_{k'<k \in \mathcal{K}_m} \{\mathbf{H}_k\}_{mm}p_{k'} + \sigma^2_k}\right),
\end{equation}
where $\sum_{m = 1}^{M}\{\mathbf{\Pi}\}_{mk} = 1$ as each user can only be allocated to at most one subcarrier in the overloaded scenario.
Note that user $k$ suffers from IUI from user $k'< k \in \mathcal{K}_m$ having a stronger channel in subcarrier $m$.
Particularly, for the strongest user in subcarrier $m$, it needs to perform $|\mathcal{K}_m|-1$ steps SIC and then decodes its own information.
As a result, IUI-free decoding is realized for this user, and its achievable rate is given by
\begin{equation}\label{Eqn:NaturalChannelSCNOMARate2}
	R_k = \sum_{m = 1}^{M}\{\mathbf{\Pi}\}_{mk}\log_2\left(1+\frac{\{\mathbf{H}_k\}_{mm}p_k}{ \sigma^2_k}\right).
\end{equation}
The resource allocation for downlink power domain NOMA is performed at the BS in a centralized manner and its task is to schedule users and to allocate power based on the CSI of all the $K$ users, i.e., $\mathbf{H}_1,\ldots,\mathbf{H}_K$, for realizing a certain system design goal.

\subsubsection{Uplink DDMA}
OTFS modulation is a recently proposed multi-carrier waveform that aims to provide reliable communication over time-varying channels~\cite{Hadani2017orthogonal,ZhiqiangLetterPartI,ShuangyangLetterPartII,WeijieLetterPartIII}. 
In contrast to the conventional OFDM waveform, OTFS modulates information symbols in the delay-Doppler (DD) domain~\cite{Yuan2023survey,Yuan2019simple}, offering a distinct approach to signal processing and transmission. 
As a result, OTFS modulation can effectively exploit the DD domain channel properties, including quasi-static, compactness, potential sparsity, and separability~\cite{Zhiqiang_magzine}, which enables numerous advanced designs in various wireless applications~\cite{Mengmeng2023P2PMIMO,Ruoxi2022ISWCS,Ruoxi2023Globecom,Shuangyang_ISAC,Weijie2021JSTSP,Chang2023Predictive}. 
Notably, the coupling between the DD domain channel and the information symbols is characterized by the ``twisted convolution''~\cite{Shuangyang2023Globecom,LSY_THP,Zhiqiang2022off}, a specialized form of convolution that incorporates an additional phase term. The convolution nature of the DD domain channel naturally provides new design challenges and opportunities for NGMA. {More specifically, the MA schemes in the DD domain naturally exhibit a non-orthogonal configuration due to the convolutional structure of the effective DD domain channel. This occurs even if the information symbols are multiplexed orthogonally at the transmitter side. As a result, the receiver observes a superposition of multiple information symbols, which are shifted due to the underlying channel delay and Doppler effects. In comparison to OFDMA, where information symbols do not overlap with each other after transmitting over static channels, DD domain MA schemes generally suffer from interference due to this non-orthogonality.
To mitigate such interference, guard spaces can be introduced between information symbols of different users at the cost of increased transmission overhead. Alternatively, one may utilize advanced receiver designs to effectively manage the IUI ~\cite{Ruoxi2022ICC}, as introduced in the following.}

Let us consider an uplink multi-user transmission leveraging OTFS waveform with $K$ users and the channel between each user and the BS has $P$ independent resolvable paths, {as shown in Fig. \ref{Fig.NGMA_Natural} (c)}.
For the sake of illustration, both the BS and $K$ users are equipped with a single antenna.
We define a DD grid of size ${\tilde M}\times {\tilde N}$ for accommodating information symbols, which corresponds to a time-frequency (TF) grid with ${\tilde M}$ subcarriers and $\tilde N$ time slots.
Besides, we consider the case of DDMA \cite{Chong2022achievable}, where each user's information symbols are placed along the whole Doppler dimension but with a specific delay, i.e., each user occupies a row in the DD grid, as shown in Fig. \ref{Fig.NGMA_Natural} (c). 
Notice that the coupling relationship between channel and information in the DD domain is convolutional in nature\cite{Hadani2017orthogonal}. 
Particularly, we shall notice that each user's data will spread to $l_{\max}$ adjacent rows due to the channel delay. Here, $l_{\max}$ denotes the maximum delay index that is computed as the ratio between the maximum delay spread and the delay resolution $T/{\tilde M}$, where $T$ denotes the duration of one time-slot (corresponding to the duration of an OFDM symbol).
Specifically, the symbol multiplexing for user $k$ can be characterized by the \textit{indicator matrix} ${\bf \Pi}_k \in {\tilde M \tilde N \times \tilde N}$, $\forall k$, associated for each user~\cite{Chong2022achievable}, which maps the information symbols to the $k$-th row of the DD grid. 
Thus, we have $D= \tilde N$ and $N =\tilde M \tilde N$.
We adopt ${\bf H}_k \in {\mathbb C}^{\tilde M \tilde N \times \tilde M \tilde N}$ denoting the DD domain channel matrix between user $k$ to the BS such that $M= \tilde M \tilde N$.
Thus, the input-output relationship in~\eqref{Eqn:NaturalChannelModel} is further specified by
\begin{equation}\label{Eqn:NaturalChannelOTFSModel}
	\mathbf{y} = \sum_{k=1}^{K} \sqrt{p_k}\mathbf{H}_k \mathbf{\Pi}_k \mathbf{s}_k + \mathbf{z}, 
\end{equation}
where $\mathbf{s}_k \in {\tilde N \times 1}$, $\forall k$, is the symbol vector of user $k$, $p_k>0$ is the transmit power of user $k$, and the detailed structure of ${\bf H}_k $ is specified in~\cite{Chong2022achievable,li2021cross}. 
Assuming that the SIC detection is adopted at the BS, i.e., the BS always first decodes information of the stronger user with a higher received signal power while treating the IUI caused by weaker users as noise.
By assuming that the users' received signal power levels are sorted in descending order, i.e., $p_k\Tr\left(\mathbf{H}_k \mathbf{\Pi}_k\mathbf{\Pi}^{\rm H}_k\mathbf{H}^{\rm H}_k\right) \ge p_{k'}\Tr\left(\mathbf{H}_{k'} \mathbf{\Pi}_{k'}\mathbf{\Pi}^{\rm H}_{k'}\mathbf{H}^{\rm H}_{k'}\right)$, $\forall k<k'$, the achievable rate of user $k$ is given by\cite{Chong2022achievable}
\begin{equation}\label{Eqn:NaturalChannelOTFSRate}
	{R_k} = {\log _2}\det \left( {{{\bf{I}}_{\tilde N}} + \frac{{{p_k}}}{{{I_k} + \sigma^2}}{{\bf{H}}_k}{{\bf{\Pi}}_k}{\bf{\Pi}}_k^{\rm{H}}{\bf{H}}_k^{\rm{H}}} \right),
\end{equation}
where the IUI faced by user $k$ is given by 
\begin{equation}\label{Eqn:NaturalChannelOTFSinterference}
I_k = \frac{1}{{\tilde N}}{\rm{Tr}}\left( {\sum\limits_{k' = k+1}^K {{p_{k'}}{{\bf{H}}_{k'}}{{\bf{\Pi}}_{k'}}{\bf{\Pi}}_{k'}^{\rm{H}}{\bf{H}}_{k'}^{\rm{H}}} } \right).
\end{equation}
%
%
It should be noted that the rate expression given in~\eqref{Eqn:NaturalChannelOTFSRate} is subjected to the specific realization of channel delay, Doppler, and fading coefficients, which is generally challenging to acquire at the user side in uplink transmissions. 
It was shown in~\cite{Chong2022achievable} that the achievable rate in~\eqref{Eqn:NaturalChannelOTFSRate} can be tightly upper-bounded by the rate determined by the effective signal-to-interference-plus-noise ratio (SINR) of each user, which is only related to the effective channel gains. 
More importantly, compared to uplink OFDMA with SIC, a modest achievable rate improvement can be observed thanks to the fact that OTFS is insensitive to delay and Doppler, while OFDMA is prone to substantial channel variability within the TF domain due to the presence of delay and Doppler~\cite{Li2020performance,Wei2020transmitter,Shuangyang2021hybrid}. This sensitivity leads to considerable fluctuations in the effective  SINR of OFDMA, and consequently decreases the achievable rate.

\subsubsection{Downlink RSMA}
RSMA offers a versatile and generalized non-orthogonal transmission, which presents a dynamic interference management framework that softly bridges SDMA and power domain NOMA, adapting to varying interference levels and specific system requirements\cite{BrunoJSAC}.

Here, we consider a narrow-band downlink single-layer linearly precoded RSMA scheme \cite{YijieRSMA} where a multi-antenna BS serves $K$ single-antenna users, {as shown in Fig. \ref{Fig.NGMA_Natural} (d)}.
In this case, the channel matrix in \eqref{Eqn:NaturalChannelModel} is defined in the spatial domain, the number of antennas at the BS is $N$, and $M = K$.
Again, in overloaded scenarios, we assume that the BS schedules only one data stream to transmit for each user, i.e., $K = D$ and $D_k=1$, $\forall k$.
The message $W_k$ for user $k$ is firstly split into two sub-messages, i.e., one common message $W_{\mathrm{c},k}$ and one private message $W_{\mathrm{p},k}$.
The common messages of all the $K$ users $W_{\mathrm{c},1}, \ldots, W_{\mathrm{c},K}$ are combined into one common message $W_{\mathrm{c}}$ and is encoded into a common data stream $s_{\mathrm{c}}$, which is decoded by all the users.
The private sub-messages of all the $K$ users $W_{\mathrm{p},1}, \ldots, W_{\mathrm{p},K}$ are independently encoded into private data streams ${s}_{\mathrm{p},1}, \ldots, {s}_{\mathrm{p},K}$, which are decoded by the corresponding users, respectively.
Then, the generated $K+1$ data streams intended for $K$ users are linearly precoded and transmitted to the channel.
For the considered downlink single-layer RSMA system, the input-output relationship in \eqref{Eqn:NaturalChannelModel} can be specified as 
\begin{equation}\label{Eqn:NaturalChannelRSMA}
	\mathbf{y} = \mathbf{H}\left(\mathbf{p}_{\mathrm{c}}s_{\mathrm{c}} + \sum_{k=1}^K \mathbf{p}_{\mathrm{p},k}{s}_{\mathrm{p},k}\right) + \mathbf{z},
\end{equation}
where $\mathbf{y} \in \mathbb{C}^{K \times 1}$ collects the received signal of all the $K$ users and $\mathbf{H} = [\mathbf{h}_1,\ldots,\mathbf{h}_K]^{\rm T} \in \mathbb{C}^{K \times N}$ collects the channel vectors between the BS and all the users.
$\mathbf{p}_{\mathrm{c}} \in \mathbb{C}^{N \times 1}$ and $\mathbf{p}_{\mathrm{p},k} \in \mathbb{C}^{N \times 1}$, respectively, denote the precoding vectors for the common data stream and the private data stream of user $k$, $\forall k \in \{1,\ldots,K\}$.
The received signal at user $k$ is given by
\begin{equation}\label{Eqn:NaturalChannelRSMA_Rxk}
	\{\mathbf{y}\}_k = \mathbf{h}_k^{\mathrm{H}} \left(\mathbf{p}_{\mathrm{c}}s_{\mathrm{c}} + \sum_{k=1}^K \mathbf{p}_{\mathrm{p},k}{s}_{\mathrm{p},k}\right) + \{\mathbf{z}\}_k.
\end{equation}

At the receiver side, each user first decodes the common data stream $s_{\mathrm{c}}$ into the common message $\widehat{W}_{\mathrm{c}}$ and performs SIC to subtract its signal from the received signal while treating the interference from all the private data streams as noise. 
Then, user $k$ decodes its private data stream ${s}_{\mathrm{p},k}$ into its private message $\widehat{W}_{\mathrm{p},k}$ by treating the remaining interference from other private data streams as noise.
After decoding the private message, user $k$ reconstructs the original message by extracting $W_{\mathrm{c},k}$ from $W_{\mathrm{c}}$ and combining $W_{\mathrm{c},k}$ with $W_{\mathrm{p},k}$ into $W_{k}$.
Based on the above procedure, the instantaneous rates for decoding the common and private data streams at user $k$ are given by 
\begin{align}
    {R}_{\mathrm{c},k} &= \log_2\left(1+\frac{|\mathbf{h}_k^{\mathrm{H}}\mathbf{p}_{\mathrm{c}}|^2}{\sum_{j=1}^{K}|\mathbf{h}_k^{\mathrm{H}}\mathbf{p}_{\mathrm{p},j}|^2 + \sigma^2_k}\right) \; \text{and} \label{Eqn:NaturalChannelRSMA_RateI}\\
    R_{\mathrm{p},k} &= \log_2\left(1+\frac{|\mathbf{h}_k^{\mathrm{H}}\mathbf{p}_{\mathrm{p},k}|^2}{\sum_{j\neq k}^{K}|\mathbf{h}_k^{\mathrm{H}}\mathbf{p}_{\mathrm{p},j}|^2 + \sigma^2_k}\right),\label{Eqn:NaturalChannelRSMA_RateII}
\end{align}
respectively.
Note that the achievable rate for the common message cannot exceed $\{\mathbf{R}\}_{\mathrm{c},k}$, such that all the users can decode $s_{\mathrm{c}}$ successfully, i.e., 
\begin{equation}
    R_{\mathrm{c}} = \min\{{R}_{\mathrm{c},1},\ldots,{R}_{\mathrm{c},K}\}.
\end{equation}
Without loss of generality, let $C_k$ denote the common rate allocation to user $k$ for $W_{\mathrm{c},k}$.
The overall achievable rate for user $k$ includes both the common rate $C_k$ and the private rate $R_{\mathrm{p},k}$, i.e., 
\begin{equation}
    R_{k} = R_{\mathrm{p},k} + C_k.
\end{equation}
The resource allocation for downlink single-layer RSMA is also performed at the BS in a centralized manner.
The resource allocation aims to design the precoding vectors and the common rate allocation based on the CSI of all the $K$ users to achieve a specific system design objective.

In summary, the domain over which the channel matrix $\mathbf{H}$ is defined significantly influences the specification of the signal generation function $\mathbf{g}(\cdot)$.
Indeed, the domain of $\mathbf{H}$ sets the foundation for the overall NGMA framework, while $\mathbf{g}(\cdot)$ specifies the particular NGMA scheme to employ.
More importantly, both $\mathbf{H}$ and $\mathbf{g}(\cdot)$ are indicative of the available system resources that can be allocated, which determine the problem formulation and establish the resource allocation design DoF for improving the system performance.


\subsection{NGMA in Reconfigurable Channels}
\par
Although the previously discussed NGMA schemes can efficiently enhance the performance of multi-user communication systems, their efficiency in practical systems is not always guaranteed, especially in challenging radio propagation environments \cite{Ding2015b,Lei2016NOMA,9679390}. In particular, due to the random and dynamic nature of radio propagation environments, some users may experience adverse channel conditions such as fast fading in high-mobility communication scenarios or severe attenuation due to the absence of line-of-sight (LoS) link caused by the existence of blockages. Under such circumstances, the QoS requirements of the users, e.g., achievable rate or SINR, might not be satisfied,  despite the deployment of OMA, NOMA, or RSMA schemes.
\par
\subsubsection{A Unified Framework for NGMA in Reconfigurable Channels}
Recently, several promising techniques have been proposed to proactively configure wireless channels, aiming to mitigate the issues mentioned earlier. By customizing the radio propagation environment, these approaches facilitate the creation of favorable wireless channel conditions for multiple access, thereby inherently overcoming the obstacles faced by NGMA communication systems. Generally, the NGMA model for reconfigurable channels is described by
\begin{equation}\label{Eqn:ReConfigChannelModel}
\mathbf{y} = \mathbf{H}\left(\mathbf{\Phi}\right)\mathbf{g}(\mathbf{s}) + \mathbf{z},
\end{equation}
where $\mathbf{\Phi}$ is the channel reconfiguration variable and its dimension is determined by the specific type of channel reconfiguration techniques. In this scenario, the method of multiple access and the resource allocation among multiple users, i.e., the vector function $\mathbf{g}(\mathbf{s})$, can be jointly designed with the channel reconfiguration variable $\mathbf{\Phi}$. Particularly, the system resources can be utilized strategically to enhance the system performance based on specific QoS requirements. Different from the fixed conditions for natural channels, the reconfigurable channel condition enabled by $\mathbf{\Phi}$ introduces greater flexibility in designing NGMA schemes. Consequently, the focus of resource allocation optimization problems shifts from merely determining an efficient strategy, $\mathbf{g}(\mathbf{s})$, for a given fixed radio propagation environment to jointly designing both the strategy $\mathbf{g}(\mathbf{s})$ and the wireless channel $\mathbf{H}\left(\mathbf{\Phi}\right)$. From a mathematical perspective, the former scenario is essentially a subset of the latter case. With this concept in mind, it is crystal clear that the potential for performance enhancement is greater when the overall channel is properly reconfigured by smartly designing $\mathbf{\Phi}$. In particular, in practical scenarios where $D \le \min\{N,M\}$, the original system resources are sufficient to accommodate all the $K$ users, OMA schemes remain to be an efficient approach to serve all the $K$ users. As such, the extra design DoF can be leveraged to establish more favorable radio propagation environments to achieve desired system design objectives. As such, reshaping each user's channel is possible such that the orthogonality among channels can be maintained, facilitating the mitigation of IUI. On the other hand, when the original system DoF of the natural channel is limited, i.e., $D > \min\{N,M\}$, the additional design DoF introduced by $\mathbf{\Phi}$ becomes crucial. It enables the creation of a more favorable radio propagation environment for NOMA or hybrid NOMA. In particular, by fine-tuning the channel gains, the user paring policy in NOMA or hybrid NOMA can be designed more flexibly and appropriately, potentially enhancing the system performance. 

        \begin{figure}[t] 
    \centering
    \includegraphics[width=5.4in]{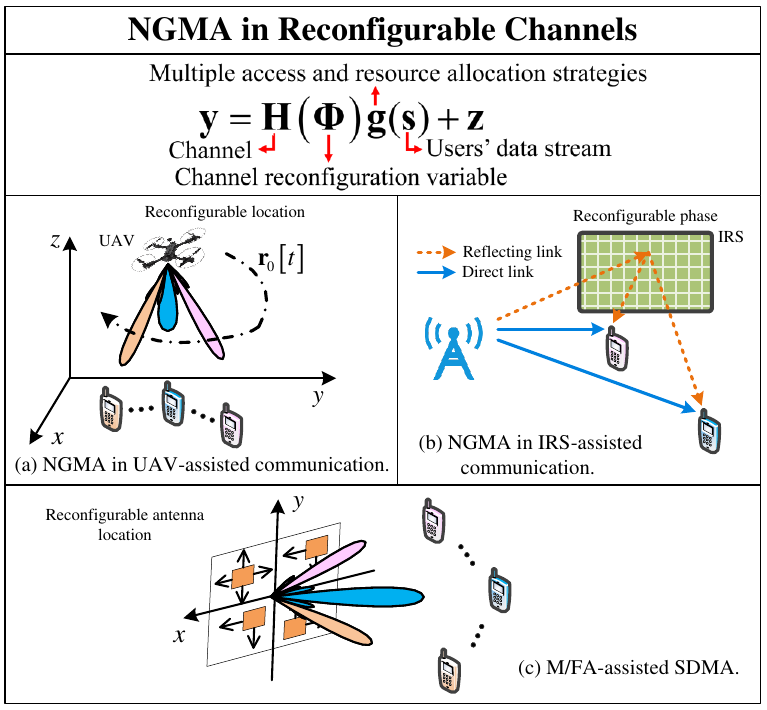}
    \caption{NGMA in reconfigurable channels.}
    \label{Fig.NGMA_Reconfig}
    \end{figure}
    
\par
In the following, we will briefly introduce several promising emerging techniques in the literature that can be exploited to shape desired radio propagation environments. 
\par
\subsubsection{NGMA in UAV-Assisted Communication} Equipping UAVs with communication transceivers introduces the capability to establish air-to-ground NGMA links, which can significantly enhance system performance. Compared to conventional cellular systems that rely on fixed terrestrial infrastructure \cite{wong2017key,you2021towards}, UAV-assisted communication systems offer on-demand connectivity by deploying UAV-mounted wireless transceivers flexibly over a target area. This flexibility is particularly useful in emergency situations, such as natural disasters and major accidents, where UAVs can be leveraged as aerial base stations to establish temporary communication links quickly and cost-effectively \cite{8663615}. Additionally, the high mobility and maneuverability of UAVs allow them to adapt their trajectories based on the environment and terrain variations, thus improving system performance. By properly optimizing the UAV trajectories, we can place new communication nodes at desired locations to facilitate certain optimal wireless channel conditions for NGMA schemes. Therefore, UAV-assisted communication systems offer a promising solution for providing connectivity in remote or hard-to-reach areas, emergency situations, and other scenarios where traditional infrastructure may be unavailable or insufficient.
\par
Next, to illustrate the above advantages, we consider a toy example where a UAV-mounted BS is equipped with a uniform planar array (UPA) composing $N_x \times N_y$ antenna elements, i.e., $N= N_xN_y$, and serves a group of $K$ single-antenna users, {as shown in Fig. \ref{Fig.NGMA_Reconfig} (a)}. In such a case, the reconfigurable channel between the UAV and user $k$ is given by \cite{8445944}
\begin{equation}  \mathbf{h}_k=\sqrt{\varrho}\left\|\mathbf{r}_0- \mathbf{r}_k \right \|^{-1}\mathbf{a}_k.
\end{equation}
Here, $\varrho=(\frac{c}{4\pi f_c})^2$ is a constant determined by the speed of light $c$ and the carrier frequency $f_c$. Moreover, variables $\mathbf{r}_{0}=[x_{0},~y_{0},~H_{0}]^{\mathrm{T}}$ and $\mathbf{r}_k=[x_k,~y_k,~0]^{\mathrm{T}}$ indicate the three-dimensional (3D) Cartesian coordinates of the UAV and user $k$, respectively, where $H_{0}$ is the fixed altitude of the UAV.
Furthermore, the term $\sqrt{\varrho} \left\|\mathbf{r}_0- \mathbf{r}_k \right \|^{-1}$ and vector $\mathbf{a}_k\in\mathbb{C}^{{N}\times 1}$ represent the channel gain between the UAV and user $k$ and the corresponding antenna array response (AAR), respectively. Specifically, vector $\mathbf{a}_k$ is expressed as \cite{8974403}
\begin{eqnarray}
\label{steeringvec}
\mathbf{a}_k\hspace*{-4mm}&&=\big ( 1,\ldots ,e^{-j\frac{2\pi bf_c}{c}\mathrm{sin}\theta_k(n_x-1)\mathrm{cos}\varphi_k},\ldots,e^{-j\frac{2\pi bf_c}{c}\mathrm{sin}\theta_k(N_x-1)\mathrm{cos}\varphi_k} \big )\notag\\
\hspace*{-4mm}&&\otimes \big ( 1,\ldots ,e^{-j\frac{2\pi bf_c}{c}\mathrm{sin}\theta_k(n_y-1)\mathrm{sin}\varphi_k},\ldots,e^{-j\frac{2\pi bf_c}{c}\mathrm{sin}\theta_k(N_y-1)\mathrm{sin}\varphi_k} \big )\notag\\
\hspace*{-4mm}&&\overset{\Delta }{=}\mathbf{a}\big(\theta_k,\varphi_k\big),
\end{eqnarray}
where $b$ is the antenna spacing, and indices $n_x$ and $n_y$ denote the rows and columns of the UPA, respectively. Moreover, variables $\theta_k$ and $\varphi_k$ represent the vertical and horizontal angle-of-departures between the UAV and user $k$, respectively. 
As can be seen from the above channel model $\mathbf{h}_k$, the channel is reconfigured by varying the location of the UAV. As such, the position of the UAV $\mathbf{r}_0$ and the corresponding AAR vector $\mathbf{a}_k$ can be regarded as the channel reconfigure variable $\mathbf{\Phi}$ in \eqref{Eqn:ReConfigChannelModel}.
\par
In the $t$-th time slot, the received signal of user $k$ is given by 
\begin{equation}\label{Eqn:UAV-NOMA_SystemModel}
    y_k[t]=\underset{\mathrm{desired~signal}}{\underbrace{\sqrt{\varrho}\left\|\mathbf{r}_0[t]- \mathbf{r}_k \right \|^{-1}\mathbf{a}^H_k[t]\mathbf{p}_k[t]s_k[t]}} +  \underset{\mathrm{IUI}}{\underbrace{\underset{ r\in\mathcal{K}\setminus \left \{ k \right \}}{\sum}\sqrt{\varrho}\left\|\mathbf{r}_0[t]- \mathbf{r}_k \right \|^{-1}\mathbf{a}^H_k[t]\mathbf{p}_r[t]s_r[t]}} + z_k[t],
\end{equation}
where $z_k[t]\sim \mathcal{CN}( 0, \sigma^2_k)$ and $\mathbf{p}_k[t]\in \mathbb{C}^{N \times 1}$ is the precoding vector for user $k$ in the $t$-th time slot, $\forall k \in \{1,\ldots,K\}$. For UAV-assisted NGMA systems, the position of the UAV, i.e., $\mathbf{r}_0[t]$, can be jointly designed with beamforming vector $\mathbf{p}_k[t]$ to facilitate the implementation of NGMA schemes \cite{8918497,8648498}. On the other hand, according to the SIC principle for NOMA, strong users will first decode the information of the weaker users to avoid the interference caused. In particular, user $k$ will first detect and remove the information signals of user $r$ ($r>k$) successively, and then the signal dedicated to user $k$ is detected by treating the
rest of the users’ signals as noise. To model this, we define a binary variable $\alpha_{k,r}[t]\in \{0, 1\}$, $\forall k\neq r \in\mathcal{K}$, to represent the SIC decoding order of users $k$ and $r$ in the $t$-th time slot. For the case $\alpha_{k,r}[t]=0$, we assume that user $k$ will perform SIC
to eliminate the signal intended for user $r$. For the case $\alpha_{k,r}[t]=1$, we assume that user $k$ will treat user $r$'s signal as noise. In addition, we should guarantee that $\alpha_{k,r}[t]+\alpha_{r,k}[t]=1$ since it is not reasonable
to mutually implement SIC at both users \cite{8114722,7973146}. The received SINR of user $k$ is given by
\begin{equation}\label{SINR_k}
\Gamma_k[t]={\frac{\frac{\varrho}{\left\|\mathbf{r}_0[t]- \mathbf{r}_k \right \|^2}\left |{\mathbf{a}^H_k[t]}\mathbf{p}_k[t]\right |^2}{\underset{ r\in\mathcal{K}\setminus \left \{ k \right \}}{\sum}\alpha_{k,r}\frac{\varrho}{\left\|\mathbf{r}_0[t]- \mathbf{r}'_k \right \|^2}{\left |{\mathbf{a}^H_k[t]}\mathbf{p}_r[t]\right |^2}+  \sigma^2_k}}.
\end{equation}
\par
As can be seen from the above equation, for UAV-assisted NGMA systems, the position of the UAV $\mathbf{r}_0$ can be jointly designed with the transmit precoding $\mathbf{p}_k$ to achieve the desired objective, including system sum rate or total power consumption of the UAV. Moreover, multiple UAVs can collaborate with each other to further improve the performance of NGMA systems. In this case, the position or trajectory of multiple UAVs can be jointly designed with the other available resources in the system. Besides, additional constraints, such as a collision avoidance constraint, should be considered for safety.
\subsubsection{NGMA in IRS-Assisted Communication} Thanks to the advances in electromagnetic meta-material, IRSs have emerged as a disruptive solution for harnessing interference in wireless communication systems \cite{QingqingIRS,9183907,XianghaoJSAC}. An IRS is a planar metasurface consisting of an array of small, passive, low-cost elements, such as phase shifters and printed dipoles, capable of reflecting incident signals with a controlled phase shift. By adaptively adjusting the phase shifts of the IRS elements according to the dynamic radio propagation environment, the wireless channel can be proactively manipulated, which introduces additional designing DoFs for resource allocation design\cite{yu2019miso,9024490}. Furthermore, the reflected signals can be strategically superimposed with the non-reflected signals constructively or destructively to enhance the desired signal power strength or mitigate detrimental interference, improving overall system performance \cite{9154337}. The relatively simple structure of IRSs allows for flexible installation on building facades and interior walls, smoothly and seamlessly integrating them into existing cellular communication systems. By properly configuring the phase shift patterns of the IRS, we can tailor different wireless channel conditions to match the characteristics of various multiple access schemes. Therefore, IRSs offer a promising solution for improving wireless communication systems, enhancing signal strength, and mitigating interference.
\par

Considering a wireless network comprising a BS, an IRS, and a set of $K$ single-antenna users, {as shown in Fig. \ref{Fig.NGMA_Reconfig} (b)}. In particular, the BS is equipped with a uniform linear array (ULA) of $N$ antenna elements, while the IRS contains $M_{\rm ps}$ phase shift elements. Without loss of generality, we assume that there exist direct links between the BS and all users. In such a case, the narrow-band reconfigurable channel between the UAV and user $k$ can be explicitly expressed as
\begin{equation}  \mathbf{h}_k=\mathbf{h}_{\mathrm{D},k}+\mathbf{F}\mathbf{\Psi}\mathbf{h}_{\mathrm{R},k},
\end{equation}
where the channel vectors $\mathbf{h}_{\mathrm{D},k}\in \mathbb{C}^{\mathit{N}\times 1}$ and $\mathbf{h}_{\mathrm{R},k}\in \mathbb{C}^{M_{\rm ps}\times 1}$ characterize the direct link between the BS and user $k$ and the link between the IRS and user $k$, respectively. In the literature, $\mathbf{h}_{\mathrm{D},k}$ and $\mathbf{h}_{\mathrm{R},k}$ can be modeled as Ricean or Rayleigh random variables based on the specific communication scenarios \cite{8910627}. Also, matrix $\mathbf{F}\in\mathbb{C}^{\mathit{N_{\rm a}}\times\mathit{M_{\rm ps}}}$ models the channel between the BS and IRS. Moreover, the diagonal matrix $\mathbf{\Psi}\in \mathbb{C}^{M_{\rm ps}\times M_{\rm ps}}$ represents the phase shift matrix of the IRS with $\{\mathbf{\Psi}\}_{mm} = e^{j2\pi\psi_m}$, where $\psi_m\in \left [ -\pi ,\pi  \right ]$, $\forall m \in \left\{1,\ldots,M_{\rm ps} \right\}$, is the phase shift introduced by the $m$-th IRS element \cite{9154252,9154337}. Alternatively, by employing active IRS, both the amplitude and the phase of the IRS element can be jointly optimized.\footnote{The literature presents a range of IRS models, each designed for specific implementations and applications, where the phase shift matrix, $\mathbf{\Psi}$, is subject to distinct constraints. For detailed information, readers are encouraged to refer \cite{10134546,10266592}.}
\par
As can be seen from the above channel model, the channel is reconfigured by varying the phase shift configuration of the IRS. As such, the phase shift matrix $\mathbf{\Psi}$ can be regarded as the channel reconfigure variable $\mathbf{\Phi}$ in \eqref{Eqn:ReConfigChannelModel}. In fact, by deploying the IRS, an additional reflecting link is established. In the case where the direct link is present, the signals coming from the two links can be combined in a constructive or destructive manner. On the other hand, when the direct link is absent, the IRS-induced link can be smartly tuned to reflect the signal from the BS to the users to establish a strong end-to-end propagation path.
\par
The received signal of user $k$ is given by
\begin{eqnarray}
y_k=\underset{\mathrm{desired~signal}}{\underbrace{(\mathbf{h}^{\rm H}_{\mathrm{D},k}+\mathbf{h}^{\rm H}_{\mathrm{R},k}\mathbf{\Psi}^{\rm H}\mathbf{F}^{\rm H})\mathbf{p}_ks_k}}+\underset{\mathrm{IUI}}{\underbrace{\underset{ r\in\mathcal{K}\setminus \left \{ k \right \}}{\sum}(\mathbf{h}^{\rm H}_{\mathrm{D},k}+\mathbf{h}^{\rm H}_{\mathrm{R},k}\mathbf{\Psi}^{\rm H}\mathbf{F}^{\rm H})\mathbf{p}_rs_r}}+z_k.
\end{eqnarray}
Therefore, by employing the SIC principle for NOMA, the received SINR of user $k$ can be expressed as
\begin{equation}\label{Eqn:NOMA_IRS_SINRk}
\Gamma_{k}=\frac{\left |(\mathbf{h}^{\rm H}_{\mathrm{D},k}+\mathbf{h}^{\rm H}_{\mathrm{R},k}\mathbf{\Psi}^{\rm H}\mathbf{F}^{\rm H})\mathbf{p}_k\right |^2}{\underset{s\in\mathcal{K}\setminus\left\{k\right\}}{\sum}\alpha_{k,s}\left |(\mathbf{h}^{\rm H}_{\mathrm{D},k}+\mathbf{h}^{\rm H}_{\mathrm{R},k}\mathbf{\Psi}^{\rm H}\mathbf{F}^{\rm H})\mathbf{p}_s\right |^2+\sigma^2_{k}},
\end{equation}
where $\alpha_{k,s}\in \mathbb{B} $ is a binary variable representing the SIC decoding order between user $k$ and user $s$.
\par
We note the aforementioned characteristics of the IRSs are particularly suitable for facilitating resource allocation design for NGMA schemes in wireless networks. In conventional downlink NOMA, users with stronger channels utilize the SIC technique to cancel co-channel interference from others. In contrast, since IRS is capable of reconfiguring user channels by shaping the reflected signal amplitudes and/or phase shifts, the user decoding order of NOMA can be permuted by adjusting the IRS reflection which further facilitates a more flexible performance trade-offs among the users. Furthermore, in scenarios where multiple IRSs are present within an NGMA system \cite{9440764}, it is possible to leverage all IRSs or select a subset to assist in transmission, optimizing the balance between system performance, signaling overhead, and computational resources needed for resource allocation. Specifically, for IRS-assisted NGMA systems, the selection policy and phase shift settings of the IRSs can be strategically coordinated with the beamforming strategy of the BS, ensuring a harmonized and efficient network operation. {On the other hand, we note that the choice between adopting OMA or NOMA depends on the DoF of the system being sufficient to accommodate all accessing users for communication or not, i.e., underload and overload cases, respectively. For IRS-aided communications, the system DoF is determined by the rank of the cascaded channel matrix. Therefore, when reconfiguring the IRS can effectively increase the rank of the cascaded channel matrix and result in an underload case, OMA might be favorable as it guarantees interference-free transmission and enables low complexity channel equalization \cite{XidongIRS}. 
For example, it has been shown in the literature \cite{BeixiongComml} that for the case of near-IRS users with symmetric rates, NOMA may perform worse than OMA.
However, when reconfiguring the IRS cannot improve the spatial DoF effectively and the system still operates in the overload regime, NOMA might be preferred since it can support all the accessing users.
In this case, the IRS should be reconfigured to enlarge the channel disparity to exploit the performance gain of NOMA over OMA \cite{PerGainWei}.}

\par
\subsubsection{M/FA-Assisted SDMA} 
The emerging holographic MIMO technique has been proposed to fully harness the spatial variation of wireless channels within a specific spatial transmitter area \cite{10243545,9264694}. Holographic MIMO surfaces consist of numerous miniature passive elements spaced at sub-wavelength distances that can manipulate the electromagnetic properties of transmitted or reflected waves. By utilizing zero-spacing continuous antenna elements, holographic MIMO can fully leverage the spatial DoFs of the spatially continuous transmitter area. However, the large number of antenna elements required for holographic MIMO presents a critical challenge for both channel estimation and data processing, hindering its practical implementation.
\par
To bridge the gap between holographic MIMO and conventional MIMO, a new MIMO concept based on the M/FA system has been proposed. In the M/FA system, each antenna element is connected to an RF chain via a flexible cable, and its physical position can be adjusted within a designated spatial region. For solid antenna elements, this is enabled by using an electromechanical device, such as a stepper motor \cite{10243545}. As for fluid antennas, changing the position of the antenna elements can be realized by employing a piston structure or electromagnetic device \cite{9264694,9650760}. Unlike conventional MIMO systems, which have a set of antenna elements mounted at fixed locations, M/FA can utilize the full spatial DoFs within the available spatial transmitter area by leveraging the flexible movement of the M/FA \cite{10318134}. In fact, by adjusting the position of antenna elements, we can vary some characteristics of the channels, e.g., the path loss between the BS and the receivers and the correlation between the antenna elements. As a result, the M/FA can be regarded as an additional channel reconfiguration based on the conventional position-fixed antenna array, which introduces extra DoFs for resource allocation design for NGMA systems \cite{10078147}. On the other hand, since M/FA requires only a small number of antenna elements to exploit the available DoFs, the computational complexity of the required signal processing is significantly reduced compared to holographic MIMO systems \cite{10146274}. Therefore, M/FA offers a promising solution for fully exploiting the spatial variation of wireless channels within a given spatial transmitter area while overcoming the challenges of holographic MIMO.
\par
We consider a multi-user wireless communication system consisting of a BS and $K$ single-antenna users. The BS is equipped with $N$ M/FA elements to serve the users, {as shown in Fig. \ref{Fig.NGMA_Reconfig} (c)}. The positions of the M/FA elements can be adjusted simultaneously within a given two-dimensional (2D) rectangular region. To facilitate the resource allocation design, the 2D rectangular region is quantized, which results in $Q$ possible positions of the M/FAs. For notational simplicity, we define a set $\mathcal{P}$ to collect all possible positions, i.e., $\mathcal{P}=\{\mathbf{r}_1,\ldots, \mathbf{r}_Q\}$. In particular, the vector $\mathbf{r}_q=[x_q,y_q]^{\rm T}$ denotes the $q$-th possible position, where variable $x_q$ and $y_q$ are the corresponding horizontal and vertical coordinates, respectively. We note that for the $n$-th M/FA element, its feasible position, denoted by $\mathbf{l}_n \in \mathbb{R}^{2 \times 1}$, is selected from $\mathcal{P}$, i.e., $\mathbf{l}_n\in\mathcal{P}$. As such, the physical channel can be reshaped by adjusting the positions of the M/FA elements. In particular, the channel between the $n$-th M/FA element and all $K$ users is defined as $\mathbf{h}_n(\mathbf{l}_n)=[h_{n,1}(\mathbf{l}_n),\ldots,h_{n,K}(\mathbf{l}_n)]^{\mathrm T}\in\mathbb{C}^{K\times 1}$, where variable $h_{n,k}(\mathbf{l}_n)\in\mathbb{C}$ is the channel coefficient between the $n$-th M/FA element and user $k$. We note that the channel vector is a function of $\mathbf{l}_n$. To facilitate notational simplicity, we define a matrix $\mathbf{C}_{n}=[\mathbf{h}_n(\mathbf{l}_1),\ldots,\mathbf{h}_{n}(\mathbf{l}_Q)]\in\mathbb{C}^{K\times Q}$ to collect the channel vectors from the $n$-th M/FA element to all $K$ users for all $Q$ feasible discrete locations. Then, channel vector $\mathbf{h}_{n}(\mathbf{l}_n)$ can be represented as follows
\begin{equation}
    \mathbf{h}_{n}(\mathbf{l}_n)=\mathbf{C}_{n}\mathbf{t}_n,
\end{equation}
where vector $\mathbf{t}_n=\big[t_n[1],\ldots,t_n[Q]\big]^{\rm T} \in \mathbb{B}^{Q \times 1}$. In particular, binary variable $t_n[q]\in\left\{0,\hspace*{1mm}1\right\}$ is the indicator to determine the position of the $n$-th M/FA element and we have $\sum_{q=1}^{Q} t_n[q] = 1$. For the considered system, the channel matrix between the BS and the $K$ users, $\mathbf{H}=[\mathbf{h}_{1}(\mathbf{l}_1),\ldots,\mathbf{h}_{N}(\mathbf{l}_{N})]\in\mathbb{C}^{K\times N}$, is then given by
\begin{equation}
    \mathbf{H}=\mathbf{C}\mathbf{T},
\end{equation}
where matrices ${\mathbf{C}}\in \mathbb{C}^{K\times NQ}$ and $\mathbf{T}\in \mathbb{C}^{NQ\times N}$ are defined as follows, respectively,
\begin{eqnarray}
\mathbf{C}=
[\mathbf{C}_{1},\ldots,\mathbf{C}_{N}]\;\text{and}\;
\mathbf{T}=  \begin{bmatrix}
    \mathbf{t}_1 & \mathbf{0}_{Q} & \mathbf{0}_{Q} & \ldots & \mathbf{0}_{Q}\\
    \mathbf{0}_{Q} & \mathbf{t}_2 & \mathbf{0}_{Q} &\ldots & \mathbf{0}_{Q}\\
    \ldots & \ldots & \ldots & \ldots & \ldots\\
    \mathbf{0}_{Q} & \mathbf{0}_{Q} & \mathbf{0}_{Q} & \hspace*{1mm}\ldots & \mathbf{t}_{N}
  \end{bmatrix}.
\end{eqnarray}
Next, we define $\hat{\mathbf{h}}_k^{\rm H}\in\mathbb{C}^{1 \times NQ}$ as the $k$-th row of $\mathbf{C}$, which denotes the channel vector between the BS and user $k$.
Then, the received signal of user $k$ is given by
\begin{equation}\label{Eqn:MFA-NGMA-SystemModel}
    y_k=\hat{\mathbf{h}}_k^{\rm H}\mathbf{T}\mathbf{P}\mathbf{s}+z_k,
\end{equation}
where $\mathbf{P}=[\mathbf{p}_1,\ldots,\mathbf{p}_K] \in \mathbb{C}^{N \times K}$ denotes the precoding matrix at the BS. For notational simplicity, we define sets $\mathcal{K}\in\{1,\ldots,K\}$, $\mathcal{N}\in\{1,\ldots, N\}$, and $\mathcal{Q}\in\{1,\ldots,Q\}$ to collect the indices of the users, M/FA elements, and candidate positions of the antenna elements, respectively.
\par
By introducing an auxiliary matrix $\mathbf{U}=\mathbf{T}\mathbf{P},\ \mathbf{U}\in\mathbb{C}^{NQ\times K}$, the received signal of user $k$ can be rewritten as
\begin{equation}
y_k=\hat{\mathbf{h}}_k^{\rm H}\mathbf{U}\mathbf{s}+z_k.
\end{equation}
Thus, the SINR of user $k$ is given by
\begin{equation}\label{Eqn:MFA-SINRk}
\Gamma_k=\frac{|\hat{\mathbf{h}}_k^H\mathbf{u}_k|^2}{\sum_{r\in\mathcal{K}\setminus\{k\}}|\hat{\mathbf{h}}_k^H\mathbf{u}_{r}|^2+\sigma_{k}^2},
\end{equation}
where $\mathbf{u}_k$ denotes the $k$-th column of $\mathbf{U}$.
\par
The distinctive features of the M/FA system are exceptionally well-suited for enhancing NGMA schemes within wireless networks.  In particular, in case some of the users suffer from unfavorable wireless channel conditions,the M/FA enables the reconfiguration of the antenna array to cultivate a more conducive radio propagation environment. For example, the positions of specific antenna elements can be modified to bypass obstructions between the antenna array and the users. Alternatively, we may assign a subset of the antenna elements to dedicatedly serve a user whose channel condition is poor. Moreover, due to the high maneuverability of the antenna elements, we can flexibly adjust the spacing between adjacent antenna elements to manipulate the correlation and coupling of the antenna elements for better interference management in NGMA.
\par
In summary, the resource allocation design of NGMA for reconfigurable channels requires the joint design of channel reconfiguring and resource allocation variables, such as joint passive and active beamforming design for IRS-assisted communications and joint trajectory and resource allocation design for UAV-assisted communications. 
Compared to natural channels, the capability of channel reconfiguration provides additional DoF for accommodating more users and further improving the system performance.
%
In Table \ref{tab:tablecRC}, we summarize the DoFs for NGMA design, limitations, and implementation complexity of different channel reconfiguration techniques. We can observe that although the channel reconfiguration techniques introduce additional flexibility in resource allocation design, they also bring new challenges. In particular, to fully exploit the potential of the UAV-assisted NGMA, the dynamic and uncertain environments have to be taken into account when designing the trajectory and transmit precoding of the UAV. Also, no-fly zones and collision avoidance should be considered for safety issues. Moreover, considering the fact that the RF chains are usually absent at the IRSs, advanced channel estimation methods should be developed to facilitate the acquisition of accurate CSI, which enables efficient IRS-assisted NGMA. Furthermore, compared to the coherence time of the wireless channels, the reconfiguration speed of the M/FA systems is significantly slower. As such, it is currently impractical to reconfigure the M/FA systems per time slot. As a result, advanced hardware should be developed to facilitate fast array reconfiguration. Also, a more delicate and suitable frame structure should be proposed to perfectly accommodate the NGMA schemes in M/FA systems.
\begin{table}[t]
  \centering
  \caption{Comparison of different channel reconfiguration techniques}
    \begin{tabular}{c|c|c}
    \hline 
     {Techniques} & {DoFs} & 
     {Limitations} \\ 
     \hline
     {UAV} & UAV position & Weather dependent, safety issue \\  
     {IRS} & Phase, amplitude & Difficult to obtain accurate CSI \\ 
     {M/FA} & Antenna position, correlation & Relatively low array reconfiguration speed \\ \hline 
    \end{tabular}%
  \label{tab:tablecRC}%
\end{table}%

\subsection{NGMA in Functional Channels}
\subsubsection{A Unified Framework for NGMA in Functional Channels}
Future cellular networks are envisioned to support various wireless functionalities, extending well beyond the scope of traditional communication to satisfy the demands of emerging applications and use cases \cite{10349846,WeijieLetterPartIII,10266619}.
This evolution is propelled by the growing need for sensing and computing within wireless networks, which makes the discussion of the performance of multiple functionalities under limited network resources more meaningful~\cite{Fan2020,Jinke2019JCAC}.
In this paper, we refer to wireless channels that must support multiple functionalities simultaneously as functional channels.
Particularly, the functional channel inevitably introduces new challenges and opportunities for NGMA. Firstly, in functional channels, the concurrent optimization of multiple functionalities is imperative, indicating that traditional multiple access strategies tailored for natural channels might not seamlessly apply.  Secondly, the integration of multiple functionalities opens up innovative design avenues for NGMA by strategically utilizing the outcomes of these functionalities. For example, in ISAC channels, the sensing results can be leveraged to predict the future CSI of the users, which helps improve the communication performance~\cite{Weijie2021JSTSP,Shuangyang2022ISAC}. Specifically, sensing-assisted communication and communication-assisted sensing transmissions are shown to outperform the communication-only and sensing-only cases~\cite{Shuangyang2022ISAC,Weijie2021JSTSP,chang2022learning}, respectively. More interestingly, it was shown in~\cite{Kobayashi2018Joint} that a careful design of the transmitted signal can improve the performance of both functionalities simultaneously. Therefore, the presence of functional channels may be beneficial for NGMA designs.
In line with~\eqref{Eqn:NaturalChannelModel}, it is convenient to discuss the NGMA designs in a functional channel according to the following model
\begin{align}\label{Eqn:ISACChannelModel_new}
    {\left[ {{\bf{y}}_{\rm{C}}^{\rm{T}},{\bf{y}}_{\rm{F}}^{\rm{T}}} \right]^{\rm{T}}} = \left[ \begin{array}{l}
{{\bf{H}}_{\rm{C}}}\\
{{\bf{H}}_{\rm{F}}}
\end{array} \right]{\bf g}\left( {\bf{s}} , {\bf{d}}\right) + {\left[ {{\bf{z}}_{\rm{C}}^{\rm{T}},{\bf{z}}_{\rm{F}}^{\rm{T}}} \right]^{\rm{T}}},
\end{align}   
where ${\bf y}_{\rm C} \in {\mathbb C}^{M_{\rm C} \times 1}$ and ${\bf y}_{\rm F}\in {\mathbb C}^{M_{\rm F} \times 1}$ are the channel observations for the communication functionality and the other functionalities, respectively,  e.g., radar sensing, and $M_{\rm C}$ and $M_{\rm F}$ are the numbers of resource elements available at receivers. Furthermore, ${\bf{s}}\in {\mathbb C}^{D_{\rm C} \times 1}$ and ${\bf{d}}\in {\mathbb C}^{D_{\rm F} \times 1}$ are the symbol vectors for communication information and the other functionalities, respectively,  e.g., computing, and $D_{\rm C}$ and $D_{\rm F}$ are the numbers of symbols required by the communication and the other functionalities, respectively. Also, $\mathbf{H}_{\mathrm{C}} \in {\mathbb C}^{M_{\mathrm{C}} \times N}$ and $\mathbf{H}_{\mathrm{F}} \in {\mathbb C}^{M_{\mathrm{F}} \times N}$ denote the communication channel and the channel for other functionalities, respectively, with $N$ being the number of resource elements available at the transmitter, while ${\bf z}_{\rm C}\in {\mathbb C}^{M_{\rm C} \times 1}$ and ${\bf z}_{\rm F}\in {\mathbb C}^{M_{\rm F} \times 1}$ are the corresponding AWGN vectors for different functionalities.
From~\eqref{Eqn:ISACChannelModel_new}, we notice that the function ${\bf g}\left(\cdot\right): {\mathbb C}^{D \times 1} \to {\mathbb C}^{N \times 1}$ in~\eqref{Eqn:ISACChannelModel_new} maps both the symbols for communications and other functionalities to system resources, where $D=D_{\rm C}+D_{\rm F}$ is the total number of symbols required by all functionalities. Different from~\eqref{Eqn:NaturalChannelModel}, the function ${\bf g}\left(\cdot\right)$ needs to be designed carefully by jointly considering the requirements of various functionalities.
{To shed light on the specific NGMA designs, we propose to discuss the importance of NGMA for two representative types of functional channels in the following, namely the ISAC channel and the JCAC channel, respectively.
Note that learning can be viewed as a special case of the computing task. For instance, in federated learning \cite{TranFL}, the local computing task at wireless terminals corresponds to updating the local neural network model, the offloading traffic to the edge server is analogous to the neural network parameters, and the computing task at the edge server mirrors the aggregating neural network models. To shed light on the specific NGMA designs, we propose to discuss the importance of NGMA for two representative types of functional channels in the following, namely the ISAC channel ~\cite{Fan2020} and the JCAC channel ~\cite{Sun2019JCAC}, respectively.}

\begin{figure}[t]
    \centering
    \includegraphics[width=5.4in]{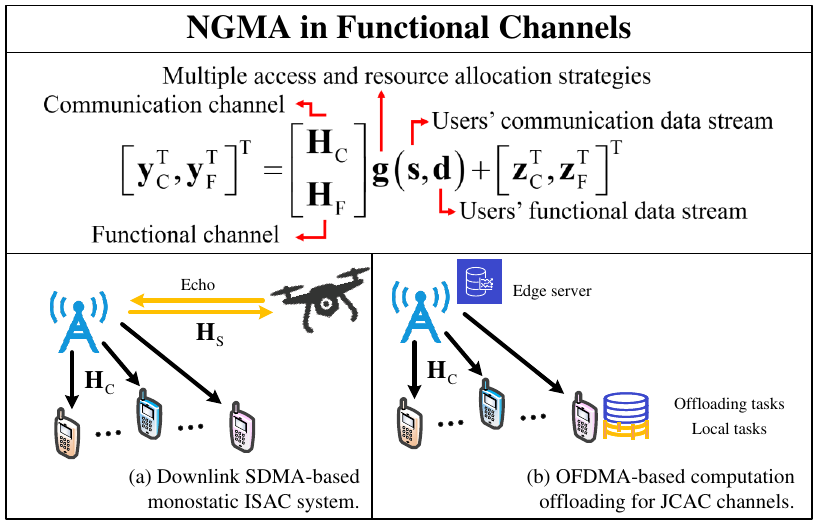}
    \caption{NGMA in functional channels.}
    \label{Fig.NGMA_Func}
\end{figure}
    
\subsubsection{NGMA for ISAC Channels}
ISAC represents a paradigm shift in the design and development of wireless networks, emphasizing providing communication and sensing functions into a single unified platform. 
Typically, the ISAC system can be described by the following general form \cite{Fan2020}
\begin{align}\label{Eqn:ISACChannelModel_new}
    {\left[ {{\bf{y}}_{\rm{C}}^{\rm{T}},{\bf{y}}_{\rm{S}}^{\rm{T}}} \right]^{\rm{T}}} = \left[ \begin{array}{l}
{{\bf{H}}_{\rm{C}}}\\
{{\bf{H}}_{\rm{F}}}
\end{array} \right]{\bf g}\left( {\bf{s}} \right) + {\left[ {{\bf{z}}_{\rm{C}}^{\rm{T}},{\bf{z}}_{\rm{S}}^{\rm{T}}} \right]^{\rm{T}}},
\end{align}   
where $\mathbf{H}_{\mathrm{C}} \in {\mathbb C}^{M_{\mathrm{C}} \times N}$ and $\mathbf{H}_{\mathrm{S}} \in {\mathbb C}^{M_{\mathrm{S}} \times N}$ denote the communication and sensing channels, respectively,  $\mathbf{y}_{\mathrm{C}}\in {\mathbb C}^{ M_{\mathrm{C}} \times 1 }$ and $\mathbf{y}_{\mathrm{S}}\in {\mathbb C}^{ M_{\mathrm{S}} \times 1 }$ denote the received signals at the communication and sensing receivers, respectively, and
$\mathbf{z}_{\mathrm{C}}\in {\mathbb C}^{ M_{\mathrm{C}} \times 1}$ and $\mathbf{z}_{\mathrm{S}}\in {\mathbb C}^{ M_{\mathrm{S}} \times  1}$ denote the corresponding communication and sensing noise vectors, respectively. 
Here, $M_{\mathrm{C}}$ and $M_{\mathrm{S}}$ denote the numbers of resource elements available at the receiver for communication and radar sensing, respectively, and  ${\bf g}\left( {\cdot} \right)$ represent the resource allocation strategy that maps the information symbols ${\bf s} \in {\mathbb C}^{D\times 1}$ to the transmitted symbols ${\bf x} \in {\mathbb C}^{N\times 1}$.
Particularly, it is important to highlight that the parameters 
the values of $M_{\mathrm{C}}$, $M_{\mathrm{S}}$, and $N$ can be different due to the available system resource elements at the transmitter and receiver for both communication and radar sensing functionalities. 

Denote by ${\bf x} \in {\mathbb C}^{N\times 1}$ the transmitted symbol vector after mapping. We shall rewrite~\eqref{Eqn:ISACChannelModel_new} by 
\begin{align}\label{Eqn:ISACChannelModel}
\mathbf{y}_{\mathrm{C}} &= \mathbf{H}_{\mathrm{C}}\mathbf{x} + \mathbf{z}_{\mathrm{C}}\;\text{and} \notag\\
\mathbf{y}_{\mathrm{S}} &= \mathbf{H}_{\mathrm{S}}\mathbf{x} + \mathbf{z}_{\mathrm{S}},
\end{align}
which aligns with the conventional ISAC models~\cite{Yifeng2023fundamental}.
From~\eqref{Eqn:ISACChannelModel_new} and~\eqref{Eqn:ISACChannelModel}, it is clear that the resource allocation problem for functional channels is to design an effective mapping function for given communication and sensing channels $\mathbf{H}_{\mathrm{C}}$ and $\mathbf{H}_{\mathrm{S}}$. However, the major difference of such a resource allocation problem in comparison to the one discussed in Section II-A is the fact that ${\bf g}\left( {\cdot} \right)$ needs to exploit the connection between $\mathbf{H}_{\mathrm{C}}$ and $\mathbf{H}_{\mathrm{S}}$ for simultaneously satisfying both the communication and sensing requirements. In the following, we present an illustrative example system for enabling dual functionality of sensing and communication.

\textbf{Downlink SDMA-based Monostatic ISAC system}: Let us examine a downlink monostatic ISAC multiple access transmission with dual-functional MIMO arrays at the BS, {as shown in Fig. \ref{Fig.NGMA_Func} (a)}. The BS is equipped with a ULA with $N$ antennas serving $K$ single-antenna users while capturing backscattered signals from targets for radar detection and estimation purposes. We consider the case where each user transmits a data stream of length $L$ denoted by ${\bf s}_k \in {\mathbb C}^{L \times 1}$, i.e., $D = K$, and we assume that $M_{\rm C}=M_{\rm S}=K$ and $N$ are the numbers available resource elements.
Furthermore, assume that the system operates over a flat fading channel with a sufficiently long coherence time. In accordance to~\eqref{Eqn:ISACChannelModel}, the underlying ISAC transmission is modeled by
\begin{align}\label{Eqn:ISACChannelModelExample2}
{{\bf{Y}}_{\rm{C}}} &= {{\bf{H}}_{\rm{C}}}{\bf{PS}} + {{\bf{Z}}_{\rm{C}}}\;\text{and} \notag\\
{{\bf{Y}}_{\rm{S}}} &= {{\bf{H}}_{\rm{S}}}{\bf{PS}} + {{\bf{Z}}_{\rm{S}}},
\end{align}
where ${\bf S}= {\left[ {{{\bf{s}}_1},{{\bf{s}}_2},...,{{\bf{s}}_K}} \right]^{\rm{T}}} \in {\mathbb C}^{K \times L}$ is the information matrix to be transmitted, ${\bf P} \in {\mathbb C}^{N \times K}$ is the underlying precoding matrix, and ${{\bf{Y}}_{\rm{C}}} = {\left[ {{\bf{y}}_{\rm{C}}^{\left( 1 \right)},{\bf{y}}_{\rm{C}}^{\left( 2 \right)},...,{\bf{y}}_{\rm{C}}^{\left( K \right)}} \right]^{\rm{T}}} \in {\mathbb C}^{K \times L}$ and ${{\bf{Y}}_{\rm{S}}} = {\left[ {{\bf{y}}_{\rm{S}}^{\left( 1 \right)},{\bf{y}}_{\rm{S}}^{\left( 2 \right)},...,{\bf{y}}_{\rm{S}}^{\left( K \right)}} \right]^{\rm{T}}}  \in {\mathbb C}^{K \times L}$ are the communication and radar received symbol matrices, respectively.
In~\eqref{Eqn:ISACChannelModelExample2}, ${{\bf{H}}_{\rm{C}}} =\left[{\bf h}_{\rm C}^{(1)},{\bf h}_{\rm C}^{(2)},...,{\bf h}_{\rm C}^{(K)}\right]^{\rm T}\in {\mathbb C}^{K \times N}$ and ${{\bf{H}}_{\rm{S}}} =\left[{\bf h}_{\rm S}^{(1)},{\bf h}_{\rm S}^{(2)},...,{\bf h}_{\rm S}^{(K)}\right]^{\rm T}\in {\mathbb C}^{K \times N}$, where ${\bf h}_{\rm C}^{(k)} \in {\mathbb C}^{N \times 1}$ and ${\bf h}_{\rm S}^{(k)}\in {\mathbb C}^{N \times 1}$ are the channel vectors for communication and sensing of  user $k$, respectively.

We propose to study the resource allocation in this example by optimizing the precoding matrix ${\bf P}$, such that the received signal at each user requires little or no equalization, while the transmitted signals at the BS mimic a certain beampattern \cite{Fan2018Toward} that is favorable for radar sensing. More specifically, by rearranging the communication input-output relation into
\begin{align}\label{Eqn:ISACChannelModelExample2_com_io}
{{\bf{Y}}_{\rm{C}}} = {\bf{S}} + \left( {{{\bf{H}}_{\rm{C}}}{\bf{P}} - {{\bf{I}}_{K}}} \right){\bf{S}} + {{\bf{Z}}_{\rm{C}}},
\end{align}
we can define the effective SINR by
\begin{align}\label{Eqn:ISACChannelModelExample2_SINR}
\Gamma_k = \frac{{\frac{1}{L}{{\left\| {{{\bf{s}}_k}} \right\|}^2}}}{{\frac{1}{L}{{\left\| {{{\left( {{\bf{h}}_{\rm{C}}^{\left( k \right)}} \right)}^{\rm{T}}}{\bf{PS}} - {{\bf{s}}_k}} \right\|}^2} + {\sigma^2_k}}}.
\end{align}
Based on \eqref{Eqn:ISACChannelModelExample2_SINR}, the achievable rate for user $k$ is calculated by 
\begin{align}\label{Eqn:ISAC_example2_rate}
{R_k} = {\log _2}\left( {1 + \Gamma_k} \right),
\end{align}
which can be adopted to evaluate communication performance. On the other hand, the radar performance is evaluated by deriving the difference between the transmit beampattern and the desired radar beampattern. Let ${\bf R}_{{\bf X}_0}=\frac{1}{L}{\bf X}_0{\bf X}_0^{\rm H}$ be the radar beampattern of interest, where ${\bf X}_0$ is the desired radar transmitted signal~\cite{Fan2018Toward}. Then, the radar performance is characterized by the MSE of the form
\begin{align}\label{Eqn:ISACChannelModelExample2_radar_MSE}
{\rm MSE} = {\frac{1}{L}{{\left\| {\bf PS}-{\bf X}_0\right\|}_{\rm F}^2}}.
\end{align}
Note that a wide range of sensing metrics exists within the field\cite{Fan2020}. However,  for the purposes of this analysis, we have deliberately selected the MSE between the transmitted and the sensing-desired signals as the metric for assessing radar performance. This choice is motivated by the MSE's relatively straightforward formulation for resource allocation and its direct influence on the actual performance of the radar system.

Based on~\eqref{Eqn:ISAC_example2_rate} and~\eqref{Eqn:ISACChannelModelExample2_radar_MSE}, we observe that the choice of $\bf P$ will directly affect the signal beampattern and the communication achievable rate. Therefore, we note that the resource allocation in the above example boils down to the design of $\bf P$ for providing a desired tradeoff between communication and radar sensing performance. Note that the desired tradeoff may be different according to the NGMA requirements. For example, in communication-centric designs, the sensing performance is less important, and therefore the resource allocation should lean towards the communication functionality, where $\bf P$ would ideally resemble a zero-forcing precoder that aims to null the interference caused by sensing signals. On the other hand, in sensing-centric designs, where the priority is on the radar sensing capabilities over communication performance, resource allocation is oriented more towards enhancing the sensing functionality. Consequently, the design of $\bf P$ should aim to align $\bf P S$ closely with the desired signal ${\bf X}_0$, evaluated by the Frobenius norm as specified in \eqref{Eqn:ISACChannelModelExample2_radar_MSE}. In summary, the resource allocation strategy for NGMA within the ISAC system framework necessitates a proper balance between communication and radar sensing functionalities, tailored according to the constraints of available resource elements. This requirement for equilibrium poses a significant difference from traditional resource allocation approaches in NGMA for natural channels, underlining the unique challenges and considerations involved in optimizing integrated systems that offer multiple functionalities.

\subsubsection{NGMA for JCAC Channels}
Mobile edge computing (MEC) is recently emerged as a potential solution to relieve the burden of backhaul links in wireless networks\cite{Yuyi2017}. Specifically, the MEC protocol aims to move the computing resources from the central network towards the network edges. In this context, both communication and computing functionalities are involved and therefore it falls in the category of JCAC transmissions~\cite{Feng2019multiantenna}.

Let us consider the NGMA design for JCAC channels, where $K$ users aim to compute their tasks exploiting the MEC protocol and transmit information to the BS simultaneously, {as shown in Fig. \ref{Fig.NGMA_Func} (a)}.
To leverage the computation resource at the BS and fulfill the constraint of latency, each user separates their tasks into two parts computed either at the edge or at the center via offloading and downloading. In this context, the channel model in \eqref{Eqn:ISACChannelModel_new} is further specified by
\begin{align}\label{Eqn:JCICChannelModel}
{\bf{y}} = {\bf{H}}\mathbf{g}\left( {{\bf{s}},{\bf{d}}} \right) + {\bf{z}},  
\end{align}
where ${\bf y} \in {\mathbb C}^{M \times 1}$ is the received signal at the BS containing information for both communication and computing, ${\bf H} \in {\mathbb C}^{M \times N}$ is the corresponding channel matrix, and  ${\bf z} \in {\mathbb C}^{M \times 1}$ is the AWGN vector. In~\eqref{Eqn:JCICChannelModel}, ${\bf s} \in {\mathbb C}^{D_{\rm C}\times 1}$ and ${\bf d} \in {\mathbb C}^{D_{\rm MEC}\times 1}$ are the information symbols to be transmitted and task symbols describing the computation tasks of all users.
Based on~\eqref{Eqn:JCICChannelModel}, we notice that ${\bf g}\left( {\cdot} \right)$ needs to satisfy both the communication and computing requirements, which requires the resource allocation taking into account both communication and computation resource, e.g., energy consumption. 
The following example provides further details on designing 
${\bf g}\left( {\cdot} \right)$.

\textbf{OFDMA-based Computation Offloading for JCAC channels}: Let us consider the computation offloading issue for JCAC transmissions, where the $K$ users communicate and offload part of their tasks to the center BS simultaneously via OFDMA while computing the rest of their tasks locally \footnote{The consideration of OFDMA serves as an illustrative example. The extension to a non-orthogonal framework can be achieved by following a similar approach as in previous sections.}.  
For ease of exposition, we ignore the transmission where users collect their task symbols after computing from the central BS via the downlink~\cite{Feng2019multiantenna}. 
Let $M$ be the number of subcarriers available in the OFDMA system and we have $N=M$. Considering  transmissions over static channels, we shall further specify~\eqref{Eqn:JCICChannelModel} based on~\eqref{Eqn:NaturalChannelOFDMA} by
\begin{align}\label{Eqn:JCICChannelModelExample}
{\bf{y}} = \sum\limits_{k = 1}^K {{{\bf{H}}_k}{{\bf{\Pi }}_k}{{\bf{\Lambda }}_k}{{\bf{x}}_{{k}}} + {\bf{z}}} ,
\end{align}
where ${{\bf{x}}_{k}} = \left[{{\bf{s}}_{{k}}^{\rm T}}, {{\bf{d}}_{{k}}^{\rm T}}\right]^{\rm T} \in {\mathbb C}^{D_{{\rm Tot},k} \times 1}$ are the transmitted symbols of user $k$, and the rest of notations follow the same definition given in~\eqref{Eqn:NaturalChannelOFDMA}. Specifically, ${{\bf{s}}_{k}}$ is a subset of the total $D_{\rm C}$ data streams $\bf s$ with cardinality 
${{D_{{\rm C},k}}}$, containing the corresponding information symbols of user $k$. 
Similarly, ${{\bf{d}}_{{k}}}$ is a subset of the total $D_{\rm MEC}$ task symbols $\bf d$ with cardinality  
${{D_{{\rm MEC},k}}}$, containing the corresponding computation task symbols of user $k$. Hence, we have $D_{{\rm Tot},k}=D_{{\rm C},k}+D_{{\rm MEC},k}$.
Following the same derivation as for~\eqref{Eqn:NaturalChannelSCOFDMA}, we shall obtain the input-output relations of user $k$ by highlighting both the communication and computing functionalities in the form of
\begin{align}\label{Eqn:JCICChannelModelExample_Com}
y_m =& \sum\limits_{k = 1}^K {\sum\limits_{{l_{\rm C}} = 1}^{{D}_{{\rm C},k}} {{{\{ {{\bf{\Pi }}_k}\} }_{m{l_{\rm C}}}}} } {\{ {{\bf{H}}_k}\} _{mm}}\sqrt {{p_{k{l_{\rm C}}}}} {\{ {{\bf{s}}_{k}}\} _{{l_{\rm C}}}} \notag\\
&+\sum\limits_{k = 1}^K{\sum\limits_{{l_{\rm MEC}} = 1}^{{D}_{{\rm MEC},k}} {{{\{ {{\bf{\Pi }}_k}\} }_{m{\left({D}_{{\rm C},k}+l_{\rm MEC}\right)}}}} } {\{ {{\bf{H}}_k}\} _{mm}}\sqrt {{p_{k{\left({D}_{{\rm C},k}+l_{\rm MEC}\right)}}}} {\{ {{\bf{d}}_{k}}\} _{{l_{\rm MEC}}}}+ z_m,\forall m.
\end{align}
Notice that entries in ${\bf s}_k$ and ${\bf d}_k$ do not share the same subcarrier.
We can characterize the communication rate of user $k$ by
\begin{align}\label{Eqn:JCICChannelRate_Com}
{R_{{\rm{C,}}k}} = \sum\limits_{{l_{\rm C}} = 1}^{D_{{\rm C},k}} {\sum\limits_{m = 1}^M {{{\{ {{\bf{\Pi }}_k}\} }_{ml_{\rm C}}}} } {\log _2}\left( {1 + \frac{{{{\left| {{{\{ {{\bf{H}}_k}\} }_{mm}}} \right|}^2}{p_{k{{l_{\rm C}}}}}}}{{{\sigma ^2}}}} \right).
\end{align}
Similarly, the achievable rate for task offloading of user $k$ is given by
\begin{align}\label{Eqn:JCICChannelRate_Computing}
{R_{{\rm{MEC}},k}} = \sum\limits_{{l_{\rm MEC}} = 1}^{{D_{{\rm MEC},k}}} {\sum\limits_{m = 1}^M {{{\{ {{\bf{\Pi }}_k}\} }_{m{\left({D_{{\rm C},k}}+{l_{\rm MEC}}\right)}}}} } {\log _2}\left( {1 + \frac{{{{\left| {{{\{ {{\bf{H}}_k}\} }_{mm}}} \right|}^2}{p_{k\left({D_{{\rm C},k}}+{l_{\rm MEC}}\right)}}}}{{{\sigma ^2}}}} \right).
\end{align}
In addition to the achievable rates, it is important to characterize the total consumed energy for each device to transmit symbols and to locally compute part of the task using the central processing
unit (CPU). According to~\cite{Feng2019multiantenna}, the total energy consumption is given by
\begin{align}\label{Eqn:JCACChannelModelExample_energy}
{E_{{\rm{tot}}}} = \sum\limits_{k = 1}^K \left({\frac{{{\varsigma _k}C_k^3L_k^3}}{{{{\tilde T}^2}}} + \sum\limits_{l = 1}^{D_{{\rm Tot},k}} {{p_{kl}}}{{T}}} \right).
\end{align}
In~\eqref{Eqn:JCACChannelModelExample_energy}, $\tilde T$ is the latency allowed for all users finishing computing the task, while $T$ is the allowed signal transmission time to meet the latency according to the computation ability of the BS that is the inverse of the total bandwidth. Furthermore, $L_k$ is the number of bits left for user $k$ to compute locally, $C_k$ is the number of CPU cycles for computing one input-bit for user $k$ that is chosen to guarantee the task completion within $\tilde T$, ${{\varsigma _k}}$ is the effective capacitance coefficient that depends on the chip architecture of user $k$.  
On the other hand, the term $\sum\nolimits_{l = 1}^{D_{{\rm Tot},k}} {{p_{kl}}}{{T}}$ denotes the average energy required for the signal transmission of user $k$.
Finally,~\eqref{Eqn:JCACChannelModelExample_energy} can be understood as the summation of each user's consumed energy including the energy for both task offloading, communication symbol transmission, and local computing.  
In fact, ${E_{{\rm{tot}}}}$ in~\eqref{Eqn:JCACChannelModelExample_energy} depends on the achievable rate ${R_{{\rm{MEC}},k}} $ given in~\eqref{Eqn:JCICChannelRate_Computing}. Let ${\tilde L}_k$ be the number of bits describing the task of user $k$. Then, in order to guarantee the computing functionality, $L_k+{R_{{\rm{MEC,}}k}} \ge {\tilde L}_k$ must hold. 
From~\eqref{Eqn:JCICChannelRate_Com},~\eqref{Eqn:JCICChannelRate_Computing}, and~\eqref{Eqn:JCACChannelModelExample_energy}, we also observe that the resource allocation for the JCAC transmission needs to consider the tradeoff between communication rate, computing efficiency, and the total energy consumption. Particularly, we observe that the computing efficiency is translated to the latency constraint as included in~\eqref{Eqn:JCACChannelModelExample_energy}. Therefore, the resource allocation problem is reduced to find the desired balance between the communication rate and the total energy consumption of all users.

\section{Resource Allocation Design for Next-Generation Multiple Access}\label{SectionIII}
In this section, we present the typical problem formulations for resource allocation design of different types of NGMA.
\subsection{Resource Allocation Design for NGMA in Natural Channel}

Resource allocation design for NGMA in natural channels can be categorized into three types according to the design orientations, i.e., rate-oriented, power-oriented, and reliability-oriented resource allocation design.
In particular, the rate-oriented resource allocation is suitable for rate-demanding applications or scenarios.
Power-oriented resource allocation intends to reduce the system power consumption, which fits scenarios where power is a limited system resource, such as for massive IoT devices.
Besides, reliability-oriented resource allocation aims to enhance the system robustness against the existence of imperfect CSI.
In this subsection, we present the problem formulations for these three types of resource allocation designs for NGMA in natural channels.
\subsubsection{Rate-oriented Resource Allocation Design \cite{Hanif2016,7812683,ZhiqiangOFDMA}}
We consider the downlink power domain NOMA system in \eqref{Eqn:NaturalChannelNOMA} and formulate the rate-oriented resource allocation design problem.
The main motivation of NOMA is accommodating more users and improving the spectral efficiency when the system spectrum resource is limited.
In particular, the power allocation for each data stream at each user and user scheduling strategy are jointly designed to maximize the system sum rate.
The system sum rate maximization problem can be formulated as the following optimization problem:
\begin{eqnarray}
\label{NCprob_1}
&&\underset{p_{k},\{\mathbf{\Pi}\}_{mk}}{\maxo} \,\, \,\, \underset{ k\in\mathcal{K}}{\sum} {R}_k\notag\\
&&\mbox{subject\hspace*{1.6mm}to}\hspace*{4mm}
\mbox{C1:}\hspace*{1mm}\sum_{k=1}^{K} p_{k} \le P_{\mathrm{max}},\notag\\
&&\hspace*{21mm}\mbox{C2:}\hspace*{1mm}
\sum_{m=1}^{M}\{\mathbf{\Pi}\}_{mk} = 1,\hspace*{1mm}\forall k,\notag\\
&&\hspace*{21mm}\mbox{C3:}\hspace*{1mm}
{R}_k \ge R_{\mathrm{min},k},\hspace*{1mm}\forall k,\notag\\
&&\hspace*{21mm}\mbox{C4:}\hspace*{1mm}
\{\mathbf{\Pi}\}_{mk} \in\left\{0, 1\right\},\hspace*{1mm}\forall m,\hspace*{1mm}\forall k,
\end{eqnarray}
\noindent
where ${R}_k$ is given by \eqref{Eqn:NaturalChannelSCNOMARate}.
C1 guarantees that the total transmit power does not exceed the power budget at the BS $P_{\mathrm{max}}$.
C2 limits that each user can only be allocated to at most one subcarrier. 
%
%
Constraint C4 is imposed to ensure variable $\{\mathbf{\Pi}\}_{mk}$ to be an indicator. 
This optimization problem is also non-convex due to the non-convex objective function (NOF), variable coupling (VC) between $p_{k}$ and $\{\mathbf{\Pi}\}_{mk}$, fractional constraint (FC) C3, and binary constraint (BC) C4.

\subsubsection{Power-oriented Resource Allocation Design \cite{DerrickEEOFDMA,DerrickFD2016,Wei2017,Lei2016NPM}}
We consider the uplink OFDMA system in \eqref{Eqn:NaturalChannelOFDMA} and formulate a power-oriented resource allocation design problem. 
The power allocation and the data stream scheduling strategy should be jointly designed to minimize the total transmit power of all users while satisfying the QoS requirement of each user.
In particular, the power minimization problem can be formulated as the following optimization problem:
\begin{eqnarray}
\label{NCprob_2}
&&\underset{p_{kd_k},\{\mathbf{\Pi}_k\}_{md_k}}{\mino} \,\, \,\, \underset{ k\in\mathcal{K}}{\sum}\sum_{d_k=1}^{D_k} p_{kd_k}\notag\\
&&\mbox{subject\hspace*{1.6mm}to}\hspace*{4mm}
\mbox{C1:}\hspace*{1mm}\sum_{d_k=1}^{D_k} p_{kd_k} \le P_{k,\mathrm{max}},\hspace*{1mm}\forall k,\notag\\
&&\hspace*{21mm}\mbox{C2:}\hspace*{1mm}
\sum_{k=1}^{K}\sum_{d_k=1}^{D_k}\{\mathbf{\Pi}_k\}_{md_k} \le 1,\hspace*{1mm}\forall m,\notag\\
&&\hspace*{21mm}\mbox{C3:}\hspace*{1mm}
{R}_k \ge R_{\mathrm{min},k},\notag\\
&&\hspace*{21mm}\mbox{C4:}\hspace*{1mm}
\{\mathbf{\Pi}_k\}_{md_k} \in\left\{0, 1\right\},\hspace*{1mm}\forall m,\hspace*{1mm}\forall k,\hspace*{1mm}\forall d_k,
\end{eqnarray}
\noindent
where ${R}_k$ is given by \eqref{Eqn:NaturalChannelOFDMARate}.
C1 guarantees that the total transmit power of user $k$ does not exceed its power budget $P_{k,\mathrm{max}}$.
Besides, the user scheduling variables must obey C2 such that at most one data stream of a user can be scheduled exclusively to one subcarrier for maintaining orthogonality in resource allocation. 
Constant $R_{\mathrm{min},k}$ in C3 denotes the minimum required data rate of user $k$ and C3 is introduced to guarantee the QoS of each user.
This optimization problem is also non-convex due to the VC between $p_{kd_k}$ and $\{\mathbf{\Pi}_k\}_{md_k}$, FC in C3, and BC in C4.

\subsubsection{Reliability-oriented Resource Allocation Design \cite{WeiProceeding,WangWorstCase}}
For the case of imperfect CSI at the transmitter (CSIT), the resource allocation design for NGMA only has the knowledge of channel estimate $\hat{\mathbf{h}}_k$, which may suffer from the channel estimation error as follows:
\begin{equation}
    {\mathbf{h}}_k = \hat{\mathbf{h}}_k + \Delta{\mathbf{h}}_k,
\end{equation}
where $\Delta{\mathbf{h}}_k$ is the channel estimation error.
The channel estimation error can be modeled by a Gaussian model $\Delta{\mathbf{h}}_k \sim \mathcal{CN}\left(\mathbf{0},\mathbf{C}^k_{\Delta}\right)$ or a bounded model $\left\|\Delta{\mathbf{h}}_k\right\| \le \delta_k$, where $\mathbf{C}^k_{\Delta}$ is the covariance matrix of the channel estimation error and $\delta_k$ is the hyper-sphere radius where $\Delta{\mathbf{h}}_k$ is located in.
A reliability-oriented resource allocation design should be considered to guarantee communication performance even in the presence of channel estimation error.
We consider the downlink RSMA system in \eqref{Eqn:NaturalChannelRSMA_Rxk} and formulate a robust resource allocation design problem concerning the worst-case achievable rates.
{color{blue}In the case of imperfect CSIT for RSMA, the resource allocation design is formulated as the following optimization problem:}
\begin{eqnarray}
\label{NCprob_3}
&&\underset{\mathbf{p}_{\mathrm{p},k},\mathbf{p}_{\mathrm{c}},C_k}{\maxo} \,\, \,\, \sum_{k=1}^{K} \left(\tilde{R}_{\mathrm{p},k} + C_k\right)\notag\\
&&\mbox{subject\hspace*{1.6mm}to}\hspace*{4mm}
\mbox{C1:}\hspace*{1mm}\sum_{k=1}^{K} C_k \le \min\left\{\tilde{R}_{\mathrm{c},1},\ldots,\tilde{R}_{\mathrm{c},K}\right\},\notag\\
&&\hspace*{21mm}\mbox{C2:}\hspace*{1mm}
\sum_{k=1}^{K}\left\|\mathbf{p}_{\mathrm{p},k}\right\|^2 + \left\|\mathbf{p}_{\mathrm{c}}\right\|^2 \le P_{\mathrm{max}},\notag\\
&&\hspace*{21mm}\mbox{C3:}\hspace*{1mm}
\tilde{R}_{\mathrm{p},k} + \tilde{R}_{\mathrm{c},k} \ge R_{\mathrm{min},k},\hspace*{1mm}\forall k,\notag\\
&& \hspace*{21mm}\mbox{C4:}\hspace*{1mm} C_k \ge 0, \forall k,
\end{eqnarray}
\noindent
where $\tilde{R}_{\mathrm{p},k}$ and $\tilde{R}_{\mathrm{c},k}$ are the worst-case achievable rate in the presence of channel estimation error and they are defined by
\begin{equation}
    \tilde{R}_{\mathrm{p},k} = \min_{\left\|\Delta{\mathbf{h}}_k\right\| \le \delta_k} {R_{\mathrm{p},k}}\;\;\text{and} \;\;
     \tilde{R}_{\mathrm{c},k} = \min_{\left\|\Delta{\mathbf{h}}_k\right\| \le \delta_k} {R_{\mathrm{c},k}},
\end{equation}
with ${R_{\mathrm{p},k}}$ and ${R_{\mathrm{c},k}}$ given by \eqref{Eqn:NaturalChannelRSMA_RateI} and \eqref{Eqn:NaturalChannelRSMA_RateII}, respectively.
Constraint C1 guarantees the decodability of the common stream for all users and constraint C2 is the total power constraint for the precoding vectors.
Constraint C3 guarantees the minimum data rate for each user, where the sum of the private and common data rates for user $k$ should not be less than its minimum required data rate $R_{\mathrm{min},k}$.
Constraint C4 is introduced to guarantee a non-negative common rate allocation to user $k$.
This optimization problem is non-convex due to the NOF, semi-infinite constraint (SMIC) in C1, and FCs in both C3 and C4.

{We note that investigating the tradeoff between spectral efficiency, power consumption, and reliability is also a potential research direction for the resource allocation design of NGMA systems.
In scenarios with perfect CSI, maximizing spectral efficiency and minimizing power consumption are generally two conflicting design objectives.
Typically, higher spectral efficiency results in increased power consumption and vice versa.
For the cases with imperfect CSI, both spectral efficiency and power consumption are constrained by communication reliability, such as outage probability.
In particular, more stringent reliability requirements lead to reduced spectral efficiency and increased power consumption.
Therefore, multi-objective optimization approaches should be adopted for exploring the Pareto-optimal tradeoff curves between these distinctive performance metrics in the resource allocation design \cite{pereira2022review}.
However, multi-objective optimization problems are often transformed into single-objective optimization problems by introducing proper auxiliary weight variables to balance different design goals or by setting specific constraints \cite{pereira2022review}.
In this work, we have directly considered a single-objective problem formulation with proper constraints for resource allocation design of NGMA systems.
For instance, in the rate-oriented resource allocation design, by changing the maximum allowable power consumption, the tradeoff between the system spectral efficiency and power consumption can be obtained, as illustrated in Fig. \ref{Fig.Simulation} in the simulation section later.}
\subsection{Resource Allocation Design for NGMA in Reconfigurable Channel}
Emerging techniques including UAV, IRS, and M/FAs enable the joint design of wireless channels and signals. These techniques bring extra DoFs for resource allocation design. However, some challenges need to be tackled before the full potential of these emerging techniques can be unlocked. In the following, for each of these techniques, we formulate a typical optimization problem to help the readers better understand this.
\subsubsection{Resource Allocation for IRS-assisted NGMA \cite{ZhiqiangOFDMA,yu2020power,wu2023globally}}
We consider a passive IRS-assisted multi-user communication system. One typical design objective for resource allocation in IRS-assisted communication systems is the maximization of the sum rate. This leads to the following optimization problem:
\begin{eqnarray}
\label{RCprob_1}
&&\underset{\mathbf{p}_k,\mathbf{\Psi},\alpha_{k,r}}{\maxo} \,\, \,\, \underset{ k\in\mathcal{K}}{\sum} \mathrm{log}_2(1+\Gamma_k) \notag\\
&&\mbox{subject\hspace*{1.6mm}to}\hspace*{4mm}
\mbox{C1:}\hspace*{1mm}\underset{k\in\mathcal{K}}{\sum }\left \| \mathbf{p}_k \right \|^2\leq P_{\mathrm{max}},\notag\\
&&\hspace*{21mm}\mbox{C2:}\hspace*{1mm}
\Big |\{\mathbf{\Psi}\}_{mm}\Big |=1, \forall m\in\left\{1,\ldots,M_{\rm ps}\right\},\notag\\
&&\hspace{21mm}\mbox{C3:}\hspace*{1mm} \alpha_{k,r}\in\{0,1\},\hspace*{1mm}\forall k\neq r \in\mathcal{K},
\notag
\\
&&\hspace{21mm}\mbox{C4:}\hspace*{1mm}\alpha_{k,r}+\alpha_{r,k}=1,\hspace*{1mm}\forall k\neq r \in\mathcal{K},
\end{eqnarray}
where $\Gamma_k$ is the SINR of user $k$, given by \eqref{Eqn:NOMA_IRS_SINRk}.
In the above optimization problem, the BS transmit beamforming vector, the IRS phase shift pattern, and the SIC decoding order are jointly optimized for the maximization of the system sum rate. The constant $P_{\mathrm{max}}$ in constraint C1 is the maximum BS transmit power. Due to the passive nature of the IRS, unit-modulus constraint (UMC) C2 is adopted to enforce each main diagonal element of the IRS phase shift matrix to have a unit gain \cite{xu2021resource,yu2019enabling}. Alternatively, by employing a discrete phase shift model, the elements of $\mathbf{\Psi}$ can be selected from a discrete feasible set \cite{wu2023globally}. Binary constraint C3 represents the candidate SIC decoding order and constraint C4 ensures that for two users, only one possible SIC decoding order is employed for both users. 

\subsubsection{Resource Allocation for UAV-assisted NGMA \cite{8663615,8445944,8644086}}
We consider a rotary-wing UAV-assisted multi-user communication system in \eqref{Eqn:UAV-NOMA_SystemModel}. One possible design goal for resource allocation in UAV-assisted communication systems is the minimization of the UAV's total power consumption. In time slot $t$, the power minimization problem is formulated as the following optimization problem:
\begin{eqnarray}
\label{RCprob_2}
&&\underset{\substack{\mathbf{p}_k[t],\mathbf{r}_0[t],\\\mathbf{v}_u[t],\alpha_{k,r}[t]}}{\mino} \,\, \,\, \hspace*{2mm}
\underset{ k\in\mathcal{K}}{\sum} \mathbf{p}^H_k[t]\mathbf{p}_k[t]+P_{\mathrm{aero}}[t]+M\cdot P_{\mathrm{circ}}\notag\\
&&\mbox{subject\hspace*{1.6mm}to}\hspace*{4mm}\mbox{C1:}\hspace*{1mm}\{\underset{k\in\mathcal{K}}{\sum} \mathbf{p}_k[t]\mathbf{p}^H_k[t]\}_{ii} \leq \mathit{P}_i,\hspace*{1mm}\forall i,\notag
\\
&&\hspace{21mm}\mbox{C2:}\hspace*{1mm}
\Gamma_k[t]\geq \Gamma_{\mathrm{req}_k},\hspace*{1mm}\forall k, \notag \\
&&\hspace*{21mm}\mbox{C3:}\hspace*{1mm} 
\left \|\mathbf{v}_u[t]-\mathbf{v}_u[t-1]\right \|\leq a_\mathrm{max}\delta_T,
\notag
\\
&&\hspace{21mm}\mbox{C4:}\hspace*{1mm} \left \|\mathbf{v}_u[t]\right \|\delta_T = \left \|\mathbf{r}_0[t] -\mathbf{r}_0[t-1]\right \|,
\notag
\\
&&\hspace{21mm}\mbox{C5:}\hspace*{1mm} \alpha_{k,r}[t]\in\{0,1\},\hspace*{1mm}\forall k\neq r \in\mathcal{K},
\notag
\\
&&\hspace{21mm}\mbox{C6:}\hspace*{1mm}\alpha_{k,r}[t]+\alpha_{r,k}[t]=1,\hspace*{1mm}\forall k\neq r \in\mathcal{K},
\end{eqnarray}
where the SINR of user $k$ in time slot $t$ $\Gamma_k[t]$ is given by \eqref{SINR_k}.
The constant $P_i$ in constraint C1 denotes the maximum transmit power of the $i$-th antenna element, $\forall i\in\left\{1,\ldots,N \right\}$, due to the limitation of the analog RF front-end. Constraint C2 indicates a pre-defined SINR threshold of user $k$ should be met to ensure the desired QoS requirement, denoted by $\Gamma_{\mathrm{req}_k}$. In addition to conventional communication-oriented performance metrics and constraints, UAV-assisted communication also introduces additional UAV-related terms and constraints for resource allocation design. First, $P_{\mathrm{aero}}[t]$ in the objective function denotes the UAV aerodynamic power and it is determined by the UAV speed $\mathbf{v}_u[t]$. In particular, $P_{\mathrm{aero}}[t]$ can be modelled as $P_{\mathrm{aero}}[t]=\frac{\sqrt{2}W_uc_1^2}{\sqrt{\left \| \mathbf{v}_u[t] \right \|^2 +\sqrt{\left \|\mathbf{v}_u[t]\right \|^4+4c_1^4}}}+c_2 V_{\mathrm{T}}^3\Big[1+c_3\Big(\frac{\left \| \mathbf{v}_u[t] \right \|}{V_{\mathrm{T}}}\Big)^2\Big]+c_4\left \| \mathbf{v}_u[t] \right \|^3$. Here, constant $V_{\mathrm{T}}$ denotes the rotor tip speed and $W_u$, $c_1$, $c_2$, $c_3$, and $c_4$ are UAV aerodynamic power consumption parameters \cite{seddon2011basic,8663615}. Moreover, constraints C3 and C4 are the inherent UAV kinetic constraints. Constraints C5 and C6 are imposed to ensure the SIC decoding order. This optimization problem is non-convex due to the NOF, FC in C2, and BC in C5.
\subsubsection{Resource Allocation for M/FA-enabled NGMA \cite{10146274,10207991}}
We consider a M/FA-enabled multi-user communication system in \eqref{Eqn:MFA-NGMA-SystemModel}. One possible common design objective of M/FA-enabled communication systems is to minimize the BS total transmit power by jointly optimizing the beamforming vectors and the position of the M/FAs. This leads to the following optimization problem:
\begin{eqnarray}
\label{RCprob_3}
&&\underset{\mathbf{P},\mathbf{T},\mathbf{U}}{\mino}\hspace*{4mm}\sum_{k\in\mathcal{K}}\left\|\mathbf{p}_k\right\|_2^2\notag\\
    &&\mbox{subject\hspace*{1.6mm}to}\hspace*{4mm}\mbox{C1:}\hspace*{1mm}\Gamma_k\geq \Gamma_{\mathrm{req}_k},\hspace*{1mm}\forall k,\notag\\
    &&\hspace*{21mm}\mbox{C2:}\hspace*{1mm} t_n[q]\in \{0,1\},\hspace*{1mm} \forall n,\hspace*{1mm} \forall q,\notag\\
    &&\hspace*{21mm}\mbox{C3:}\hspace*{1mm} \sum_{q=1}^Qt_n[q]=1,\hspace*{1mm}\forall n,\notag\\
    &&\hspace*{21mm}\mbox{C4:}\hspace*{1mm}\mathbf{U}=\mathbf{T}\mathbf{P},
\end{eqnarray}
where the SINR of user $k$ $\Gamma_k$ is given by \eqref{Eqn:MFA-SINRk}.
The constant $\Gamma_{\mathrm{req}_k}$ in constraint C1 denotes the minimum required SINR of user $k$ to ensure a satisfying communication service. Constraints C2 and C3 are imposed to represent the position selection of M/FA elements. {Equality constraint (EC) C4 is imposed to ensure the successful recovery of the beamforming and antenna selection policy after optimization.} This optimization problem is non-convex due to the FC C1, BC C2, and EC C4. 
\subsection{Resource Allocation Design for NGMA in Functional Channels}
The resource allocation design for functional channels departs from the objectives for conventional communication-only channels, given the unique integrated nature, where communication functionality and other functionalities coexist. Particularly, the objective function for functional channels necessitates a careful tradeoff between communication performance and other performance of interest. Depending on the specific system requirements and objectives, the objective function in NGMA can be broadly classified into three categories. Specifically, the communication-centric design may be the most straightforward design that aims to optimize the communication performance subject to a reasonable performance requirement of the other functionalities. In contrast to the communication-centric design, the resource allocation design can focus on the performance of other functionalities while maintaining a reasonable communication performance. This design is referred to as the function-centric design. According to the above two design criteria, the resource allocation design for NGMA in functional channels is detailed as follows.

\subsubsection{Communication-Centric Resource Allocation \cite{Jinke2019JCAC,Fan2020,10207991}}
We study the communication-centric resource allocation based on the downlink SDMA-based monostatic ISAC system given in Section II-C.  
In the communication-centric resource allocation, the design objective is to optimize the communication performance while satisfying certain constraints required by the radar sensing functionality. Considering~\eqref{Eqn:ISACChannelModelExample2_SINR}, the corresponding optimization problem can be formulated as follows:
\begin{eqnarray}
\label{FCprob_1}
&&\underset{\mathbf{P}\in\mathbb{C}^{{N}\times{K}}}{\maxo} \,\, \,\,\sum\limits_{k = 1}^K {{{\log }_2}\left( {1 + \Gamma_k} \right)}\notag\\
&&\mbox{subject\hspace*{1.6mm}to}\hspace*{4mm}
\mbox{C1:}\hspace*{1mm}\frac{1}{L}{\rm{Tr}}\left( {{{\bf{S}}^{\rm{H}}}{{\bf{P}}^{\rm{H}}}{\bf{PS}}} \right)\leq P_{\mathrm{max}},\notag\\
&&\hspace*{21mm}\mbox{C2:}\hspace*{1mm}
\frac{1}{L}\left\| {{\bf{PS}} - {{\bf{X}}_0}} \right\|_{\rm{F}}^2  \le \delta,
\end{eqnarray}
where $\Gamma_k$ is given by \eqref{Eqn:ISACChannelModelExample2_SINR}, $\delta$ is a pre-determined parameter characterizing the required sensing quality, and $P_{\max}$ is the power constraint. As mentioned before, the radar sensing quality is evaluated by the similarity between the transmitted beampattern and the desired beampattern for radar sensing, which becomes the optimization problem constraint C2 \cite{xu2022integrated}. 
Compared to the resource allocation problem for conventional communication-oriented channels, the size of the solution set for the sum rate optimization problem is reduced due to the requirement of radar sensing quality. This optimization problem is non-convex due to the NOF.
\subsubsection{Function-Centric Resource Allocation \cite{Fan2018Toward,Fan2020,Feng2019multiantenna}}
We study the function-centric resource allocation for JCAC transmissions discussed in Section II-C. In the considered problem, the design objective is to
minimize energy consumption while satisfying the constraint on the communication rate.
The corresponding optimization problem can be formulated as follows:
\begin{eqnarray}
\label{FCprob_2}
&&\underset{{p_{kl}},\{\mathbf{\Pi_k}\}_{ml},L_k}{\mino}\,\, \,\,{E_{\rm{tot}}}\notag\\
&&\mbox{subject\hspace*{1.6mm}to}\hspace*{4mm}
\mbox{C1:} R_{{\rm C},k} \ge R_{\min,k}, \forall k, \notag\\
&&\hspace*{21mm}\mbox{C2:}\hspace*{1mm}
R_{{\rm MEC},k}+L_k \ge{\tilde L}_k, \forall k ,\notag\\
&&\hspace*{21mm}\mbox{C3:}\hspace*{1mm}
p_{kl} \ge 0,  \forall k , \forall l \in \{1,\ldots, D_{{\rm Tot},k}\} ,\notag\\
&&\hspace*{21mm}\mbox{C4:}\hspace*{1mm}
0 \le L_k \le {\tilde L}_k,  \forall k ,\notag\\
&&\hspace*{21mm}\mbox{C5:}\hspace*{1mm}
\{\mathbf{\Pi}_k\}_{ml} \in\left\{0, 1\right\},\hspace*{1mm}\forall m,\hspace*{1mm}\forall k,\hspace*{1mm}\forall l \in \{1,\ldots, D_{{\rm Tot},k}\},\notag\\
&&\hspace*{21mm}\mbox{C6:}\hspace*{1mm}
\sum_{k=1}^{K}\sum_{l_{\rm C}=1}^{D_{{\rm C},k}}\{\mathbf{\Pi}_k\}_{m{l_{\rm C}}} + \sum_{k=1}^{K}\sum_{l_{\rm MEC}=1}^{D_{{\rm MEC},k}}\{\mathbf{\Pi}_k\}_{m{\left(D_{{\rm C},k}+l_{\rm MEC}\right)}} \le 1,\hspace*{1mm}\forall m.
\end{eqnarray} 
{Note that while the value of $L_k$ needs to be an integer by definition, it is treated as a continuous variable here. However, the performance loss is in general negligible when $L_k$ is sufficiently large~\cite{Feng2019multiantenna}.} In~\eqref{FCprob_2}, C1 states that the communication rate of each user needs to satisfy the corresponding minimum constraint; C2 is the constraint on the offloading rate such that the 
the whole task of each user can be computed either locally at the user side or at the BS;
C3 is the power constraint; C4 is the length requirements of the offloading bit sequences; C5 and C6 together ensure that the communication symbol and task symbol can be allocated to at most one subcarrier in the transmission.

{Note that NGMA systems might operate in dynamic and uncertain environments, which may introduce dynamic channel conditions, user mobilities, and evolving network dynamics. These factors in practical deployment scenarios would introduce channel uncertainties and dynamic system parameters for the resource allocation design of NGMA systems. The proposed reliability-oriented resource allocation design in \eqref{NCprob_3} can adapt to these channel uncertainties by assuming a certain channel estimation error model and adopting robust resource allocation design approaches. On the other hand, when the system parameters are dynamic, such as the variant number of accessing users due to user mobility or variant network topology, the resource allocation needs to be redesigned, regardless of the type of NGMA schemes, channels, or design objectives involved.}

\section{Optimization Tools for Resource Allocation Design}

In this section, we present a variety of optimization strategies tailored for solving the optimization challenges of NGMA systems outlined in Section \ref{SectionIII}. Specifically, we explore several globally optimal methods designed to address a specific form of non-convexity. These include semidefinite relaxation (SDR), branch-and-bound (BnB), and monotonic optimization (MO). {The resource allocation issues within NGMA systems are inherently complex and characterized by diverse forms of non-convexities. To achieve a globally optimal solution, it is often necessary to leverage a synergistic blend of various global optimization techniques. We note that these optimal approaches help establish the performance upper bounds for NGMA systems and serve as performance benchmarks for evaluating related suboptimal approaches. While these methods allow us to acquire globally optimal solutions for optimization problems with certain types of non-convexities, the globally optimal approaches, e.g., BnB and MO, typically require a significant amount of computational resources. To strike a balance between optimality and complexity, we then introduce a few low-complexity suboptimal approaches such as successive convex approximation (SCA) and block coordinate descent (BCD).} Moreover, for each approach, we will provide a clear explanation of the basic ideas behind these approaches and interpret the flow and key steps in detail. Also, we will take some optimization problems listed in Section \ref{SectionIII} as examples and briefly discuss how these approaches can be employed to tackle the optimization problems.
\begin{table*}[t]
		\caption{Comparison of different resource allocation design problems from an optimization perspective.}
		\label{tab:table2}
		\newcommand{\tabincell}[2]{\begin{tabular}{@{}#1@{}}#2\end{tabular}}
		\centering
\begin{tabular}{c|c|c|c|c}
\hline
		\textbf{Problem} & \textbf{Challenges} & \textbf{Methods} & \textbf{Optimality} & \textbf{Complexity}\\
\hline
   	\eqref{NCprob_1} & NOF, VC, FC, BC & BnB, SCA & Locally optimal & Exponentially high\\
   	\eqref{NCprob_2} & VC, FC, BC & BnB & Globally optimal & Exponentially high\\
   	\eqref{NCprob_3} & NOF, SMIC, FC & SCA, S-procedure\cite{boyd2004convex} & Locally optimal & Moderate\\
   	\eqref{RCprob_1} & NOF, UMC, FC & BCD, SCA, SDR, manifold optimization\cite{absil2009optimization} & Locally optimal & Moderate\\			
   	\eqref{RCprob_2} & NOF, FC, BC & MO, SDR & Globally optimal & Exponentially high\\
   	\eqref{RCprob_3} & FC, BC, EC & BnB, bilinear transformation\cite{6698281} & Locally optimal & Exponentially high\\
   	\eqref{FCprob_1} & NOF & SCA & Locally optimal & Low\\
   	\eqref{FCprob_2} & NOF FC & MO & Globally optimal & Exponentially high\\
\hline
\end{tabular}
\end{table*}
\subsection{Global Optimization Approaches}
\label{globaloptimization}
\subsubsection{Semidefinite Relaxation Approach}
When the transmitter is equipped with multiple antennas and the targeted receivers are equipped with single antennas, the corresponding transmitter beamforming design problem, e.g., the problems in \eqref{RCprob_1}, \eqref{RCprob_2}, and \eqref{RCprob_3}, can be equivalently transformed as semidefinite programming (SDP) problems. We take the optimization problem in \eqref{RCprob_2} as an example. To deliver the basic idea, we assume that the UAV trajectory and SIC decoding policy are already determined by employing effective approaches, e.g., MO-based approach or BnB-based approach, and focusing on the beamforming design. In particular, by defining the beamforming matrix $\mathbf{P}_k[t]\in {\mathbb C}^{N\times N}$, $\mathbf{P}_k[t]=\mathbf{p}_k[t]\mathbf{p}_k^H[t]$, \eqref{RCprob_2} with respect to $\mathbf{P}_k[t]$ can be equivalently represented as the following SDP problem:
\begin{eqnarray}
\label{RCprob_2SDP}
&&\underset{\substack{\mathbf{P}_k[t]}}{\mino} \,\, \,\, \hspace*{3mm}
\underset{ k\in\mathcal{K}}{\sum} \mathrm{Tr}(\mathbf{P}_k[t])+P_{\mathrm{aero}}[t]+M\cdot P_{\mathrm{circ}}\notag\\
&&\mbox{subject\hspace*{1.6mm}to}\hspace*{4mm}\mbox{C1:}\hspace*{1mm}\{\underset{k\in\mathcal{K}}{\sum} \mathbf{P}_k[t]\}_{ii} \leq \mathit{P}_i,\hspace*{1mm}\forall i,\notag
\\
&&\hspace{21mm}\mbox{C2:}\hspace*{1mm}
\frac{\frac{\varrho}{\left\|\mathbf{r}'_0[t]- \mathbf{r}'_k \right \|^2+H_0^2}\mathrm{Tr}(\mathbf{a}_k^H[t]\mathbf{P}_\mathit{k}[t]\mathbf{a}_k[t])}{\underset{ r\in\mathcal{K}\setminus \left \{ k \right \}}{\sum}{\alpha_{k,r}[t]\frac{\varrho}{\left\|\mathbf{r}'_0[t]- \mathbf{r}'_k \right \|^2+H_0^2}\mathrm{Tr}(\mathbf{a}_k^H[t]\mathbf{P}_\mathit{r}[t]\mathbf{a}_k[t]})+\sigma ^2_{n_k}[t]}\geq \Gamma_{\mathrm{req}_k},\hspace*{1mm}\forall k, \notag \\
&&\hspace*{21mm}\mbox{C3-C6},
\notag
\\
&&\hspace{21mm}\mbox{C7:}\hspace*{1mm} \mathbf{P}_k[t]\succeq\mathbf{0},\hspace*{1mm}\forall k,
\notag
\\
&&\hspace{21mm}\mbox{C8:}\hspace*{1mm} \mathrm{Rank}(\mathbf{P}_k[t])= 1,\hspace*{1mm}\forall k.
\end{eqnarray}
We note that the original quadratic function with respect to the beamforming vector is now replaced by an affine function with respect to the beamforming matrix $\mathbf{P}_k[t]$ \cite{9723093}. This is achieved by replacing the beamforming vector to be optimized with the beamforming matrix $\mathbf{P}_k[t]$ where $\mathbf{P}_k[t]=\mathbf{p}_k[t]\mathbf{p}_k^H[t]$. To efficiently recover the beamforming vector, two additional constraints, i.e., a semidefinite constraint  C7 and a rank constraint C8, are imposed. {Although the rank constraint is highly non-convex, we can exploit the popular SDR approach by dropping the rank constraint. In case the unit-rank constraint is the only non-convex constraint of the problem at hand, we can then optimally solve the resulting rank constraint-relaxed version of the original optimization problem in polynomial time.} The tightness of the SDR has been proved in the literature, e.g., \cite{DerrickEERobust}. Moreover, the SDR-based algorithm entails a polynomial complexity and its convergence is guaranteed \cite{yu2020irs}. In the following, we briefly explain the key steps. First, by exploiting the duality theory \cite{boyd2004convex}, we can express the Lagrange function of the considered optimization problem with respect to the beamforming matrix $\mathbf{P}_k[t]$. Then, based on the Lagrange function, we can formulate the dual problem of the original optimization problem and express the Karush-Kuhn-Tucker (KKT) conditions with respect to $\mathbf{P}_k[t]$. Subsequently, by analyzing the KKT conditions, we can conclude that there is always an optimal solution $\mathbf{P}_k[t]$ satisfying $\mathrm{Rank}(\mathbf{P}_k[t])\leq 1$.
\subsubsection{Branch-and-Bound Approach} 
In NGMA systems, some resource allocation optimization problems, such as the subcarrier assignment or SIC decoding strategy design, involve binary optimization variables \cite{7812683,xu2023joint}. This results in non-convex binary integer programming (BIP) problems, cf. \eqref{NCprob_1} and \eqref{RCprob_2}, which are challenging to solve. On the other hand, since the available subcarriers or the feasible SIC decoding strategies are countable and finite in an NGMA system, the number of corresponding binary optimization variables is limited. Based on this observation, we can exploit enumeration-based approaches to traverse all possible binary variable-induced schemes and obtain the optimal solution. One popular and systematic enumeration-based approach is the BnB approach \cite{lawler1966branch}. We take the subcarrier assignment and power allocation optimization problem in \eqref{NCprob_2} as an example and rewrite it here
\begin{figure}[t] 
\centering
\includegraphics[width=3.2in]{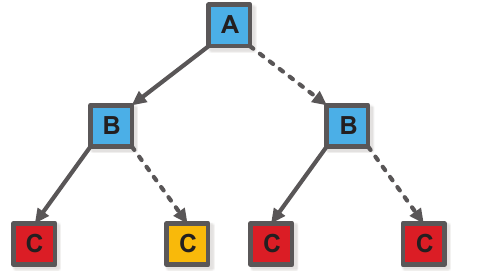}
\caption{Illustration of a BnB-based tree structure for solving a BIP problem involving two binary optimization variables. The solid-line arrows and dashed-line arrows denote the two cases where the optimization variable is set to $0$ and $1$, respectively. The yellow-colored and red-colored nodes denote the optimal and non-optimal solutions to the BIP problem, respectively. Here, block A denotes the root node of the search tree, and blocks B and C denote the subnodes of the search tree.}
\label{Fig.BnB_Algorithm}
\end{figure}

\begin{eqnarray}
\label{Solution_NCprob_2}
&&\underset{ p_{kd_k},\{\mathbf{\Pi}_k\}_{md_k}}{\mino} \,\, \,\, \hspace*{2mm}\underset{ k\in\mathcal{K}}{\sum}\sum_{d_k=1}^{D_k} p_{kd_k}\\
&&\hspace*{2mm}\mbox{subject\hspace*{1.6mm}to}\hspace*{6mm}
\mbox{C1-C3},\notag\\
&&\hspace*{25mm}\mbox{C4:}\hspace*{1mm}\{\mathbf{\Pi}_k\}_{md_k}\in\left\{0,1\right\},\hspace{1mm}\forall k, \forall m, \forall d_k.\notag
\end{eqnarray}
We note that the above optimization problem can be optimally solved by employing the BnB approach. Next, we explain the basic concept and the key steps of the BnB approach. Specifically, the fundamental idea behind the BnB approach is to construct a search tree and check the branches of the tree to find the best solution. This is achieved by iteratively applying the three steps of partitioning, branching, and bounding until the optimal solution is obtained when upper and lower bounds are merged \cite{lawler1966branch,9669263}. Initially, the full feasible set is mapped onto the root node of the search tree. According to a pre-defined partitioning rule, we partition each node as two subnodes and the two subnodes represent two subsets of the full feasible set. In this case, both subsets contain one specific binary variable that is already set as $0$ and $1$, respectively. Accordingly, for each subnode, we have to solve one subproblem associated with a subset and obtain the corresponding objective function value. Then, among the objective function values of all subnodes, we select the one with the minimum value as the upper bound for the original optimization problem in \eqref{Solution_NCprob_2}. As for the lower bound of \eqref{Solution_NCprob_2}, we can replace the binary constraint C4 by a convex constraint with a set of continuous variables, i.e., $0\leq\{\mathbf{\Pi}_k\}_{md_k}\leq 1$. We note that in the convex constraint, those variables already set as $0$ or $1$ in the previous iteration of BnB will not be relaxed as continuous variables. Then, such a convex optimization problem, i.e., the relaxed version of \eqref{Solution_NCprob_2} with the aforementioned convex constraint, can be optimally solved by employing bilinear optimization approaches, e.g., McCormick envelopes \cite{mitsos2009mccormick}. As such, the resulting objective function value of the relaxed version of \eqref{Solution_NCprob_2} serves as a lower bound of the original optimization problem in \eqref{Solution_NCprob_2}. To simplify the traversal in each iteration, if the objective function value of one subproblem is larger than the upper bound, the associated subnode will be discarded from the tree. Along with the expansion of the search tree, the original feasible set is progressively partitioned into more subsets and the BnB approach will not be terminated until the gap between the upper bound and the lower bound is smaller than a pre-defined threshold. It has been proved in \cite{horst2013global} that as long as the number of binary variables is finite, the BnB approach converges to the globally optimal solution of the BIP problem in a limited number of iterations. Yet, we also note that the BnB-based algorithm entails an exponentially high complexity \cite{horst2013global}. For the considered optimization problem in \eqref{Solution_NCprob_2}, there are in total $DM$ binary optimization variables, where $\sum_{k=1}^{K} D_k = D$. Hence, we need to construct an $(DM+1)$-tier search tree comprising at most $2^{DM}$ subnodes in the last tier. This is illustrated in Fig. \ref{Fig.BnB_Algorithm} where the optimum of a BIP problem involving 2 binary optimization variables can be obtained by constructing a $3$-tier search tree with $4$ subnodes in the last tier. 
\subsubsection{Monotonic Optimization Approach}

Although resource allocation designs for NGMA systems often encounter non-convexity, most resulting optimization problems do preserve monotonicity \cite{tuy2000monotonic}. In this case, their objective function monotonically increases or decreases relative to the optimization variables over the feasible set, cf. \eqref{NCprob_1} and \eqref{RCprob_1}. This property provides us with a promising means to solve this kind of problem, that is, to iteratively reduce an upper bound on the monotonic objective function until the maximum over the feasible set is obtained. In the following, we take the optimization problem in \eqref{NCprob_1} as an example. To deliver the basic idea, we assume that the given user scheduling policy is already determined by employing enumeration-based approaches and focusing on the power allocation design. In this case, we replace the channel matrix  $\{\mathbf{H}_k\}_{mm}$ in the objective function of \eqref{NCprob_1} by a scalar channel counterpart $h_k\in\mathbb{C}$ and thus the optimization problem in \eqref{NCprob_1} is degenerated as follows
\begin{eqnarray}
\label{solution_NCprob_1}
&&\underset{p_{k}}{\maxo} \,\, \,\, \underset{ k\in\mathcal{K}}{\sum} \log_2\left(1+\frac{\left|h_k\right|^2p_k}{\sum_{k'<k \in \mathcal{K}} \left|h_k\right|^2p_{k'} + \sigma^2_k}\right)\notag\\
&&\mbox{subject\hspace*{1.6mm}to}\hspace*{4mm}
\mbox{C1},\mbox{C3}.
\end{eqnarray}
We note that for \eqref{solution_NCprob_1}, the objective function is monotonically increasing with respect to the SINR term $\frac{\left|h_k\right|^2p_k}{\sum_{k'<k \in \mathcal{K}} \left|h_k\right|^2p_{k'} + \sigma^2_k}$. Moreover, constraints C1 and C3 define a normal set and a conormal set \cite{zhang2013monotonic}, respectively. As a result, we define an optimization variable $\zeta_k$, which satisfies $1\leq\zeta_k\leq 1+\frac{\left|h_k\right|^2p_k}{\sum_{k'<k \in \mathcal{K}} \left|h_k\right|^2p_{k'} + \sigma^2_k}$. Then, the optimization problem in \eqref{solution_NCprob_1} can be equivalently reformulated as a canonical form of MO problem which is given as follows
\begin{figure*}[t] 
    \centering
    \hspace*{-8mm}\includegraphics[width=6.8in]{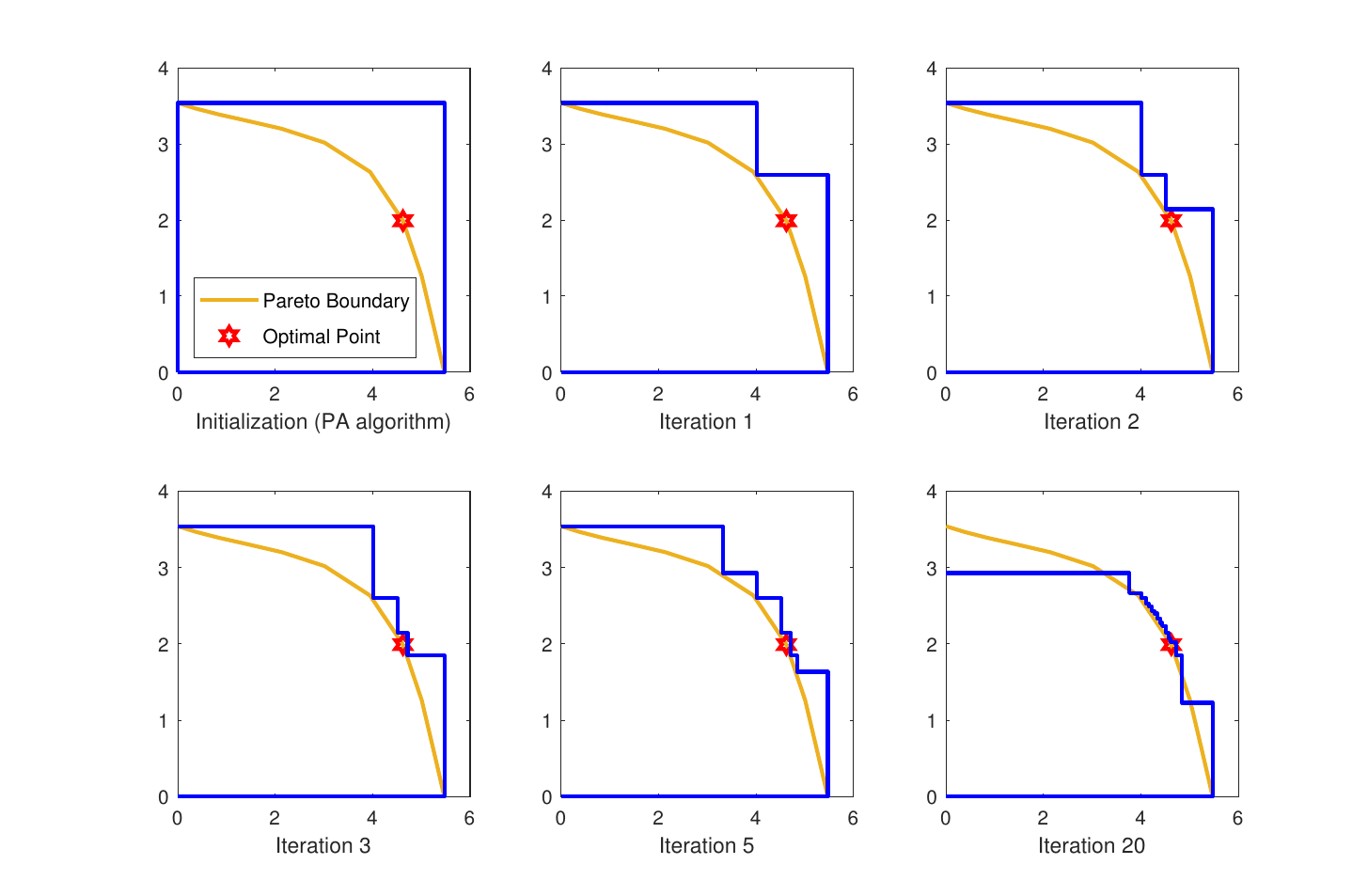}
    \caption{Illustration of a few snapshots of the polyblock outer approximation algorithm to produce the optimum of a MO problem involving two optimization variables $x_1$ and $x_2$. The blue-colored polyblock encloses the feasible set of the MO problem and the dashed line denotes the boundary of the feasible set which can be either convex or non-convex. The red-colored star denotes the globally optimal solution to the considered problem.}
    \label{Fig.PA_Algorithm}
\end{figure*}
\begin{eqnarray}
\label{solution_NCprob_2}
&&\underset{p_{k},\zeta_k}{\maxo} \,\, \,\, \underset{ k\in\mathcal{K}}{\sum} \log_2\left(\zeta_k\right)\notag\\
&&\mbox{subject\hspace*{1.6mm}to}\hspace*{4mm}\zeta_k\in\mathcal{F},
\end{eqnarray}
where feasible set $\mathcal{F}$ is defined as $\mathcal{F}=\mathcal{G}\cap\mathcal{H}$ and the normal set $\mathcal{G}$ and the conormal set $\mathcal{H}$ are given by, respectively, 
\begin{eqnarray}
&&\mathcal{G}\overset{\Delta }{=}\left \{ p_{k}\hspace*{1mm}|\hspace*{1mm}p_{k}\in\mathcal{U} \right \},\\
&&\mathcal{H}\overset{\Delta }{=} \left \{ \zeta_k\hspace*{1mm}|\hspace*{1mm}\zeta_k\in\mathcal{V} \right\},
\end{eqnarray}
where sets $\mathcal{U}$ and $\mathcal{V}$ are spanned by constraints C1 and C3, respectively. By exploiting MO theory, we can develop MO-based algorithms, e.g., two-layer polyblock outer approximation algorithm \cite{bjornson2013optimal,8648498,9423667}, to obtain the globally optimal solution to \eqref{solution_NCprob_2}. Next, we introduce the key steps of the polyblock outer approximation algorithm. To start with, we construct a polyblock $\mathcal{P}$ that contains the feasible set $\mathcal{F}$. As such, the vertex of $\mathcal{P}$ serves as an upper bound of the objective function of \eqref{solution_NCprob_2}. Then, in each iteration of the outer layer of the algorithm, we continuously shrink $\mathcal{P}$ by removing a cone that does not belong to the feasible set $\mathcal{F}$. However, as the feasible set $\mathcal{F}$ is in general unknown in advance and difficult to characterize, we exploit the bisection projection search in the inner layer of the algorithm to find the projection of a vertex on the upper boundary of the feasible set \cite{zhang2013monotonic}. In particular, we check the feasibility of an optimization problem subject to constraints C1 and C3 for a given projection factor. The two-layer algorithm is not terminated until the gap between the vertex of the current block $\mathcal{P}$ and its projection on the upper boundary of the feasible set $\mathcal{F}$ is smaller than a pre-defined threshold. When the pre-defined threshold is infinitely small, we can guarantee to find the globally optimal solution of \eqref{solution_NCprob_2}. A more detailed convergence proof can be found in \cite{zhang2013monotonic}. We note that the MO-based algorithm entails a significantly high computational complexity, which scales exponentially with the number of vertices. Besides, the computational complexity of the two-layer polyblock outer approximation algorithm increases exponentially with the number of users. A graphical example of the polyblock outer approximation algorithm is given in Fig. \ref{Fig.PA_Algorithm}. As can be observed from the figure, the polyblock outer approximation algorithm finds the optimum after sequentially shrinking the polyblock roughly $20$ times.

\subsection{Low-Complexity Suboptimal Approaches}

Although the optimization approaches introduced in Section \ref{globaloptimization} can be employed to optimally solve the considered optimization problem with a certain type of non-convex constraints, most of them, e.g., BnB and MO approaches, entail high computational complexity. As such, the required computation resources may not be affordable in practical NGMA systems to perform real-time resource allocation. To overcome this issue, in the following, we introduce several low-complexity optimization approaches to facilitate the real-time design of the NGMA systems.
\subsubsection{Successive Convex Approximation Approach} 
Wireless communication systems have become increasingly complex, necessitating the joint allocation of various wireless resources. This complexity has led to more sophisticated system models and made non-convex resource allocation optimization problems more prevalent. Often, the non-convexity in practical optimization challenges arises from specific parts of the objective function or constraints, while the rest of the optimization problem remains convex. Recognizing this, the sequential/successive convex approximation (SCA) approach was developed. The core idea behind the SCA approach is straightforward: approximating non-convex functions with tractable convex ones and iteratively solving a series of approximated convex optimization problems with differentiable and smooth objective functions until convergence is achieved.

It has been shown in the literature via simulations that the SCA approach can find a solution that is close to the global optimum in a computationally efficient manner. In the following, we briefly explain how the SCA approach is employed to tackle the resource allocation optimization problem in the NGMA systems. We take the optimization problem in \eqref{FCprob_1} as an example and rewrite it here:
\begin{eqnarray}
\label{solution_FCprob_1}
&&\underset{\mathbf{P}\in\mathbb{C}^{{N}\times{K}}}{\maxo} \,\, \,\,\sum\limits_{k = 1}^K {\log }_2\left( 1 + \frac{{\frac{1}{L}{{\left\| {{{\bf{s}}_k}} \right\|}^2}}}{{\frac{1}{L}{{\left\| {{{\left( {{\bf{h}}_{\rm{C}}^{\left( k \right)}} \right)}^{\rm{T}}}{\bf{PS}} - {{\bf{s}}_k}} \right\|}^2} + {\sigma_k^2}}} \right)\notag\\
&&\mbox{subject\hspace*{1.6mm}to}\hspace*{4mm}
\mbox{C1:}\hspace*{1mm}\frac{1}{L}{\rm{Tr}}\left( {{{\bf{S}}^{\rm{H}}}{{\bf{P}}^{\rm{H}}}{\bf{PS}}} \right)\leq P_{\mathrm{max}},\notag\\
&&\hspace*{21mm}\mbox{C2:}\hspace*{1mm}
\frac{1}{L}\left\| {{\bf{PS}} - {{\bf{X}}_0}} \right\|_{\rm{F}}^2  \le \delta.
\end{eqnarray}
We note that constraints C1 and C2 are two convex functions with respect to the optimization variable $\mathbf{P}$. The non-convexity of \eqref{solution_FCprob_1} originates from the objective function. To show the basic idea of the SCA, we define a function $f_k(\mathbf{P})$ as follows
\begin{equation}
f_k(\mathbf{P})\overset{\Delta}{=}{\log }_2\left( 1 + \frac{{\frac{1}{L}{{\left\| {{{\bf{s}}_k}} \right\|}^2}}}{{\frac{1}{L}{{\left\| {{{\left( {{\bf{h}}_{\rm{C}}^{\left( k \right)}} \right)}^{\rm{T}}}{\bf{PS}} - {{\bf{s}}_k}} \right\|}^2} + {N_0}}} \right).
\end{equation}
\begin{figure}[t] 
\centering
\includegraphics[width=3.4in]{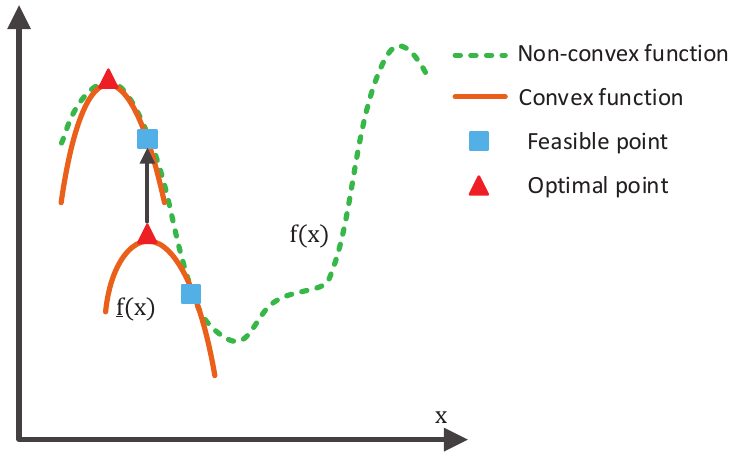}
\caption{Illustration of the basic concept of an SCA approach for a maximization problem. 
}
\label{Fig.SCA_Algorithm}
\end{figure}
\par
Instead of directly solving the above non-convex problem, we construct a surrogate function, e.g., a global underestimator, for the objective function in \eqref{solution_FCprob_1} as follows
\begin{eqnarray}
\label{SCA_prob2}
&&\underline{f}_k(\mathbf{P},\mathbf{P}^{(j)})\overset{\Delta }{=}f_k(\mathbf{P}^{(j)})+\mathrm{Tr}\Big(\big[\nabla_{\mathbf{P}}f_k(\mathbf{P}^{(j)})\big]^H(\mathbf{P}-\mathbf{P}^{(j)})\Big)
\leq f_k(\mathbf{P}),
\end{eqnarray}
where $j$ is the iteration index and $\mathbf{P}^{(j)}$ is a feasible point in the $j$-th iteration. Then, in the $j$-th iteration of the SCA approach, a lower bound of the problem in \eqref{solution_FCprob_1} can be obtained by solving the following optimization problem 
\begin{eqnarray}
\label{solution_FCprob_2}
&&\underset{\mathbf{P}\in\mathbb{C}^{{N}\times{K}}}{\maxo} \,\, \,\,\sum\limits_{k = 1}^K \underline{f}_k(\mathbf{P},\mathbf{P}^{(j)})\notag\\
&&\mbox{subject\hspace*{1.6mm}to}\hspace*{4mm}
\mbox{C1},\mbox{C2}.
\end{eqnarray}
By applying the SCA approach, the lower bound for \eqref{solution_FCprob_1} is gradually tightened. It has been proved in \cite{dinh2010local} that the SCA approach is assured to provide a locally optimal solution to the original problem in \eqref{solution_FCprob_1}. Also, since in each iteration of the SCA approach, a convex optimization problem is solved, the objective function value of \eqref{solution_FCprob_1} converges efficiently. In Fig. \ref{Fig.SCA_Algorithm}, we illustrate the basic idea behind the SCA approach. As can observed from the figure, for different feasible initial points, the SCA approach may converge to the globally optimal solution or a locally optimal solution. This becomes a drawback when compared with the global optimization approaches where the convergence point is independent of initial points. On the other hand, the computational complexity of the SCA approach is of polynomial time, due to the fact that it requires solving only a sequence of convex optimization problems. This efficiency highlights SCA's practicality in managing the complex resource allocation challenges in wireless communication systems. 
\subsubsection{Block Coordinate Descent Approach} 
\begin{figure}[t] 
\centering
\includegraphics[width=3.4in]{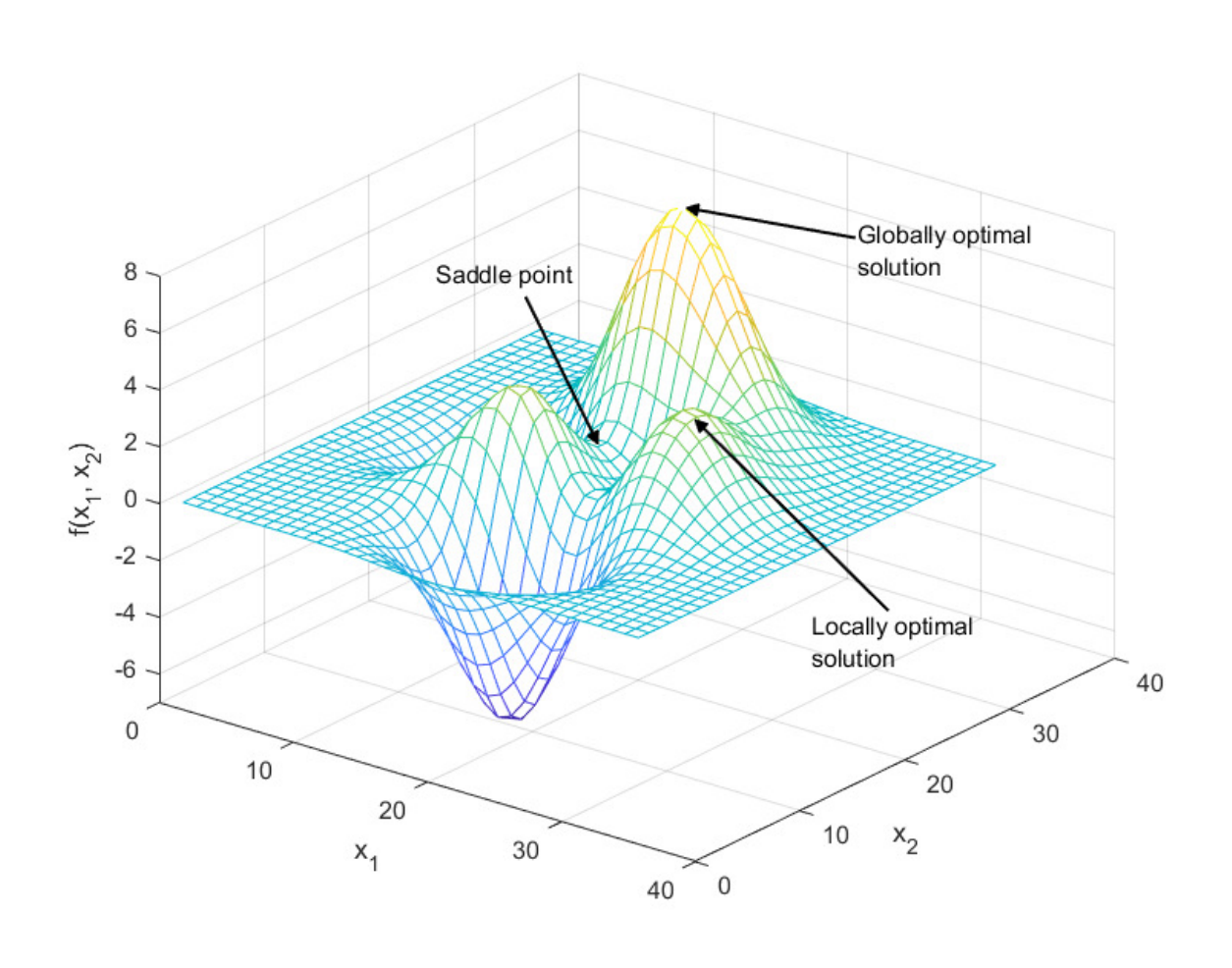}
\caption{Illustration of the globally optimal solution, locally optimal solution, and saddle point of a maximization problem to which the BCD-based algorithm potentially converges.}
\label{Fig.BCD_Algorithm}
\end{figure}
To enhance the performance of NGMA systems, available wireless resources such as beamforming policy, subcarrier assignment, and other components in the systems should be jointly designed. Yet, this inevitably leads to the coupling between various optimization variables, which brings new a challenge for the design of NGMA systems. To circumvent this challenge, the BCD approach has been proposed in the literature \cite{bezdek2002some}. The fundamental idea of the BCD approach is to divide the feasible set into several disjoint subsets and solve the corresponding subproblems in an alternating manner. Next, we take the optimization problem in \eqref{RCprob_1} as an example to show how the BCD approach works. In particular, we focus on the following optimization problem
\begin{eqnarray}
\label{solution_RCprob_1}
&&\underset{\mathbf{p}_k,\mathbf{\Psi},\alpha_{k,r}}{\maxo} \,\,\,\, \underset{ k\in\mathcal{K}}{\sum} \mathrm{log}_2(1+\Gamma_k) \notag\\
&&\mbox{subject\hspace*{1.6mm}to}\hspace*{4mm}
\mbox{C1:}\hspace*{1mm}\underset{k\in\mathcal{K}}{\sum }\left \| \mathbf{p}_k \right \|^2\leq P_{\mathrm{max}},\notag\\
&&\hspace*{21mm}\mbox{C2:}\hspace*{1mm}
\Big |\{\mathbf{\Psi}\}_{mm}\Big |=1, \forall m\in\left\{1,\ldots,M_{\rm ps}\right\},\notag\\
&&\hspace{21mm}\mbox{C3:}\hspace*{1mm} \alpha_{k,r}\in\{0,1\},\hspace*{1mm}\forall k\neq r \in\mathcal{K},
\notag
\\
&&\hspace{21mm}\mbox{C4:}\hspace*{1mm}\alpha_{k,r}+\alpha_{r,k}=1,\hspace*{1mm}\forall k\neq r \in\mathcal{K},
\end{eqnarray}
where $\Gamma_k$ is given by \eqref{Eqn:NOMA_IRS_SINRk}.
In the above optimization problem, the beamforming vectors $\mathbf{p}_k$, IRS phase shift pattern $\mathbf{\Psi}$, and SIC decoding order $\alpha_{k,r}$ are coupled with each other. As a result, we employ the BCD approach and partition optimization variables into three blocks, i.e., $\left\{\mathbf{p}_k \right\}$, $\left\{\mathbf{\Psi}\right\}$, and $\left\{\alpha_{k,r}\right\}$. Accordingly, the optimization problem in \eqref{RCprob_1} is decomposed into three subproblems as follows
\begin{eqnarray}
\mbox{Subproblem 1:}
&&\underset{\mathbf{p}_k}{\maxo} \,\, \,\, \underset{ k\in\mathcal{K}}{\sum} \mathrm{log}_2(1+\Gamma_k) \notag\\
&&\mbox{subject\hspace*{1.6mm}to}\hspace*{4mm}
\mbox{C1}.
\end{eqnarray}
\begin{eqnarray}
\mbox{Subproblem 2:}
&&\underset{\mathbf{\Psi}}{\maxo} \,\, \,\, \underset{ k\in\mathcal{K}}{\sum} \mathrm{log}_2(1+\Gamma_k) \notag\\
&&\mbox{subject\hspace*{1.6mm}to}\hspace*{4mm}
\mbox{C2}.
\end{eqnarray}
\begin{eqnarray}
\mbox{Subproblem 3:}
&&\underset{\alpha_{k,r}}{\maxo} \,\, \,\, \underset{ k\in\mathcal{K}}{\sum} \mathrm{log}_2(1+\Gamma_k) \notag\\
&&\mbox{subject\hspace*{1.6mm}to}\hspace*{4mm}
\mbox{C3},\mbox{C4}.
\end{eqnarray}
Subsequently, in each iteration of the BCD approach, one subproblem is tackled with the other two blocks fixed. In case the subproblems are still non-convex, other optimization approaches such as SCA or SDR will be used. It has been shown in \cite{bezdek2002some} that the BCD approach is guaranteed to converge to a stationary point in a finite number of iterations \eqref{solution_RCprob_1}, if all subproblems are convex with respect to the corresponding optimization variables. However, since the BCD approach overcomes the variable coupling by simply discarding the joint optimality of the original optimization problem, it can yield a range of outcomes: from globally optimal solutions to locally optimal solutions or even saddle points. Furthermore, the BCD method's effectiveness is heavily influenced by the choice of the initial point and the sequence in which the blocks are optimized.  As a result, the BCD approach may result in poor performance for practical NGMA systems. On the other hand, as the BCD approach tackles each subproblem with a subset of the original feasible space, it usually requires significantly fewer computational resources compared to that for solving the original optimization problem, which facilitates the real-time design of practical NGMA systems. In Fig. \ref{Fig.BCD_Algorithm}, we consider a toy example where a non-convex maximization problem involving two coupled optimization variables is solved by employing the BCD approach. As can be observed from the figure, for different initial points and different orders of the blocks, the BCD algorithm may converge to saddle point, locally optimal solution, or globally optimal solution \cite{bezdek2002some}.

\section{Simulations and Discussions}
\begin{figure}[t] 
    \centering
    \includegraphics[width=2.4in]{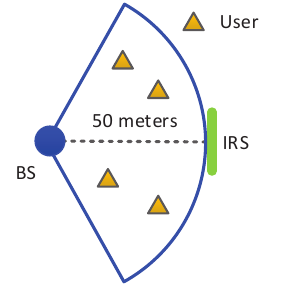}
    \caption{Simulation setup of the IRS-assisted NGMA system.}
    \label{Fig.Simulation_setup}
    \end{figure}

In this section, we present some simulation results to validate the effectiveness of the introduced resource allocation design approaches and discuss the physical insights. As an illustrative example,  we consider the resource allocation design for an IRS-assisted NOMA system in \eqref{RCprob_1}. In particular, the BS is equipped with $N=8$ antennas to serve a sector of a cell with a radius of $50$ m. There are $K=4$ single-antenna users randomly and uniformly distributed in the sector. To assist in the information transmission, a passive IRS is deployed $50 \text{m}$ away from the BS, cf. Fig. \ref{Fig.Simulation_setup}. The IRS is composed of $M_{\rm ps}=16$ phase shift elements. From a practical perspective, we assume that each IRS element employs a 2-bit phase shifter, i.e., the discrete phase shift values are from the set $\{0,\frac{\pi}{2},\pi,\frac{3\pi}{2}\}$. All channels in the considered system are assumed to be Rician distributed with a path loss exponent of $3$ and a Rician factor of $1$. The noise variances of all users are set to $\sigma_k^2=-117$ dBm. We focus on solving the optimization problem in \eqref{RCprob_1}. In particular, we propose a two-layer optimization algorithm to obtain the globally optimal of the considered problem. In the inner layer, for a given decoding order, we solve \eqref{RCprob_1} for the maximization of the system sum rate. In this case, the corresponding optimization problem in \eqref{RCprob_1} is a mixed integer programming problem and is solved optimally by employing the BnB algorithm. In the outer layer, since there are $K=4$ users, the total number of the corresponding decoding order combinations is $K!=24$. As such, we evaluate the objective function value for each possible decoding order and choose the order that yields the maximum objective function value across all options. Due to the exponentially high complexity of the proposed optimal scheme, we also develop a low-complexity suboptimal scheme capitalizing on alternating optimization, big-M \cite{SunFullDuplex}, and SCA methods.
\par
For comparison, we also consider three baseline schemes. For baseline scheme 1, we randomly select a SIC decoding order from the set and jointly optimize the beamforming vectors at the BS and the phase shift matrix at the IRS. For baseline scheme 2, we randomly generate the phase shift matrix at the IRS and jointly optimize the beamforming vectors and the SIC decoding order. For baseline scheme 3, we assume that there is no IRS deployed in the considered system and jointly optimize beamforming vectors and the SIC decoding order.
    \begin{figure}[t] 
    \centering
    \includegraphics[width=4.5in]{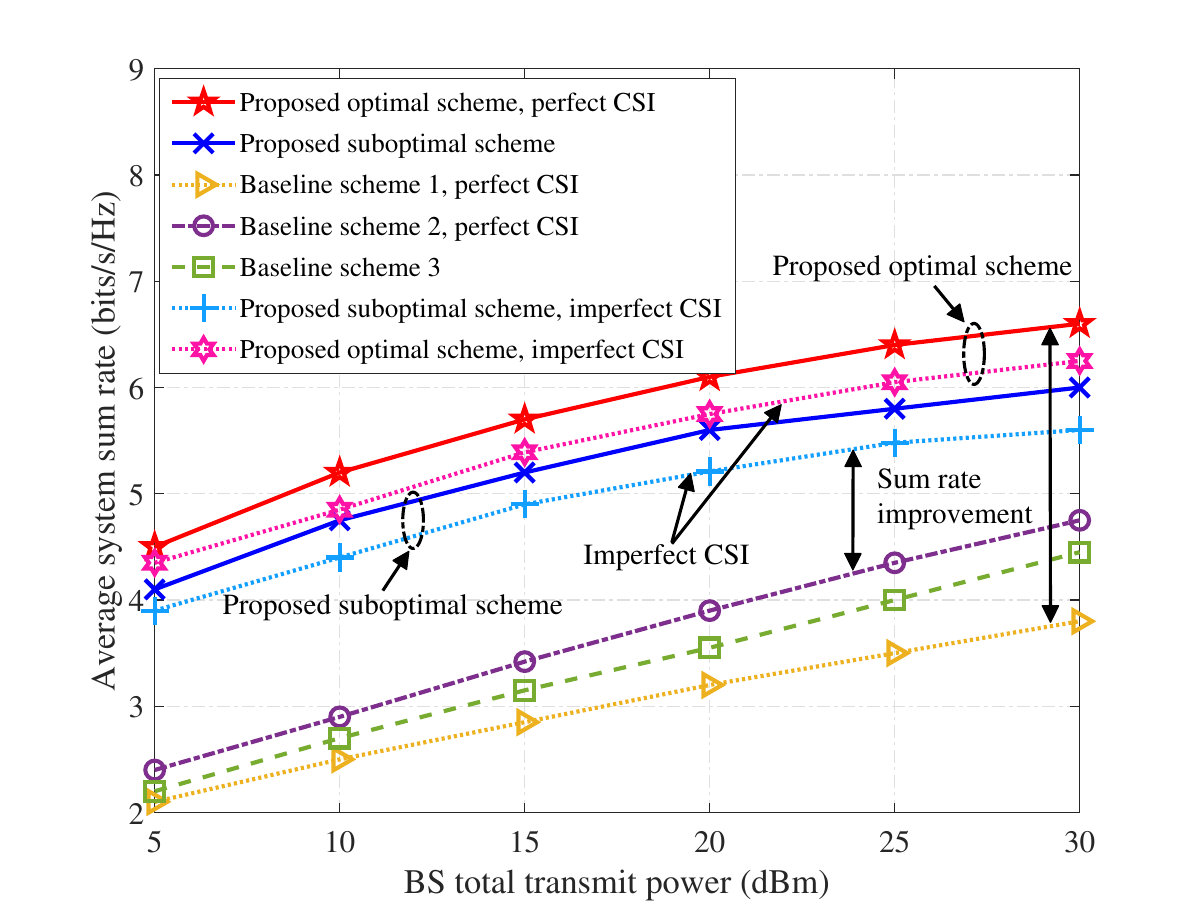}
    \caption{Average system sum rate (bits/s/Hz) versus the maximum transmit power of the BS $P_{\mathrm{max}}$ (dBm) for different resource allocation schemes.}
    \label{Fig.Simulation}
    \end{figure}
\par
Fig. \ref{Fig.Simulation} illustrates the average system sum rate (bits/s/Hz) versus the maximum transmit power of the BS $P_{\mathrm{max}}$ (dBm). The figure shows that both the proposed and baseline schemes exhibit a consistent increase in the average system sum rate with an increase in the maximum transmit power budget. This phenomenon can be attributed to the proposed optimization framework, which effectively enhances the SINRs of users by providing them with additional transmit power. This leads to a significant improvement in the overall system sum rate. Moreover, the figure also highlights that the proposed optimal scheme outperforms the suboptimal scheme. This is because the optimal scheme jointly optimizes all available wireless resources, thereby achieving better results. However, both the optimal and suboptimal schemes exhibit a remarkable sum rate gain compared to the three baseline schemes. This can be attributed to the fact that the three baseline schemes employ simple implementation methods, which often come at the cost of performance. Therefore, the results clearly confirm the effectiveness of the proposed optimal and suboptimal optimization algorithms. Overall, the figure provides valuable insights into the performance of different wireless resource allocation schemes and highlights the importance of optimizing all available resources for achieving optimal system performance.
\par
{On the other hand, to reveal the relationship between spectral efficiency, power consumption, and reliability, we also evaluate the performance of the proposed optimal and suboptimal schemes for the case with imperfect CSI. In particular, we assume that the multi-path fading channels, i.e., the channel between the BS and users and the channel between the IRS and users, vary within two known regions and adopt a bounded channel uncertainty model with known boundaries $\delta_{\mathrm{D},k}^2$ and $\delta_{\mathrm{R},k}^2$ to capture their impact on resource allocation design, respectively. For ease of presentation, we define the maximum normalized estimation errors of the CSI of the users as $\upsilon^2_{\mathrm{D},k}=\frac{\delta_{\mathrm{D},k}^2}{\left \| \mathbf{h}_{\mathrm{D},k} \right \|^2}$ and
$\upsilon^2_{\mathrm{R},k}=\frac{\delta_{\mathrm{R},k}^2}{\left \| \mathbf{h}_{\mathrm{R},k} \right \|^2}$, where $\upsilon_{\mathrm{D},k}^2=\upsilon^2_{\mathrm{R},k}=\upsilon^2=10\%$, $\forall k\in\mathcal{K}$. To improve the reliability of the NGMA systems, we exploit the $\mathcal{S}$-procedure \cite{cvx} to develop a robust resource allocation optimization framework. As can be observed from Fig. \ref{Fig.Simulation}, compared to the schemes with perfect CSI, the schemes with imperfect CSI only suffer from a slight performance loss. Yet, even with $10\%$ CSI uncertainty, both the optimal and suboptimal schemes outperform the three baseline schemes with perfect CSI. This indicates that the proposed scheme is robust against CSI uncertainties.}

\section{Future Research Directions}
This section outlines research directions regarding the resource allocation design for NGMA.
\subsection{Task-oriented Resource Allocation Design for NGMA}
The current resource allocation design for NOMA, i.e., rate-oriented, power-oriented, or reliability-oriented, aims at delivering information bits to more users without unfolding the meaning of the message or the goal of information exchange.
Recently, the burgeoning advancements in AI technologies and their widespread applications have sparked significant interest in semantics and task-oriented communications. This emerging paradigm seeks to optimize network connectivity to ensure the efficient accomplishment of intelligent tasks \cite{QinSemantic}, instead of merely transmitting information bits.
For instance, ChatGPT, a leading-edge large language model trained on a vast corpus of unlabeled text using the generative pre-trained transformer architecture, has demonstrated remarkable proficiency in generating content.
Leveraging ChatGPT's extensive knowledge base at the transceiver, it becomes conceivable to utilize a handful of prompts or semantic bits to facilitate the transmission of a video. This approach has the potential to surpass the traditional compression ratio bounds for lossless source coding, illustrating a significant leap in communication efficiency and effectiveness.
Integrating semantics into wireless communications transforms the conventional role of the network from a mere ``bit-pipe", which blindly supports higher-layer intelligence, into an active participant of networked intelligence. This evolved role embodies the capability to comprehend the significance of information bits in relation to the underlying task. It further extends to the ability to extract and deliver information that is relevant to the task at hand\cite{QinSemantic}.
In terms of the resource allocation design for NGMA of task-oriented communication, the design objectives are not limited to the radio resource efficiency but also tasks-related performance metrics, such as edge inference accuracy \cite{JiaweiJSAC} or task success probability\cite{HaijunSemantic}.
Additionally, the system resources are not limited to radio resources but also include the semantic compression ratio, task urgency, memory, and age of information, among others.
Moreover, machine learning plays a vital role in the task-oriented resource allocation design for NGMA, as task-related metrics are often challenging to model and compute.
Task-oriented user scheduling would be important in multi-agent systems\cite{ZhuoMultiAgent} where multiple agents intend to complete a cooperative task.
Task-oriented resource allocation design for NGMA has been rarely discussed in the literature and we believe this direction will become more important and attractive.

\subsection{Machine Learning-Enabled Resource Allocation Design for NGMA} 
{Recent advancements in technology have provided new solutions for resource allocation design in NGMA, among which, the application of machine learning and artificial intelligence has demonstrated great potential ~\cite{Djigal2022machine}. Indeed, conventional mathematical-based optimization methods often suffer from poor scalability and high computational complexity as the size of the solution set increases, thus preventing their direct applications in large-scale networks, especially for delay-sensitive and data-intensive applications.
In contrast, machine learning-based optimizations do not rely on explicit analytical models and therefore can provide a much better complexity scaling and improved efficiency. 
In addition, machine learning is suitable for resource allocation in complex networks, where historical data can be exploited to predict unknown performance metrics intelligently. More importantly, the neural network can be designed to be robust against channel fluctuations~\cite{bolukbasi2017adaptive}, which is particularly valuable for resource allocation in dynamic networks.}
In fact, machine learning has already shown promising performance in the design of NGMA systems~\cite{Guan2018DL,Kowshik2023DL,Huang2020DL}, including the resource allocation design. For instance, in~\cite{Huang2020DL}, an effective communication deep neural network was proposed which consists of several convolutional layers and multiple hidden layers. Relying on the training algorithms, the proposed neural network can address the power allocation problem over challenging channel conditions with extremely complex spatial stricture limitations.
A hot-booting Q-learning-based power allocation for downlink NOMA was presented in~\cite{Liang2018reinforcement}, which does not rely on the explicit representation of the jamming and channel parameters.  
Furthermore, better performance can be observed using the proposed scheme in comparison to the standard Q-learning-based strategy.
The above two examples have underscored the effectiveness of machine learning in resource allocations for natural channels. 
Machine learning-based resource allocation has also been adopted for reconfigurable channels and functional channels.
In~\cite{Minghui2023DL}, a deep learning-based RSMA scheme was proposed for RIS-assisted Tera-Hertz (THz) massive MIMO transmissions. Specifically, the proposed scheme consists of passive precoding at the RIS, analog active precoding, and RSMA digital active precoding at the BS. Furthermore, a CSI acquisition network was also devised to acquire accurate CSI. The proposed scheme enables a higher spectral efficiency compared to the conventional designs with lower signaling overhead.
A predictive beamforming-aided ISAC transmission was presented in~\cite{chang2022learning}, where a versatile unsupervised deep learning network was devised. Specifically, the network was realized by a long short-term memory (LSTM) network with historically estimated channels as inputs. Furthermore, the resource allocation problem was formulated to maximize the system sum rate while making sure the Cram{\'{e}}r–Rao lower bound (CRLB) is smaller than a threshold, which is solved by the deep learning network. Numerical results were presented and shown to approach the genie-aided upper bound. {Furthermore, in~\cite{Chang2021location}, a location-aware predictive beamforming for UAV communications was developed. This scheme showed a strong resilience to unexpected link failures, attributable to the powerful deep learning neural networks, which also demonstrated a significant communication performance improvement compared to the conventional optimization methods.}

Rather than directly applying machine learning to design the resource allocation strategy, machine learning can also be applied as a tool to solve the optimization problem for resource allocation.
An example of such design is the neural optimization machine (NOM)~\cite{Chen2022NOM}, which leverages the neural network as a surrogate model and optimizes it with and without constraints. 
Compared to conventional optimization techniques, NOM can be applied to general objective functions, e.g., linear and quadratic programming. Furthermore, NOM is flexible, where arbitrary neural network architectures and activation functions can be adopted. More importantly, NOM is capable of identifying multiple local minima without a significant increase in computational cost, even as the dimensionality of the design variables grows.
These compelling attributes position NOM as a valuable tool for addressing the complex optimization challenges in resource allocation. The exploration of NOM within the context of NGMA is still in its infancy, presenting a fascinating avenue for future research.

\subsection{Distributed Resource Allocation Design for NGMA} 
Most resource allocation designs for NGMA have been implemented in a centralized manner.
For example, in the uplink NGMA system, each user is required to report to a central controller for system information collection, and the optimized resource allocation strategy is then distributed to each user.
This centralized approach ensures high-quality resource allocation strategies and superior system performance. However, it comes with significant overhead, which escalates dramatically as the system scales, making it unsustainable for systems with a large number of users or multiple cells.
Moreover, in highly dynamic environments, such as UAV and vehicular communication systems, obtaining complete and real-time knowledge of dynamic users becomes a non-trivial challenge.
Therefore, a distributed resource allocation design for NGMA is highly desired.
The framework for distributed resource allocation must be developed based on the system model to identify the role of different terminals and establish the information exchange protocol among them.
The distributed resource allocation problem for NGMA should then be formulated in accordance with a specific design goal. Distributed optimization approaches, such as the alternating direction method of multipliers (ADMM) \cite{ChangDistributed} or a game theory-based approach \cite{TengjiaoGameTheory}, are very promising in the development of distributed resource allocation design algorithms.
Moreover, deep reinforcement learning (DRL) \cite{HaoDRL} and multi-agent reinforcement learning (MARL) \cite{YuanweiUAV} are capable of providing a distributed but intelligent resource allocation design for dynamic NGMA system, where the closed-form design objective function is usually impossible to obtain.
According to the DRL and MARL frameworks, the long-term resource allocation problem is formulated as a Markov decision process (MDP), which accommodates the dynamics and uncertainty inherent in NGMA systems. In this setup, each user or BS is considered a learning agent, and each resource allocation solution is treated as an action undertaken by the user or BS. This perspective allows for an interactive and adaptive approach for resource management, where decisions are made based on the current state of the system and its projected future states.
Each agent runs an action network to discover its best strategy according to its local observations by ignoring the other agents, and hence information exchanges between agents and computational burdens at each agent are substantially reduced.
Distributed resource allocation design for NGMA is still in its infancy and is a promising research direction and invoking DRL and MARL to dynamic NGMA systems provides a promising solution for intelligent resource allocation design.

\section{Conclusions}
This article provided a comprehensive overview of the system models, problem formulations, and potential optimization approaches for the resource allocation design of NGMA across three types of channels: natural channels, reconfigurable channels, and functional channels.
For each type of channel, we proposed a unified framework for NGMA by formulating the essential input-output relationship, emphasizing the system resources and design DoF, and highlighting the key performance metrics for resource allocation design.
Diverse NGMA system models applicable to these three channels were methodically discussed, including OFDMA, NOMA, RSMA, DDMA, IRS-assisted NGMA, UAV-assisted NGMA, M/FA-enabled NGMA, NGMA-based ISAC, and NGMA-enabled JCAC.
Even within a specific channel type and NGMA scheme, a variety of resource allocation designs can be formulated based on the available system information (whether CSI is perfect or imperfect), QoS requirements, and design objectives.
To cover more resource allocation designs, we formulated different types of resource allocation design problems for different NGMA system models, which are mainly categorized as rate-oriented, power-oriented, and reliability-oriented designs.
Corresponding optimization tools for solving the formulated resource allocation design problems were presented, including both the global optimization approaches and low-complexity suboptimal approaches. This comprehensive toolkit provides a versatile framework for tackling the diverse challenges inherent in NGMA system optimization.
Our vision for future research directions on resource allocation design for NGMA was briefly discussed, including machine learning-enabled resource allocation design approaches, distributed resource allocation design, and task-oriented NGMA and resource allocation design. This forward-looking perspective underscores the evolving nature of NGMA systems and the potential for innovative methodologies to significantly enhance their efficiency and effectiveness.
A simulation example of resource allocation design for IRS-assisted NOMA systems was provided and discussed, which highlights the importance of resource allocation designs for improving system performance.
We believe that resource allocation design will continue to be of paramount importance in multiple access strategies within future wireless cellular networks.
Emerging technologies would pose new challenges and interesting research topics for resource allocation design, which deserves continuous research efforts invested in this area.
We hope this article serves as a starter pack for junior researchers studying resource allocation design for NGMA, inspiring further exploration of compelling questions and directions in this burgeoning area of research.

\bibliographystyle{IEEEtran}
\bibliography{NOMA}

\begin{thebibliography}{100}
\providecommand{\url}[1]{#1}
\csname url@samestyle\endcsname
\providecommand{\newblock}{\relax}
\providecommand{\bibinfo}[2]{#2}
\providecommand{\BIBentrySTDinterwordspacing}{\spaceskip=0pt\relax}
\providecommand{\BIBentryALTinterwordstretchfactor}{4}
\providecommand{\BIBentryALTinterwordspacing}{\spaceskip=\fontdimen2\font plus
\BIBentryALTinterwordstretchfactor\fontdimen3\font minus
  \fontdimen4\font\relax}
\providecommand{\BIBforeignlanguage}[2]{{%
\expandafter\ifx\csname l@#1\endcsname\relax
\typeout{** WARNING: IEEEtran.bst: No hyphenation pattern has been}%
\typeout{** loaded for the language `#1'. Using the pattern for}%
\typeout{** the default language instead.}%
\else
\language=\csname l@#1\endcsname
\fi
#2}}
\providecommand{\BIBdecl}{\relax}
\BIBdecl

\bibitem{IMT-2030}
``{ITU-R} framework for {IMT-2030},'' ITU-R WP 5D, Tech. Rep., Jul. 2023.

\bibitem{Verdu1999}
S.~Verdu and S.~Shamai, ``Spectral efficiency of {CDMA} with random
  spreading,'' \emph{IEEE Trans. Inf. Theory}, vol.~45, no.~2, pp. 622--640,
  Mar. 1999.

\bibitem{YangCDMA}
L.-L. Yang and L.~Hanzo, ``Multicarrier {DS-CDMA}: A multiple access scheme for
  ubiquitous broadband wireless communications,'' \emph{IEEE Commun. Mag.}, pp.
  116--124, Oct. 2003.

\bibitem{Wong1999}
C.~Y. Wong, R.~Cheng, K.~Lataief, and R.~Murch, ``Multiuser {OFDM} with
  adaptive subcarrier, bit, and power allocation,'' \emph{IEEE J. Select. Areas
  Commun.}, vol.~17, no.~10, pp. 1747--1758, Oct. 1999.

\bibitem{Strohmer2003}
T.~Strohmer and S.~Beaver, ``Optimal {OFDM} design for time-frequency
  dispersive channels,'' \emph{IEEE Trans. Commun.}, vol.~51, no.~7, pp.
  1111--1122, July 2003.

\bibitem{Kwan_AF_2010}
D.~W.~K. Ng and R.~Schober, ``Cross-layer scheduling for {OFDMA}
  amplify-and-forward relay networks,'' \emph{IEEE Trans. Veh. Technol.},
  vol.~59, no.~3, pp. 1443--1458, Mar. 2010.

\bibitem{DerrickEEOFDMA}
D.~W.~K. Ng, E.~S. Lo, and R.~Schober, ``Energy-efficient resource allocation
  in {OFDMA} systems with large numbers of base station antennas,'' \emph{IEEE
  Trans. Wireless Commun.}, vol.~11, no.~9, pp. 3292--3304, Sep. 2012.

\bibitem{DerrickEESWIPT}
------, ``Wireless information and power transfer: Energy efficiency
  optimization in {OFDMA} systems,'' \emph{IEEE Trans. Wireless Commun.},
  vol.~12, no.~12, pp. 6352--6370, Dec. 2013.

\bibitem{DerrickLimitedBackhaul}
------, ``Energy-efficient resource allocation in multi-cell {OFDMA} systems
  with limited backhaul capacity,'' \emph{IEEE Trans. Wireless Commun.},
  vol.~11, no.~10, pp. 3618--3631, Oct. 2012.

\bibitem{ZhiqiangOFDMA}
Z.~Wei, Y.~Cai, Z.~Sun, D.~W.~K. Ng, J.~Yuan, M.~Zhou, and L.~Sun, ``Sum-rate
  maximization for {IRS}-assisted {UAV OFDMA} communication systems,''
  \emph{IEEE Trans. Wireless Commun.}, vol.~20, no.~4, pp. 2530--2550, Apr.
  2021.

\bibitem{yang2017noma}
Q.~Yang, H.-M. Wang, D.~W.~K. Ng, and M.~H. Lee, ``{NOMA} in downlink {SDMA}
  with limited feedback: Performance analysis and optimization,'' \emph{IEEE J.
  Select. Areas Commun.}, vol.~35, no.~10, pp. 2281--2294, Oct. 2017.

\bibitem{Ngo2011}
H.~Q. Ngo, T.~Q. Duong, and E.~G. Larsson, ``Uplink performance analysis of
  multicell {MU-MIMO} with zero-forcing receivers and perfect {CSI},'' in
  \emph{IEEE Swedish Communication Technologies Workshop (Swe-CTW)}, Oct. 2011,
  pp. 40--45.

\bibitem{Tse2005}
D.~Tse and P.~Viswanath, \emph{Fundamentals of wireless communication}.\hskip
  1em plus 0.5em minus 0.4em\relax Cambridge university press, 2005.

\bibitem{Ding2015b}
Z.~Ding, Y.~Liu, J.~Choi, Q.~Sun, M.~Elkashlan, C.~L. I, and H.~V. Poor,
  ``Application of non-orthogonal multiple access in {LTE} and {5G} networks,''
  \emph{IEEE Commun. Mag.}, vol.~55, no.~2, pp. 185--191, Feb. 2017.

\bibitem{LiuSWIPT}
Y.~Liu, Z.~Ding, M.~Elkashlan, and H.~V. Poor, ``Cooperative non-orthogonal
  multiple access with simultaneous wireless information and power transfer,''
  \emph{IEEE J. Select. Areas Commun.}, vol.~34, no.~4, pp. 938--953, Apr.
  2016.

\bibitem{Lei2016NOMA}
L.~Lei, D.~Yuan, C.~K. Ho, and S.~Sun, ``Power and channel allocation for
  non-orthogonal multiple access in {5G} systems: Tractability and
  computation,'' \emph{IEEE Trans. Wireless Commun.}, vol.~15, no.~12, pp.
  8580--8594, Dec. 2016.

\bibitem{9679390}
Z.~Ding, D.~{Xu}, and R.~S. H.~V. Poor, ``Hybrid {NOMA} offloading in
  multi-user {MEC} networks,'' \emph{IEEE Trans. Wireless Commun.}, vol.~21,
  no.~7, pp. 5377--5391, Jul. 2022.

\bibitem{MUDCDMA}
S.~Moshavi, ``Multi-user detection for {DS-CDMA} communications,'' \emph{IEEE
  Commun. Mag.}, vol.~34, no.~10, pp. 124--136, Oct. 1996.

\bibitem{Hanzo2003}
L.~Hanzo, L.-L. Yang, E.~Kuan, and K.~Yen, \emph{Single- and Multi-Carrier
  {DS-CDMA}: Multi-USer Detection, Space-Time Spreading, Synchronisation,
  Standards and Networking}.\hskip 1em plus 0.5em minus 0.4em\relax John Wiley
  \& Sons, Aug. 2003.

\bibitem{PerGainWei}
Z.~Wei, L.~Yang, D.~W.~K. Ng, J.~Yuan, and L.~Hanzo, ``On the performance gain
  of {NOMA} over {OMA} in uplink communication systems,'' \emph{IEEE Trans.
  Commun.}, vol.~68, no.~1, pp. 536--568, Jan. 2020.

\bibitem{XuMassiveMIMONOMA}
C.~Xu, Y.~Hu, C.~Liang, J.~Ma, and L.~Ping, ``Massive {MIMO}, non-orthogonal
  multiple access and interleave division multiple access,'' \emph{IEEE
  Access}, vol.~5, pp. 14\,728--14\,748, Jul. 2017.

\bibitem{DingSignalAlignment}
Z.~Ding, R.~Schober, and H.~V. Poor, ``A general {MIMO} framework for {NOMA}
  downlink and uplink transmission based on signal alignment,'' \emph{IEEE
  Trans. Wireless Commun.}, vol.~15, no.~6, pp. 4438--4454, Jun. 2016.

\bibitem{wei2018multibeam}
Z.~Wei, L.~Zhao, J.~Guo, D.~W.~K. Ng, and J.~Yuan, ``Multi-beam {NOMA} for
  hybrid mmwave systems,'' \emph{IEEE Trans. Commun.}, vol.~67, no.~2, pp.
  1705--1719, Feb. 2019.

\bibitem{WeiBeamWidthControl}
Z.~{Wei}, D.~W. {Kwan Ng}, and J.~{Yuan}, ``{NOMA} for hybrid mmwave
  communication systems with beamwidth control,'' \emph{IEEE J. Select. Topics
  Signal Process.}, vol.~13, no.~3, pp. 567--583, Jun. 2019.

\bibitem{Wei2017}
Z.~Wei, D.~W.~K. Ng, J.~Yuan, and H.~M. Wang, ``Optimal resource allocation for
  power-efficient {MC-NOMA} with imperfect channel state information,''
  \emph{IEEE Trans. Commun.}, vol.~65, no.~9, pp. 3944--3961, Sep. 2017.

\bibitem{Razavi2012}
R.~Razavi, M.~Al-Imari, M.~A. Imran, R.~Hoshyar, and D.~Chen, ``On receiver
  design for uplink low density signature {OFDM (LDS-OFDM)},'' \emph{IEEE
  Trans. Commun.}, vol.~60, no.~11, pp. 3499--3508, Nov. 2012.

\bibitem{WeijieSCMA}
W.~Yuan, N.~Wu, C.~Yan, Y.~Li, X.~Huang, and L.~Hanzo, ``A low-complexity
  energy-minimization-based {SCMA} detector and its convergence analysis,''
  \emph{IEEE Trans. Veh. Technol.}, vol.~67, no.~12, pp. 12\,398--12\,403, Dec.
  2018.

\bibitem{WeijieSCMAII}
W.~Yuan, N.~Wu, A.~Zhang, X.~Huang, Y.~Li, and L.~Hanzo, ``Iterative receiver
  design for {FTN} signaling aided sparse code multiple access,'' \emph{IEEE
  Trans. Wireless Commun.}, vol.~19, no.~2, pp. 915--928, Feb. 2020.

\bibitem{ShanzhiPDMA}
S.~Chen, B.~Ren, Q.~Gao, S.~Kang, S.~Sun, and K.~Niu, ``Pattern division
  multiple access—a novel nonorthogonal multiple access for fifth-generation
  radio networks,'' \emph{IEEE Trans. Veh. Technol.}, vol.~66, no.~4, pp.
  3185--3196, Apr. 2017.

\bibitem{BrunoJSAC}
B.~Clerckx, Y.~Mao, E.~A. Jorswieck, J.~Yuan, D.~J. Love, E.~Erkip, and
  D.~Niyato, ``A primer on rate-splitting multiple access: Tutorial, myths, and
  frequently asked questions,'' \emph{IEEE J. Select. Areas Commun.}, vol.~41,
  no.~5, pp. 1265--1308, May 2023.

\bibitem{YijieRSMA}
Y.~Mao, O.~Dizdar, B.~Clerckx, R.~Schober, P.~Popovski, and H.~V. Poor,
  ``Rate-splitting multiple access: Fundamentals, survey, and future research
  trends,'' \emph{IEEE Commun. Surveys Tuts. Mag.}, vol.~24, no.~4, pp.
  2073--2126, Fourthquater 2022.

\bibitem{Access2015}
``Study on downlink multiuser supersition transmission ({MUST}) for {LTE}
  ({R}elease 13),'' 3GPP TR 36.859, Tech. Rep., Dec. 2015.

\bibitem{YuanweiNGMA}
Y.~Liu, S.~Zhang, X.~Mu, Z.~Ding, R.~Schober, N.~Al-Dhahir, E.~Hossain, and
  X.~Shen, ``Evolution of {NOMA} toward next generation multiple access
  ({NGMA}) for {6G},'' \emph{IEEE J. Select. Areas Commun.}, vol.~40, no.~4,
  pp. 1037--1071, Apr. 2022.

\bibitem{XianghaoJSAC}
X.~Yu, D.~Xu, Y.~Sun, D.~W.~K. Ng, and R.~Schober, ``Robust and secure wireless
  communications via intelligent reflecting surfaces,'' \emph{IEEE J. Select.
  Areas Commun.}, vol.~38, no.~11, pp. 2637--2652, Nov. 2020.

\bibitem{9183907}
D.~{Xu}, Y.~{Sun}, D.~W.~K. {Ng}, and R.~{Schober}, ``Resource allocation for
  {IRS}-assisted full-duplex cognitive radio systems,'' \emph{IEEE Trans.
  Commun.}, vol.~68, no.~12, pp. 7376--7394, Dec. 2020.

\bibitem{8663615}
Y.~{Zeng}, J.~{Xu}, and R.~{Zhang}, ``Energy minimization for wireless
  communication with rotary-wing {UAV},'' \emph{IEEE Trans. Wireless Commun.},
  vol.~18, no.~4, pp. 2329--2345, Apr. 2019.

\bibitem{8644086}
D.~{Xu}, Y.~{Sun}, D.~W.~K. {Ng}, and R.~{Schober}, ``Robust resource
  allocation for {UAV} systems with {UAV} jittering and user location
  uncertainty,'' in \emph{Proc. IEEE Global Commun. Conf. (GC Wkshps)}, Abu
  Dhabi, United Arab Emirates, Dec. 2018, pp. 1--6.

\bibitem{LipengMovableAntenna}
L.~Zhu, W.~Ma, and R.~Zhang, ``Movable-antenna array enhanced beamforming:
  Achieving full array gain with null steering,'' \emph{IEEE Commun. Lett.},
  vol.~27, no.~12, pp. 3340--3344, Dec. 2023.

\bibitem{10318134}
W.~K. New, K.-K. Wong, H.~Xu, K.-F. Tong, C.-B. Chae, and Y.~Zhang, ``Fluid
  antenna system enhancing orthogonal and non-orthogonal multiple access,''
  \emph{IEEE Commun. Lett.}, vol.~28, no.~1, pp. 218--222, Jan. 2024.

\bibitem{Fan2020}
F.~Liu, C.~Masouros, A.~P. Petropulu, H.~Griffiths, and L.~Hanzo, ``Joint radar
  and communication design: Applications, state-of-the-art, and the road
  ahead,'' \emph{IEEE Trans. Commun.}, vol.~68, no.~6, pp. 3834--3862, Jun.
  2020.

\bibitem{Yuyi2017}
Y.~Mao, C.~You, J.~Zhang, K.~Huang, and K.~B. Letaief, ``A survey on mobile
  edge computing: The communication perspective,'' \emph{IEEE Commun. Surveys
  Tuts. Mag.}, vol.~19, no.~4, pp. 2322--2358, Fourthquarter 2017.

\bibitem{YandongCST}
Y.~Shi, L.~Lian, Y.~Shi, Z.~Wang, Y.~Zhou, L.~Fu, L.~Bai, J.~Zhang, and
  W.~Zhang, ``Machine learning for large-scale optimization in {6G} wireless
  networks,'' \emph{IEEE Commun. Surveys Tuts. Mag.}, vol.~25, no.~4, pp.
  2088--2132, Fourthquarter 2023.

\bibitem{Shuangyang_FTN_NOMA}
S.~Li, Z.~Wei, W.~Yuan, J.~Yuan, B.~Bai, D.~W.~K. Ng, and L.~Hanzo,
  ``Faster-than-{Nyquist} asynchronous {NOMA} outperforms synchronous {NOMA},''
  \emph{IEEE J. Sel. Areas Commun.}, vol.~40, no.~4, pp. 1128--1145, Apr. 2022.

\bibitem{WeiProceeding}
Z.~Wei, X.~Yu, D.~W.~K. Ng, and R.~Schober, ``Resource allocation for
  simultaneous wireless information and power transfer systems: A tutorial
  overview,'' \emph{Proceedings of the IEEE}, vol. 110, no.~1, pp. 127--149,
  Jan. 2022.

\bibitem{QingqingIRS}
Q.~Wu and R.~Zhang, ``Intelligent reflecting surface enhanced wireless network
  via joint active and passive beamforming,'' \emph{IEEE Trans. Wireless
  Commun.}, vol.~18, no.~11, pp. 5394--5409, Nov. 2019.

\bibitem{YuanxinUAVNOMA}
Y.~Cai, Z.~Wei, S.~Hu, C.~Liu, D.~W.~K. Ng, and J.~Yuan, ``Resource allocation
  and {3D} trajectory design for power-efficient {IRS}-assisted {UAV-NOMA}
  communications,'' \emph{IEEE Trans. Wireless Commun.}, vol.~21, no.~12, pp.
  10\,315--10\,334, Dec. 2022.

\bibitem{ZhiqiangISAC}
Z.~Wei, F.~Liu, C.~Liu, Z.~Yang, D.~W.~K. Ng, and R.~Schober, ``Integrated
  sensing, navigation, and communication for secure uav networks with a mobile
  eavesdropper,'' \emph{IEEE Trans. Wireless Commun.}, pp. 1--1, early access,
  2023.

\bibitem{Al-Imari2011LDSOFDM}
M.~Al-Imari, M.~A. Imran, R.~Tafazolli, and D.~Chen, ``Subcarrier and power
  allocation for {LDS-OFDM} system,'' in \emph{Proc. IEEE Veh. Techn. Conf.},
  May 2011, pp. 1--5.

\bibitem{BayestehSCMA}
A.~Bayesteh, H.~Nikopour, M.~Taherzadeh, H.~Baligh, and J.~Ma, ``Low complexity
  techniques for {SCMA} detection,'' in \emph{Proc. IEEE Global Commun. Conf.},
  2015, pp. 1--6.

\bibitem{BigueshMMSE}
M.~{Biguesh} and A.~B. {Gershman}, ``Training-based {MIMO} channel estimation:
  a study of estimator tradeoffs and optimal training signals,'' \emph{IEEE
  Trans. Signal Process.}, vol.~54, no.~3, pp. 884--893, Mar. 2006.

\bibitem{zhao2017multiuser}
L.~Zhao, D.~W.~K. Ng, and J.~Yuan, ``Multi-user precoding and channel
  estimation for hybrid millimeter wave systems,'' \emph{IEEE J. Select. Areas
  Commun.}, vol.~35, no.~7, pp. 1576--1590, Jul. 2017.

\bibitem{Sun2019}
Z.~{Sun}, Z.~{Wei}, L.~{Yang}, J.~{Yuan}, X.~{Cheng}, and L.~{Wan},
  ``Exploiting transmission control for joint user identification and channel
  estimation in massive connectivity,'' \emph{IEEE Trans. Commun.}, vol.~67,
  no.~9, pp. 6311--6326, Sep. 2019.

\bibitem{ZengHybridNOMA}
M.~Zeng, A.~Yadav, O.~A. Dobre, and H.~V. Poor, ``Energy-efficient joint
  user-{RB} association and power allocation for uplink hybrid {NOMA-OMA},''
  \emph{IEEE Internet Things J.}, vol.~6, no.~3, pp. 5119--5131, Jun. 2019.

\bibitem{Zhiqiang_magzine}
Z.~Wei, W.~Yuan, S.~Li, J.~Yuan, G.~Bharatula, R.~Hadani, and L.~Hanzo,
  ``Orthogonal time-frequency space modulation: A promising next-generation
  waveform,'' \emph{IEEE Wireless Commun.}, vol.~28, no.~4, pp. 136--144, Aug.
  2021.

\bibitem{Chong2022achievable}
R.~Chong, S.~Li, J.~Yuan, and D.~W.~K. Ng, ``Achievable rate upper-bounds of
  uplink multiuser {OTFS} transmissions,'' \emph{IEEE Wireless Commun. Lett.},
  vol.~11, no.~4, pp. 791--795, Apr. 2022.

\bibitem{Hadani2017orthogonal}
R.~{Hadani}, S.~{Rakib}, M.~{Tsatsanis}, A.~{Monk}, A.~J. {Goldsmith}, A.~F.
  {Molisch}, and R.~{Calderbank}, ``Orthogonal time frequency space
  modulation,'' in \emph{Proc. 2017 IEEE Wireless Commun. Net. Conf.}, Mar.
  2017, pp. 1--6.

\bibitem{ZhiqiangLetterPartI}
Z.~Wei, S.~Li, W.~Yuan, R.~Schober, and G.~Caire, ``Orthogonal time frequency
  space modulation—{Part I}: Fundamentals and challenges ahead,'' \emph{IEEE
  Commun. Lett.}, vol.~27, no.~1, pp. 4--8, Jan. 2023.

\bibitem{ShuangyangLetterPartII}
S.~Li, W.~Yuan, Z.~Wei, R.~Schober, and G.~Caire, ``Orthogonal time frequency
  space modulation—{Part II}: Transceiver designs,'' \emph{IEEE Commun.
  Lett.}, vol.~27, no.~1, pp. 9--13, Jan. 2023.

\bibitem{WeijieLetterPartIII}
W.~Yuan, Z.~Wei, S.~Li, R.~Schober, and G.~Caire, ``Orthogonal time frequency
  space modulation—{Part III}: {ISAC} and potential applications,''
  \emph{IEEE Communications Letters}, vol.~27, no.~1, pp. 14--18, Jan. 2023.

\bibitem{Yuan2023survey}
W.~Yuan, S.~Li, Z.~Wei, Y.~Cui, J.~Jiang, H.~Zhang, and P.~Fan, ``New delay
  {Doppler} communication paradigm in {6G} era: A survey of orthogonal time
  frequency space {(OTFS)},'' \emph{China Commun.}, vol.~20, no.~6, pp. 1--25,
  Jun. 2023.

\bibitem{Yuan2019simple}
W.~{Yuan}, Z.~{Wei}, J.~{Yuan}, and D.~W.~K. {Ng}, ``A simple variational
  {Bayes} detector for orthogonal time frequency space {(OTFS)} modulation,''
  \emph{IEEE Trans Veh. Technol.}, vol.~69, no.~7, pp. 7976--7980, Jul. 2020.

\bibitem{Mengmeng2023P2PMIMO}
M.~Liu, S.~Li, Z.~Wei, and B.~Bai, ``Near optimal hybrid digital-analog
  beamforming for point-to-point {MIMO-OTFS} transmissions,'' in \emph{IEEE
  Wireless Commun.Netw. Conf. Workshop}, 2023, pp. 1--6.

\bibitem{Ruoxi2022ISWCS}
R.~Chong, M.~Mohammadi, H.~Q. Ngo, S.~L. Cotton, and M.~Matthaiou, ``On the
  spectral efficiency of {MMSE}-based {MIMO OTFS} systems,'' in \emph{Int.
  Symp. Wireless Commun. Sys. (ISWCS)}, 2022, pp. 1--6.

\bibitem{Ruoxi2023Globecom}
R.~Chong, M.~Mohammadi, H.-Q. Ngo, S.~Cotton, and M.~Matthaiou, ``How to
  combine {OTFS} and {OFDM} modulations in massive {MIMO}?'' in \emph{IEEE
  Globe Commun. Conf.}, 2023, pp. 1--6.

\bibitem{Shuangyang_ISAC}
S.~Li, W.~Yuan, C.~Liu, Z.~Wei, J.~Yuan, B.~Bai, and D.~W.~K. Ng, ``A novel
  {ISAC} transmission framework based on spatially-spread orthogonal time
  frequency space modulation,'' \emph{IEEE J. Sel. Areas Commun.}, vol.~40,
  no.~6, pp. 1854--1872, Jun. 2022.

\bibitem{Weijie2021JSTSP}
W.~Yuan, Z.~Wei, S.~Li, J.~Yuan, and D.~W.~K. Ng, ``Integrated sensing and
  communication-assisted orthogonal time frequency space transmission for
  vehicular networks,'' \emph{IEEE J. Sel. Top. Signal Process.}, vol.~15,
  no.~6, pp. 1515--1528, Apr. 2021.

\bibitem{Chang2023Predictive}
C.~Liu, S.~Li, W.~Yuan, X.~Liu, and D.~W.~K. Ng, ``Predictive precoder design
  for {OTFS}-enabled {URLLC}: A deep learning approach,'' \emph{IEEE J. Sel.
  Areas Commun.}, vol.~41, no.~7, pp. 2245--2260, May 2023.

\bibitem{Shuangyang2023Globecom}
S.~Li, W.~Yuan, Z.~Wei, J.~Yuan, B.~Bai, and G.~Caire, ``On the pulse shaping
  for delay-{Doppler} communications,'' in \emph{IEEE Globe Commun. Conf.},
  2023, pp. 1--6.

\bibitem{LSY_THP}
S.~Li, J.~Yuan, P.~Fitzpatrick, T.~Sakurai, and G.~Caire, ``Delay-{Doppler}
  domain {Tomlinson-Harashima} precoding for {OTFS}-based downlink {MU-MIMO}
  transmissions: Linear complexity implementation and scaling law analysis,''
  \emph{IEEE Trans. Commun.}, vol.~71, no.~4, pp. 2153--2169, Apr. 2023.

\bibitem{Zhiqiang2022off}
Z.~Wei, W.~Yuan, S.~Li, J.~Yuan, and D.~W.~K. Ng, ``Off-grid channel estimation
  with sparse {Bayesian} learning for {OTFS} systems,'' \emph{IEEE Trans.
  Wireless Commun.}, vol.~21, no.~9, pp. 7407--7426, Mar. 2022.

\bibitem{Ruoxi2022ICC}
R.~Chong, S.~Li, W.~Yuan, and J.~Yuan, ``Outage analysis for {OTFS}-based
  single user and multi-user transmissions,'' in \emph{IEEE Int. Conf. Commun.
  Workshops (ICC Workshops)}, 2022, pp. 746--751.

\bibitem{li2021cross}
S.~Li, W.~Yuan, Z.~Wei, and J.~Yuan, ``Cross domain iterative detection for
  orthogonal time frequency space modulation,'' \emph{IEEE Trans. Wireless
  Commun.}, vol.~21, no.~4, pp. 2227--2242, Apr. 2022.

\bibitem{Li2020performance}
S.~Li, J.~Yuan, Z.~Wei, B.~Bai, and D.~W.~K. Ng, ``Performance analysis of
  coded {OTFS} systems over high-mobility channels,'' \emph{IEEE Trans.
  Wireless Commun.}, vol.~20, no.~9, pp. 6033--6048, Sep. 2021.

\bibitem{Wei2020transmitter}
Z.~Wei, W.~Yuan, S.~Li, J.~Yuan, and D.~W.~K. Ng, ``Transmitter and receiver
  window designs for orthogonal time-frequency space modulation,'' \emph{IEEE
  Trans. Commun.}, vol.~69, no.~4, pp. 2207--2223, Jan. 2021.

\bibitem{Shuangyang2021hybrid}
S.~Li, W.~Yuan, Z.~Wei, J.~Yuan, B.~Bai, D.~W.~K. Ng, and Y.~Xie, ``Hybrid
  {MAP} and {PIC} detection for {OTFS} modulation,'' \emph{IEEE Trans. Veh.
  Technol.}, vol.~70, no.~7, pp. 7193--7198, May 2021.

\bibitem{wong2017key}
V.~W. Wong, R.~Schober, D.~W.~K. Ng, and L.-C. Wang, \emph{Key Technologies for
  {5G} Wireless Systems}.\hskip 1em plus 0.5em minus 0.4em\relax Cambridge
  University Press, 2017.

\bibitem{you2021towards}
X.~{You} \emph{et~al.}, ``Towards {6G} wireless communication networks: Vision,
  enabling technologies, and new paradigm shifts,'' \emph{Science China
  Information Sciences}, vol.~64, no.~1, pp. 1--74, Jan. 2021.

\bibitem{8445944}
Y.~{Sun}, D.~W.~K. {Ng}, D.~{Xu}, L.~{Dai}, and R.~{Schober}, ``Resource
  allocation for solar powered {UAV} communication systems,'' in \emph{Proc.
  IEEE Intern. Workshop on Signal Process. Advances in Wireless Commun.},
  Kalamata, Greece, Jun. 2018, pp. 1--5.

\bibitem{8974403}
D.~{Xu}, Y.~{Sun}, D.~W.~K. {Ng}, and R.~{Schober}, ``Multiuser {MISO} {UAV}
  communications in uncertain environments with no-fly zones: Robust trajectory
  and resource allocation design,'' \emph{IEEE Trans. Commun.}, vol.~68, no.~5,
  pp. 3153--3172, May 2020.

\bibitem{8918497}
Y.~Zeng, Q.~Wu, and R.~Zhang, ``Accessing from the sky: A tutorial on {UAV}
  communications for {5G} and beyond,'' \emph{Proc. IEEE}, vol. 107, no.~12,
  pp. 2327--2375, 2019.

\bibitem{8648498}
Y.~{Sun}, D.~{Xu}, D.~W.~K. {Ng}, L.~{Dai}, and R.~{Schober}, ``Optimal
  {3D}-trajectory design and resource allocation for solar-powered {UAV}
  communication systems,'' \emph{IEEE Trans. Commun.}, vol.~67, no.~6, pp.
  4281--4298, Jun. 2019.

\bibitem{8114722}
Y.~Liu, Z.~Qin, M.~Elkashlan, Z.~Ding, A.~Nallanathan, and L.~Hanzo,
  ``Nonorthogonal multiple access for {5G} and beyond,'' \emph{Proc. IEEE},
  vol. 105, no.~12, pp. 2347--2381, 2017.

\bibitem{7973146}
Z.~Ding, X.~Lei, G.~K. Karagiannidis, R.~Schober, J.~Yuan, and V.~K. Bhargava,
  ``A survey on non-orthogonal multiple access for {5G} networks: Research
  challenges and future trends,'' \emph{IEEE J. Sel. Areas Commun.}, vol.~35,
  no.~10, pp. 2181--2195, 2017.

\bibitem{yu2019miso}
X.~{Yu}, D.~{Xu}, and R.~{Schober}, ``{MISO} wireless communication systems via
  intelligent reflecting surfaces,'' in \emph{Proc. IEEE Int. Conf. Commun.
  China (ICCC)}, Changchun, China, May 2019, pp. 1--6.

\bibitem{9024490}
D.~{Xu}, X.~{Yu}, Y.~{Sun}, D.~W.~K. {Ng}, and R.~{Schober}, ``Resource
  allocation for secure {IRS}-assisted multiuser {MISO} systems,'' in
  \emph{Proc. IEEE Global Commun. Conf. (GLOBECOM) Wkshps}, Waikoloa, HI, USA,
  Dec., 2019, pp. 1--6.

\bibitem{9154337}
X.~{Yu}, D.~{Xu}, and R.~{Schober}, ``Optimal beamforming for {MISO}
  communications via intelligent reflecting surfaces,'' in \emph{Proc. IEEE
  Intern. Workshop on Signal Process. Advances in Wireless Commun.}, Atlanta,
  GA, USA, May 2020, pp. 1--5.

\bibitem{8910627}
Q.~{Wu} and R.~{Zhang}, ``Towards smart and reconfigurable environment:
  Intelligent reflecting surface aided wireless network,'' \emph{IEEE Commun.
  Mag.}, vol.~58, no.~1, pp. 106--112, Jan. 2020.

\bibitem{9154252}
D.~{Xu}, X.~{Yu}, and R.~{Schober}, ``Resource allocation for intelligent
  reflecting surface-assisted cognitive radio networks,'' in \emph{Proc. IEEE
  Intern. Workshop on Signal Processing Advances in Wireless Commun. (SPAWC)},
  Atlanta, GA, USA, May 2020, pp. 1--5.

\bibitem{10134546}
G.~{Zhou}, C.~{Pan}, H.~{Ren}, D.~{Xu}, Z.~{Zhang}, J.~{Wang}, and
  R.~{Schober}, ``A framework for transmission design for active {RIS}-aided
  communication with partial {CSI},'' \emph{IEEE Trans. Wireless Commun.},
  vol.~23, no.~1, pp. 305--320, 2024.

\bibitem{10266592}
D.~{Xu}, X.~{Yu}, S.~{Song}, D.~W.~K. {Ng}, and R.~{Schober}, ``Multi-objective
  optimization for active {IRS}-assisted communication systems,'' in \emph{IEEE
  Int. Mediterr.Conf. Commun. and Netw. (MeditCom)}, Dubrovnik, Croatia, Sept.
  2023, pp. 340--345.

\bibitem{9440764}
Y.~Cheng, K.~H. Li, Y.~Liu, K.~C. Teh, and G.~K. Karagiannidis,
  ``Non-orthogonal multiple access ({NOMA}) with multiple intelligent
  reflecting surfaces,'' \emph{IEEE Trans. Wireless Commun.}, vol.~20, no.~11,
  pp. 7184--7195, Nov. 2021.

\bibitem{XidongIRS}
X.~Mu, Y.~Liu, L.~Guo, J.~Lin, and N.~Al-Dhahir, ``Capacity and optimal
  resource allocation for {IRS}-assisted multi-user communication systems,''
  \emph{IEEE Trans. Commun.}, vol.~69, no.~6, pp. 3771--3786, Jun. 2021.

\bibitem{BeixiongComml}
B.~Zheng, Q.~Wu, and R.~Zhang, ``Intelligent reflecting surface-assisted
  multiple access with user pairing: {NOMA} or {OMA}?'' vol.~24, no.~4, pp.
  753--757, Apr. 2020.

\bibitem{10243545}
W.~Ma, L.~Zhu, and R.~Zhang, ``{MIMO} capacity characterization for movable
  antenna systems,'' \emph{IEEE Trans. Wireless Commun. (Early Access)}, pp.
  1--1, 2023.

\bibitem{9264694}
K.-K. Wong, A.~Shojaeifard, K.-F. Tong, and Y.~Zhang, ``Fluid antenna
  systems,'' \emph{IEEE Trans. Wireless Commun.}, vol.~20, no.~3, pp.
  1950--1962, Mar. 2021.

\bibitem{9650760}
K.-K. Wong and K.-F. Tong, ``Fluid antenna multiple access,'' \emph{IEEE Trans.
  Wireless Commun.}, vol.~21, no.~7, pp. 4801--4815, Jul. 2022.

\bibitem{10078147}
K.-K. Wong, K.-F. Tong, Y.~Chen, Y.~Zhang, and C.-B. Chae, ``Opportunistic
  fluid antenna multiple access,'' \emph{IEEE Trans. Wireless Commun.},
  vol.~22, no.~11, pp. 7819--7833, Nov. 2023.

\bibitem{10146274}
K.-K. Wong, W.~K. New, X.~Hao, K.-F. Tong, and C.-B. Chae, ``Fluid antenna
  system—part {I}: Preliminaries,'' \emph{IEEE Commun. Lett.}, vol.~27,
  no.~8, pp. 1919--1923, Aug. 2023.

\bibitem{10349846}
D.~{Xu}, Y.~{Xu}, Z.~{Wei}, S.~{Song}, and D.~W.~K. {Ng}, ``Sensing-enhanced
  secure communication: Joint time allocation and beamforming design,'' in
  \emph{IEEE Int. Symp. Model. Optim. Mob., Ad hoc, Wireless Netw. (WiOpt)},
  Singapore, Aug. 2023, pp. 673--680.

\bibitem{10266619}
Y.~{Xu}, L.~{Xie}, D.~{Xu}, and S.~{Song}, ``Fundamental limits and base
  station selection for collaborative sensing in perceptive mobile networks,''
  in \emph{IEEE Int. Mediterr.Conf. Commun. and Netw. (MeditCom)}, Dubrovnik,
  Croatia, Sept. 2023, pp. 97--102.

\bibitem{Jinke2019JCAC}
J.~Ren, Y.~He, G.~Yu, and G.~Y. Li, ``Joint communication and computation
  resource allocation for cloud-edge collaborative system,'' in \emph{Proc.
  IEEE Wireless Commun. and Networking Conf.}, 2019, pp. 1--6.

\bibitem{Shuangyang2022ISAC}
S.~Li, W.~Yuan, C.~Liu, Z.~Wei, J.~Yuan, B.~Bai, and D.~W.~K. Ng, ``A novel
  {ISAC} transmission framework based on spatially-spread orthogonal time
  frequency space modulation,'' \emph{IEEE J. Sel. Areas Commun.}, vol.~40,
  no.~6, pp. 1854--1872, Jun. 2022.

\bibitem{chang2022learning}
C.~Liu, W.~Yuan, S.~Li, X.~Liu, H.~Li, D.~W.~K. Ng, and Y.~Li, ``Learning-based
  predictive beamforming for integrated sensing and communication in vehicular
  networks,'' \emph{IEEE J. Sel. Areas Commun.}, vol.~40, no.~8, pp.
  2317--2334, Jun. 2022.

\bibitem{Kobayashi2018Joint}
M.~Kobayashi, G.~Caire, and G.~Kramer, ``Joint state sensing and communication:
  Optimal tradeoff for a memoryless case,'' in \emph{IEEE Int. Symp. Inf.
  Theory (ISIT)}, Aug. 2018, pp. 111--115.

\bibitem{TranFL}
N.~H. Tran, W.~Bao, A.~Zomaya, M.~N.~H. Nguyen, and C.~S. Hong, ``Federated
  learning over wireless networks: Optimization model design and analysis,'' in
  \emph{IEEE INFOCOM 2019}, May 2019, pp. 1387--1395.

\bibitem{Sun2019JCAC}
S.~Mao, S.~Leng, and Y.~Zhang, ``Joint communication and computation resource
  optimization for {NOMA}-assisted mobile edge computing,'' in \emph{IEEE Int.
  Conf. Commun. (ICC)}, 2019, pp. 1--6.

\bibitem{Yifeng2023fundamental}
Y.~Xiong, F.~Liu, Y.~Cui, W.~Yuan, T.~X. Han, and G.~Caire, ``On the
  fundamental tradeoff of integrated sensing and communications under
  {Gaussian} channels,'' \emph{IEEE Trans. Inf. Theory}, vol.~69, no.~9, pp.
  5723--5751, Sept. 2023.

\bibitem{Fan2018Toward}
F.~Liu, L.~Zhou, C.~Masouros, A.~Li, W.~Luo, and A.~Petropulu, ``Toward
  dual-functional radar-communication systems: Optimal waveform design,''
  \emph{IEEE Trans. Signal Process.}, vol.~66, no.~16, pp. 4264--4279, Aug.
  2018.

\bibitem{Feng2019multiantenna}
F.~Wang, J.~Xu, and Z.~Ding, ``Multi-antenna {NOMA} for computation offloading
  in multiuser mobile edge computing systems,'' \emph{IEEE Trans. Commun.},
  vol.~67, no.~3, pp. 2450--2463, Mar. 2019.

\bibitem{Hanif2016}
M.~F. Hanif, Z.~Ding, T.~Ratnarajah, and G.~K. Karagiannidis, ``A
  minorization-maximization method for optimizing sum rate in the downlink of
  non-orthogonal multiple access systems,'' \emph{IEEE Trans. Signal Process.},
  vol.~64, no.~1, pp. 76--88, Jan. 2016.

\bibitem{7812683}
Y.~{Sun}, D.~W.~K. {Ng}, Z.~{Ding}, and R.~{Schober}, ``Optimal joint power and
  subcarrier allocation for full-duplex multicarrier non-orthogonal multiple
  access systems,'' \emph{IEEE Trans. Commun.}, vol.~65, no.~3, pp. 1077--1091,
  Mar. 2017.

\bibitem{DerrickFD2016}
D.~W.~K. Ng, Y.~Wu, and R.~Schober, ``Power efficient resource allocation for
  full-duplex radio distributed antenna networks,'' \emph{IEEE Trans. Wireless
  Commun.}, vol.~15, no.~4, pp. 2896--2911, Apr. 2016.

\bibitem{Lei2016NPM}
L.~Lei, D.~Yuan, and P.~V{\"{a}}rbrand, ``On power minimization for
  non-orthogonal multiple access {(NOMA)},'' \emph{IEEE Commun. Lett.},
  vol.~20, no.~12, pp. 2458--2461, Dec. 2016.

\bibitem{WangWorstCase}
J.~Wang and D.~P. Palomar, ``Worst-case robust {MIMO} transmission with
  imperfect channel knowledge,'' \emph{IEEE Trans. Signal Process.}, vol.~57,
  no.~8, pp. 3086--3100, Aug. 2009.

\bibitem{pereira2022review}
J.~L.~J. Pereira, G.~A. Oliver, M.~B. Francisco, S.~S. Cunha~Jr, and G.~F.
  Gomes, ``A review of multi-objective optimization: methods and algorithms in
  mechanical engineering problems,'' \emph{Archives of Computational Methods in
  Engineering}, vol.~29, no.~4, pp. 2285--2308, 2022.

\bibitem{yu2020power}
X.~{Yu}, D.~{Xu}, D.~W.~K. {Ng}, and R.~{Schober}, ``Power-efficient resource
  allocation for multiuser {MISO} systems via intelligent reflecting
  surfaces,'' in \emph{Proc. IEEE Global Commun. Conf. (GLOBECOM)}, Taipei,
  Taiwan, Dec. 2020, pp. 1--6.

\bibitem{wu2023globally}
Y.~Wu, D.~Xu, D.~W.~K. Ng, R.~Schober, and W.~Gerstacker, ``Globally optimal
  resource allocation design for irs-assisted multiuser networks with discrete
  phase shifts,'' in \emph{Proc. IEEE Intern. Commun. Conf.}, Rome, Italy, May
  2023, pp. 1--6.

\bibitem{xu2021resource}
D.~{Xu}, X.~{Yu}, V.~{Jamali}, D.~W.~K. {Ng}, and R.~{Schober}, ``Resource
  allocation for large {IRS}-assisted {SWIPT} systems with non-linear energy
  harvesting model,'' in \emph{Proc. IEEE Wireless Commun. Netw. Conf. (WCNC)},
  Nanjing, China, Mar. 2021, pp. 1--7.

\bibitem{yu2019enabling}
X.~{Yu}, D.~{Xu}, and R.~{Schober}, ``Enabling secure wireless communications
  via intelligent reflecting surfaces,'' in \emph{Proc. IEEE Global Commun.
  Conf. (GLOBECOM)}, Waikoloa, HI, USA, Dec. 2019, pp. 1--6.

\bibitem{seddon2011basic}
J.~M. {Seddon} and S.~{Newman}, \emph{Basic Helicopter Aerodynamics}.\hskip 1em
  plus 0.5em minus 0.4em\relax John Wiley \& Sons, 2011, vol.~40.

\bibitem{10207991}
D.~{Xu}, C.~{Liu}, S.~{Song}, and D.~W.~K. {Ng}, ``Integrated sensing and
  communication in coordinated cellular networks,'' in \emph{IEEE Stat. Signal
  Processing Workshop (SSP)}, Hanoi, Vietnam, Jul. 2023, pp. 90--94.

\bibitem{xu2022integrated}
D.~{Xu}, A.~{Khalili}, X.~{Yu}, D.~W.~K. {Ng}, and R.~{Schober}, ``Integrated
  sensing and communication in distributed antenna networks,'' in \emph{Proc.
  IEEE Int. Conf. Commun. Wkshps. (ICC Wkshps)}, Rome, Italy, May 2023, pp.
  1--6.

\bibitem{boyd2004convex}
S.~Boyd and L.~Vandenberghe, \emph{Convex Optimization}.\hskip 1em plus 0.5em
  minus 0.4em\relax Cambridge University Press, 2004.

\bibitem{absil2009optimization}
P.-A. {Absil} \emph{et~al.}, \emph{Optimization Algorithms on Matrix
  Manifolds}.\hskip 1em plus 0.5em minus 0.4em\relax Princeton University
  Press, 2009.

\bibitem{6698281}
U.~{Rashid}, H.~D. {Tuan}, H.~H. {Kha}, and H.~H. {Nguyen}, ``Joint
  optimization of source precoding and relay beamforming in wireless {MIMO}
  relay networks,'' \emph{IEEE Trans. Commun.}, vol.~62, no.~2, pp. 488--499,
  Feb. 2014.

\bibitem{9723093}
D.~{Xu}, X.~{Yu}, D.~W.~K. {Ng}, and R.~{Schober}, ``Resource allocation for
  active {IRS}-assisted multiuser communication systems,'' in \emph{Proc.
  Fifty-Fifth Asilomar Conf. Signals, Systems and Computers (ASILOMAR)},
  Pacific Grove, CA, USA, Oct. 2021, pp. 113--119.

\bibitem{DerrickEERobust}
D.~W.~K. Ng, E.~S. Lo, and R.~Schober, ``Robust beamforming for secure
  communication in systems with wireless information and power transfer,''
  \emph{IEEE Trans. Wireless Commun.}, vol.~13, no.~8, pp. 4599--4615, Aug.
  2014.

\bibitem{yu2020irs}
X.~{Yu}, D.~{Xu}, D.~W.~K. {Ng}, and R.~{Schober}, ``{IRS}-assisted green
  communication systems: Provable convergence and robust optimization,''
  \emph{IEEE Trans. Commun.}, vol.~69, no.~9, pp. 6313--6329, Sept. 2021.

\bibitem{xu2023joint}
Y.~Xu, D.~Xu, L.~Xie, and S.~Song, ``Joint {BS} selection, user association,
  and beamforming design for network integrated sensing and communication,'' in
  \emph{Proc. IEEE Global Commun. Conf.}, 2023, pp. 1--6.

\bibitem{lawler1966branch}
E.~L. {Lawler} and D.~E. {Wood}, ``Branch-and-bound methods: A survey,''
  \emph{Operations research}, vol.~14, no.~4, pp. 699--719, 1966.

\bibitem{9669263}
D.~{Xu}, V.~{Jamali}, X.~{Yu}, D.~W.~K. {Ng}, and R.~{Schober}, ``Optimal
  resource allocation design for large {IRS}-assisted {SWIPT} systems: A
  scalable optimization framework,'' \emph{IEEE Trans. Commun.}, vol.~70,
  no.~2, pp. 1423--1441, Feb. 2022.

\bibitem{mitsos2009mccormick}
A.~{Mitsos}, B.~{Chachuat}, and P.~I. {Barton}, ``Mccormick-based relaxations
  of algorithms,'' \emph{SIAM J. Optim.}, vol.~20, no.~2, pp. 573--601, 2009.

\bibitem{horst2013global}
R.~{Horst} and H.~{Tuy}, \emph{Global Optimization: Deterministic
  Approaches}.\hskip 1em plus 0.5em minus 0.4em\relax Springer Science \&
  Business Media, 2013.

\bibitem{tuy2000monotonic}
H.~Tuy, ``Monotonic optimization: Problems and solution approaches,''
  \emph{SIAM Journal on Optimization}, vol.~11, no.~2, pp. 464--494, 2000.

\bibitem{zhang2013monotonic}
Y.~{Zhang}, L.~{Qian}, and J.~{Huang}, ``Monotonic optimization in
  communication and networking systems,'' \emph{Foundations and
  Trends{\textregistered} in Networking}, vol.~7, no.~1, pp. 1--75, 2013.

\bibitem{bjornson2013optimal}
E.~{Bj{\"o}rnson} and E.~{Jorswieck}, ``Optimal resource allocation in
  coordinated multi-cell systems,'' \emph{Foundations and
  Trends{\textregistered} in Communications and Information Theory}, vol.~9,
  no. 2--3, pp. 113--381, 2013.

\bibitem{9423667}
X.~{Mu}, Y.~{Liu}, L.~{Guo}, J.~{Lin}, and R.~{Schober}, ``Joint deployment and
  multiple access design for intelligent reflecting surface assisted
  networks,'' \emph{IEEE Trans. Wireless Commun.}, vol.~20, no.~10, pp.
  6648--6664, Oct. 2021.

\bibitem{dinh2010local}
Q.~T. Dinh and M.~Diehl, ``Local convergence of sequential convex programming
  for nonconvex optimization,'' in \emph{Recent Advances in Optimization and
  its Applications in Engineering}.\hskip 1em plus 0.5em minus 0.4em\relax
  Springer, 2010, pp. 93--102.

\bibitem{bezdek2002some}
J.~C. {Bezdek} and R.~J. {Hathaway}, ``Some notes on alternating
  optimization,'' in \emph{AFSS Int. Conf. on Fuzzy Systems}.\hskip 1em plus
  0.5em minus 0.4em\relax Springer, 2002, pp. 288--300.

\bibitem{SunFullDuplex}
Y.~{Sun}, D.~W.~K. {Ng}, J.~{Zhu}, and R.~{Schober}, ``Robust and secure
  resource allocation for full-duplex {MISO} multicarrier {NOMA} systems,''
  \emph{IEEE Trans. Commun.}, vol.~66, no.~9, pp. 4119--4137, Sep. 2018.

\bibitem{cvx}
M.~Grant and S.~Boyd, ``{CVX}: Matlab software for disciplined convex
  programming, version 2.1,'' \url{http://cvxr.com/cvx}, Mar. 2014.

\bibitem{QinSemantic}
D.~Gündüz, Z.~Qin, I.~E. Aguerri, H.~S. Dhillon, Z.~Yang, A.~Yener, K.~K.
  Wong, and C.-B. Chae, ``Beyond transmitting bits: Context, semantics, and
  task-oriented communications,'' \emph{IEEE J. Select. Areas Commun.},
  vol.~41, no.~1, pp. 5--41, Jan. 2023.

\bibitem{JiaweiJSAC}
J.~Shao, Y.~Mao, and J.~Zhang, ``Learning task-oriented communication for edge
  inference: An information bottleneck approach,'' \emph{IEEE J. Select. Areas
  Commun.}, vol.~40, no.~1, pp. 197--211, Jan. 2022.

\bibitem{HaijunSemantic}
H.~Zhang, H.~Wang, Y.~Li, K.~Long, and A.~Nallanathan, ``{DRL}-driven dynamic
  resource allocation for task-oriented semantic communication,'' \emph{IEEE
  Trans. Commun.}, vol.~71, no.~7, pp. 3992--4004, Jul. 2023.

\bibitem{ZhuoMultiAgent}
Z.~Sun, Z.~Yu, B.~Guo, B.~Yang, Y.~Zhang, and D.~W.~K. Ng, ``Integrated sensing
  and communication for effective multi-agent cooperation systems,'' \emph{IEEE
  Commun. Mag.}, pp. 1--7, early access, 2024.

\bibitem{Djigal2022machine}
H.~Djigal, J.~Xu, L.~Liu, and Y.~Zhang, ``Machine and deep learning for
  resource allocation in multi-access edge computing: A survey,'' \emph{IEEE
  Commun. Surv. Tut.}, vol.~24, no.~4, pp. 2449--2494, Aug. 2022.

\bibitem{bolukbasi2017adaptive}
T.~Bolukbasi, J.~Wang, O.~Dekel, and V.~Saligrama, ``Adaptive neural networks
  for efficient inference,'' in \emph{Int. Conf. Machine Learning}.\hskip 1em
  plus 0.5em minus 0.4em\relax PMLR, 2017, pp. 527--536.

\bibitem{Guan2018DL}
G.~Gui, H.~Huang, Y.~Song, and H.~Sari, ``Deep learning for an effective
  nonorthogonal multiple access scheme,'' \emph{IEEE Trans. Veh. Technol.},
  vol.~67, no.~9, pp. 8440--8450, Jul. 2018.

\bibitem{Kowshik2023DL}
A.~K. Kowshik, A.~H. Raghavendra, S.~Gurugopinath, and S.~Muhaidat, ``Deep
  learning-based signal detection for rate-splitting multiple access under
  generalized {Gaussian} noise,'' \emph{IEEE Open J. Veh. Technol.}, vol.~4,
  pp. 257--270, Jan. 2023.

\bibitem{Huang2020DL}
H.~Huang, Y.~Yang, Z.~Ding, H.~Wang, H.~Sari, and F.~Adachi, ``Deep
  learning-based sum data rate and energy efficiency optimization for
  {MIMO-NOMA} systems,'' \emph{IEEE Trans. Wireless Commun.}, vol.~19, no.~8,
  pp. 5373--5388, May 2020.

\bibitem{Liang2018reinforcement}
L.~Xiao, Y.~Li, C.~Dai, H.~Dai, and H.~V. Poor, ``Reinforcement learning-based
  {NOMA} power allocation in the presence of smart jamming,'' \emph{IEEE Trans.
  Veh. Technol.}, vol.~67, no.~4, pp. 3377--3389, Dec. 2018.

\bibitem{Minghui2023DL}
M.~Wu, Z.~Gao, Y.~Huang, Z.~Xiao, D.~W.~K. Ng, and Z.~Zhang, ``Deep
  learning-based rate-splitting multiple access for reconfigurable intelligent
  surface-aided tera-hertz massive mimo,'' \emph{IEEE J. Sel. Areas Commun.},
  vol.~41, no.~5, pp. 1431--1451, Jan. 2023.

\bibitem{Chang2021location}
C.~Liu, W.~Yuan, Z.~Wei, X.~Liu, and D.~W.~K. Ng, ``Location-aware predictive
  beamforming for {UAV} communications: A deep learning approach,'' \emph{IEEE
  Wireless Commun. Lett.}, vol.~10, no.~3, pp. 668--672, Dec. 2021.

\bibitem{Chen2022NOM}
J.~Chen and Y.~Liu, ``Neural optimization machine: A neural network approach
  for optimization,'' \emph{arXiv preprint arXiv:2208.03897}, 2022.

\bibitem{ChangDistributed}
Z.~Chang, Z.~Wang, X.~Guo, C.~Yang, Z.~Han, and T.~Ristaniemi, ``Distributed
  resource allocation for energy efficiency in {OFDMA} multicell networks with
  wireless power transfer,'' \emph{IEEE J. Select. Areas Commun.}, vol.~37,
  no.~2, pp. 345--356, Feb. 2019.

\bibitem{TengjiaoGameTheory}
T.~He, K.-W. Chin, S.~Soh, and Z.~Zhang, ``A novel distributed resource
  allocation scheme for wireless-powered cognitive radio internet of things
  networks,'' \emph{IEEE Internet Things J.}, vol.~8, no.~20, pp.
  15\,486--15\,499, Oct. 2021.

\bibitem{HaoDRL}
H.~Ye, G.~Y. Li, and B.-H.~F. Juang, ``Deep reinforcement learning based
  resource allocation for {V2V} communications,'' \emph{IEEE Trans. Veh.
  Technol.}, vol.~68, no.~4, pp. 3163--3173, Apr. 2019.

\bibitem{YuanweiUAV}
J.~Cui, Y.~Liu, and A.~Nallanathan, ``Multi-agent reinforcement learning-based
  resource allocation for {UAV} networks,'' \emph{IEEE Trans. Wireless
  Commun.}, vol.~19, no.~2, pp. 729--743, Feb. 2020.

\end{thebibliography}

\end{document}